\RequirePackage{graphicx}
\RequirePackage{xspace}

\documentclass[%
 superscriptaddress,
preprint,
showpacs,preprintnumbers,
nofootinbib,
 amsmath,amssymb,
aps,
prd,
floatfix,
]{revtex4-1}

\usepackage{color} 

\usepackage{amsmath} 

\begin{document}

\newcommand{\cczeropi}{\ensuremath{\text{CC}0\pi}\xspace}
\newcommand{\cconepi}{\ensuremath{\text{CC}1\pi}\xspace}
\newcommand{\ccother}{\ensuremath{\text{CC}_\text{other}}\xspace}
\newcommand{\nc}{\ensuremath{\text{NC}}\xspace}
\newcommand{\oofv}{\ensuremath{\text{OOFV}}\xspace}


\title{Measurement of double-differential muon neutrino charged-current interactions on C$_8$H$_8$ without pions in the final state using the T2K off-axis beam}


\newcommand{\INSTEE}{\affiliation{University of Bern, Albert Einstein Center for Fundamental Physics, Laboratory for High Energy Physics (LHEP), Bern, Switzerland}}
\newcommand{\INSTFE}{\affiliation{Boston University, Department of Physics, Boston, Massachusetts, U.S.A.}}
\newcommand{\INSTD}{\affiliation{University of British Columbia, Department of Physics and Astronomy, Vancouver, British Columbia, Canada}}
\newcommand{\INSTGA}{\affiliation{University of California, Irvine, Department of Physics and Astronomy, Irvine, California, U.S.A.}}
\newcommand{\INSTI}{\affiliation{IRFU, CEA Saclay, Gif-sur-Yvette, France}}
\newcommand{\INSTGB}{\affiliation{University of Colorado at Boulder, Department of Physics, Boulder, Colorado, U.S.A.}}
\newcommand{\INSTFG}{\affiliation{Colorado State University, Department of Physics, Fort Collins, Colorado, U.S.A.}}
\newcommand{\INSTFH}{\affiliation{Duke University, Department of Physics, Durham, North Carolina, U.S.A.}}
\newcommand{\INSTBA}{\affiliation{Ecole Polytechnique, IN2P3-CNRS, Laboratoire Leprince-Ringuet, Palaiseau, France }}
\newcommand{\INSTEF}{\affiliation{ETH Zurich, Institute for Particle Physics, Zurich, Switzerland}}
\newcommand{\INSTEG}{\affiliation{University of Geneva, Section de Physique, DPNC, Geneva, Switzerland}}
\newcommand{\INSTDG}{\affiliation{H. Niewodniczanski Institute of Nuclear Physics PAN, Cracow, Poland}}
\newcommand{\INSTCB}{\affiliation{High Energy Accelerator Research Organization (KEK), Tsukuba, Ibaraki, Japan}}
\newcommand{\INSTED}{\affiliation{Institut de Fisica d'Altes Energies (IFAE), The Barcelona Institute of Science and Technology, Campus UAB, Bellaterra (Barcelona) Spain}}
\newcommand{\INSTEC}{\affiliation{IFIC (CSIC \& University of Valencia), Valencia, Spain}}
\newcommand{\INSTEI}{\affiliation{Imperial College London, Department of Physics, London, United Kingdom}}
\newcommand{\INSTGF}{\affiliation{INFN Sezione di Bari and Universit\`a e Politecnico di Bari, Dipartimento Interuniversitario di Fisica, Bari, Italy}}
\newcommand{\INSTBE}{\affiliation{INFN Sezione di Napoli and Universit\`a di Napoli, Dipartimento di Fisica, Napoli, Italy}}
\newcommand{\INSTBF}{\affiliation{INFN Sezione di Padova and Universit\`a di Padova, Dipartimento di Fisica, Padova, Italy}}
\newcommand{\INSTBD}{\affiliation{INFN Sezione di Roma and Universit\`a di Roma ``La Sapienza'', Roma, Italy}}
\newcommand{\INSTEB}{\affiliation{Institute for Nuclear Research of the Russian Academy of Sciences, Moscow, Russia}}
\newcommand{\INSTHA}{\affiliation{Kavli Institute for the Physics and Mathematics of the Universe (WPI), The University of Tokyo Institutes for Advanced Study, University of Tokyo, Kashiwa, Chiba, Japan}}
\newcommand{\INSTCC}{\affiliation{Kobe University, Kobe, Japan}}
\newcommand{\INSTCD}{\affiliation{Kyoto University, Department of Physics, Kyoto, Japan}}
\newcommand{\INSTEJ}{\affiliation{Lancaster University, Physics Department, Lancaster, United Kingdom}}
\newcommand{\INSTFC}{\affiliation{University of Liverpool, Department of Physics, Liverpool, United Kingdom}}
\newcommand{\INSTFI}{\affiliation{Louisiana State University, Department of Physics and Astronomy, Baton Rouge, Louisiana, U.S.A.}}
\newcommand{\INSTJ}{\affiliation{Universit\'e de Lyon, Universit\'e Claude Bernard Lyon 1, IPN Lyon (IN2P3), Villeurbanne, France}}
\newcommand{\INSTHB}{\affiliation{Michigan State University, Department of Physics and Astronomy,  East Lansing, Michigan, U.S.A.}}
\newcommand{\INSTCE}{\affiliation{Miyagi University of Education, Department of Physics, Sendai, Japan}}
\newcommand{\INSTDF}{\affiliation{National Centre for Nuclear Research, Warsaw, Poland}}
\newcommand{\INSTFJ}{\affiliation{State University of New York at Stony Brook, Department of Physics and Astronomy, Stony Brook, New York, U.S.A.}}
\newcommand{\INSTGJ}{\affiliation{Okayama University, Department of Physics, Okayama, Japan}}
\newcommand{\INSTCF}{\affiliation{Osaka City University, Department of Physics, Osaka, Japan}}
\newcommand{\INSTGG}{\affiliation{Oxford University, Department of Physics, Oxford, United Kingdom}}
\newcommand{\INSTBB}{\affiliation{UPMC, Universit\'e Paris Diderot, CNRS/IN2P3, Laboratoire de Physique Nucl\'eaire et de Hautes Energies (LPNHE), Paris, France}}
\newcommand{\INSTGC}{\affiliation{University of Pittsburgh, Department of Physics and Astronomy, Pittsburgh, Pennsylvania, U.S.A.}}
\newcommand{\INSTFA}{\affiliation{Queen Mary University of London, School of Physics and Astronomy, London, United Kingdom}}
\newcommand{\INSTE}{\affiliation{University of Regina, Department of Physics, Regina, Saskatchewan, Canada}}
\newcommand{\INSTGD}{\affiliation{University of Rochester, Department of Physics and Astronomy, Rochester, New York, U.S.A.}}
\newcommand{\INSTHC}{\affiliation{Royal Holloway University of London, Department of Physics, Egham, Surrey, United Kingdom}}
\newcommand{\INSTBC}{\affiliation{RWTH Aachen University, III. Physikalisches Institut, Aachen, Germany}}
\newcommand{\INSTFB}{\affiliation{University of Sheffield, Department of Physics and Astronomy, Sheffield, United Kingdom}}
\newcommand{\INSTDI}{\affiliation{University of Silesia, Institute of Physics, Katowice, Poland}}
\newcommand{\INSTEH}{\affiliation{STFC, Rutherford Appleton Laboratory, Harwell Oxford,  and  Daresbury Laboratory, Warrington, United Kingdom}}
\newcommand{\INSTCH}{\affiliation{University of Tokyo, Department of Physics, Tokyo, Japan}}
\newcommand{\INSTBJ}{\affiliation{University of Tokyo, Institute for Cosmic Ray Research, Kamioka Observatory, Kamioka, Japan}}
\newcommand{\INSTCG}{\affiliation{University of Tokyo, Institute for Cosmic Ray Research, Research Center for Cosmic Neutrinos, Kashiwa, Japan}}
\newcommand{\INSTGI}{\affiliation{Tokyo Metropolitan University, Department of Physics, Tokyo, Japan}}
\newcommand{\INSTF}{\affiliation{University of Toronto, Department of Physics, Toronto, Ontario, Canada}}
\newcommand{\INSTB}{\affiliation{TRIUMF, Vancouver, British Columbia, Canada}}
\newcommand{\INSTG}{\affiliation{University of Victoria, Department of Physics and Astronomy, Victoria, British Columbia, Canada}}
\newcommand{\INSTDJ}{\affiliation{University of Warsaw, Faculty of Physics, Warsaw, Poland}}
\newcommand{\INSTDH}{\affiliation{Warsaw University of Technology, Institute of Radioelectronics, Warsaw, Poland}}
\newcommand{\INSTFD}{\affiliation{University of Warwick, Department of Physics, Coventry, United Kingdom}}
\newcommand{\INSTGE}{\affiliation{University of Washington, Department of Physics, Seattle, Washington, U.S.A.}}
\newcommand{\INSTGH}{\affiliation{University of Winnipeg, Department of Physics, Winnipeg, Manitoba, Canada}}
\newcommand{\INSTEA}{\affiliation{Wroclaw University, Faculty of Physics and Astronomy, Wroclaw, Poland}}
\newcommand{\INSTH}{\affiliation{York University, Department of Physics and Astronomy, Toronto, Ontario, Canada}}

\INSTEE
\INSTFE
\INSTD
\INSTGA
\INSTI
\INSTGB
\INSTFG
\INSTFH
\INSTBA
\INSTEF
\INSTEG
\INSTDG
\INSTCB
\INSTED
\INSTEC
\INSTEI
\INSTGF
\INSTBE
\INSTBF
\INSTBD
\INSTEB
\INSTHA
\INSTCC
\INSTCD
\INSTEJ
\INSTFC
\INSTFI
\INSTJ
\INSTHB
\INSTCE
\INSTDF
\INSTFJ
\INSTGJ
\INSTCF
\INSTGG
\INSTBB
\INSTGC
\INSTFA
\INSTE
\INSTGD
\INSTHC
\INSTBC
\INSTFB
\INSTDI
\INSTEH
\INSTCH
\INSTBJ
\INSTCG
\INSTGI
\INSTF
\INSTB
\INSTG
\INSTDJ
\INSTDH
\INSTFD
\INSTGE
\INSTGH
\INSTEA
\INSTH

\author{K.\,Abe}\INSTBJ
\author{C.\,Andreopoulos}\INSTEH\INSTFC
\author{M.\,Antonova}\INSTEB
\author{S.\,Aoki}\INSTCC
\author{A.\,Ariga}\INSTEE
\author{S.\,Assylbekov}\INSTFG
\author{D.\,Autiero}\INSTJ
\author{M.\,Barbi}\INSTE
\author{G.J.\,Barker}\INSTFD
\author{G.\,Barr}\INSTGG
% ...

\collaboration{The T2K Collaboration}\noaffiliation





\date{\today}

\begin{abstract}
We report the measurement of muon neutrino charged-current interactions on carbon without pions in the final state
at the T2K beam energy using 5.734$\times10^{20}$ protons on target. For the first time the measurement
is reported as a flux-integrated, double-differential cross-section in muon kinematic variables ($\cos\theta_\mu$, $p_\mu$), 
without correcting for events where a pion is produced and then absorbed by final state interactions.
Two analyses are performed with different selections, background evaluations and cross-section extraction
methods to demonstrate the robustness of the results against biases due to model-dependent assumptions.
The measurements compare favorably with recent models which include nucleon-nucleon correlations
but, given the present precision, the measurement does not solve the degeneracy between different models. The data 
also agree with Monte Carlo simulations which use effective parameters that are tuned to external data to describe the nuclear effects.
The total cross-section in the full phase space is $\sigma = (0.417 \pm 0.047 \text{(syst)} \pm 0.005 \text{(stat)})\times 10^{-38} \text{cm}^2$~$\text{nucleon}^{-1}$ and the cross-section integrated in the region of phase space with largest efficiency and 
best signal-over-background ratio ($\cos\theta_\mu>0.6$ and $p_\mu > 200$ MeV) is
$\sigma = (0.202 \pm 0.0359 \text{(syst)} \pm 0.0026 \text{(stat)}) \times 10^{-38} \text{cm}^2$ ~$\text{nucleon}^{-1}$.
\end{abstract}

\pacs{13.15.+g,25.30.Pt}

\maketitle


\section{Introduction\label{sec:intro}}

Accelerator-driven neutrino oscillation measurements~\cite{PhysRevD.74.072003,Abe:2015awa,Aguilar-Arevalo:2013pmq}
 make use of neutrino beams with energies of a few GeV or lower, at which one of the main interaction processes of neutrinos with nuclei is the charged current quasi-elastic scattering (CCQE) process.
Muon neutrinos ($\nu_\mu$) interact with a bound nucleon $N$ to produce a muon and final state nucleon $N^{\prime}$ through the exchange of a W boson ($\nu_{\mu} + N \rightarrow \mu^- + N^{\prime}$). This interaction is exploited in long-baseline neutrino oscillation experiments for the signal events with which to measure the neutrino appearance and disappearance probabilities as a function of neutrino energy. It is therefore of primary importance for the CCQE process to be well-modeled.

Over the past ten years, a complicated experimental and theoretical picture has emerged regarding CCQE interactions on nuclear targets.
The K2K experiment noted that the outgoing kinematics of the muon were not consistent with the prediction of 
a neutrino interaction on a single nucleon (1p1h) in a relativistic Fermi Gas (RFG) nuclear model~\cite{Smith:1972xh,Smith:1975}.
The nucleon axial mass was found to be $1.014 \pm 0.014$~GeV from neutrino scattering data on deuterium as well as pion electroproduction data~\cite{Kuzmin:2007kr,Bodek:2007ym} but K2K measured $M_\text{A}^\text{QE}$ to be $1.20 \pm 0.12$~GeV for interactions on a water target~\cite{Gran:2006jn}.
MiniBooNE also reported a similar anomaly on mineral oil (CH$_2$), with a large dataset of neutrinos ($M_\text{A}^\text{QE}$ of $1.35\pm 0.17$~GeV~\cite{AguilarArevalo:2010zc}) and of anti-neutrinos~\cite{Aguilar-Arevalo:2013dva}, and MINOS using iron as a target~\cite{Adamson:2014pgc}. Both experiments also noted a discrepancy at the lowest values of momentum transfer ($Q^2 <0.2$~GeV$^2$).
SciBooNE reported similar results in~\cite{AlcarazAunion:2009ku}. The previous T2K off-axis CCQE measurement~\cite{Abe:2014iza} is also in agreement with a large $M_\text{A}^\text{QE}$. The T2K on-axis measurement~\cite{Abe:2015oar} has large systematics uncertainties and is compatible with different values of  $M_\text{A}^\text{QE}$, depending on whether only one muon track is reconstructed, or both the muon and proton tracks.
Other datasets on nuclear targets from the NOMAD (carbon target) and MINERvA (hydrocarbon target) experiments are in agreement with an $M_\text{A}^\text{QE}$ of 1~GeV. At beam energies of 3--100 GeV, the NOMAD experiment has reported an $M_\text{A}^\text{QE}$ of $1.05\pm 0.06$~GeV~\cite{Lyubushkin:2008pe}. 
The MINER$\nu$A ($E_\nu \sim 3.5~\textrm{GeV}$) experiment has also measured the CCQE cross-section
with only a muon and a proton in the final state~\cite{Walton:2014esl} which is consistent with the RFG model and $M_\text{A}^\text{QE} \sim 1$~GeV. %
Interestingly, previous MINER$\nu$A CCQE measurements which use muon information and the calorimetric 
recoil energy with both neutrinos~\cite{Fiorentini:2013ezn} and anti-neutrinos~\cite{Fields:2013zhk} prefer a transverse enhancement model, suggesting the presence of meson exchange currents~\cite{Bodek:2011ps}. 
The measurements of the neutrino interaction rate are sensitive to the convolution of neutrino cross-section and flux. 
MiniBooNE and MINER$\nu$A are working to improve
the flux modeling and thus apply a more precise flux correction for the previously mentioned cross-section measurements.
The anomalies measured in neutrino interactions by modern experiments, using relatively heavy nuclei as targets, may be explained by the contribution of nuclear effects. These were not needed in the models tuned on bubble-chamber data on deuterium. Various different implementations of such nuclear effects have been proposed~\cite{Martini:2009, Nieves:2012,sf,ankowski_SF,butkevich_2009,eff-sf,leitner_2009,madrid_2003,meucci_2003,lovato_2013,pandey_2014}. In experimental measurements the effects of nucleon and hadron initial and final state interactions with the nucleus cannot typically be disentangled from the fundamental neutrino-interaction cross-section. The phenomenological interpretation of modern measurements is therefore particularly complicated.
Given the discrepancies between the available predictions or their incompleteness in the description of such nuclear effects, it is particularly important to provide experimental measurements which are, as much as possible, model-independent in order to reduce the modeling systematic uncertainties and to produce results that are useful for comparison with all the present and future models.

The Tokai-to-Kamioka (T2K) experiment has a suite of neutrino detectors placed in a neutrino beam with energy peaked at $E_\nu = 0.6$~GeV. %
This paper describes the measurement of the CCQE-like neutrino interaction cross-section, by selecting events without pions in the final sample, with plastic scintillator (C$_8$H$_8$) as the target material using the ND280 off-axis near detector in the T2K beam. Particular care has been taken to avoid model-dependent corrections to the data. Two analyses, which follow different approaches to measure the  double-differential cross-sections as a function of muon momentum and angle, are presented.

\section{Experimental apparatus\label{sec:expt}}
The T2K long-baseline neutrino oscillation experiment~\cite{Abe2011106} uses a beam of
muon neutrinos to study the appearance of electron neutrinos ($\nu_{\mu} \rightarrow \nu_{e}$) and to measure or constrain the PMNS mixing angles $\theta_{13}$ and $\theta_{23}$, the mass splitting $|\Delta m^{2}_{32}|$ and the CP-violating phase $\delta_{CP}$. The neutrinos are produced at the Japan Proton Accelerator Research Complex (J-PARC) in Ibaraki prefecture on the eastern coast of Japan, and travel 295 km through the Earth before reaching the far detector, Super-Kamiokande~\cite{Fukuda:2002uc}, in Gifu prefecture. A complex of near detectors located 280~meters from the proton beam target
is used to characterize the neutrino beam before oscillation, reducing the systematic uncertainties, and to study neutrino interactions as in the measurement reported here. 

\subsection{Neutrino beam}
\label{beam}
T2K uses a conventional neutrino beam, in which the muon neutrinos are produced by the decay of charged pions and kaons. Protons are first accelerated to 30 GeV by a sequence of three accelerators in J-PARC, then extracted to the neutrino beamline where they are directed onto a graphite target. The resulting collisions produce hadrons---predominantly charged pions---which travel inside a 96~m-long decay volume where they decay in flight into muons and muon neutrinos. A set of three magnetic horns is used to focus the positively-charged hadrons and defocus the negative particles, enhancing the neutrino component of the beam while reducing its contamination by anti-neutrinos.
At the end of the decay volume, a beam dump stops the muons and remaining hadrons, leaving an almost pure muon neutrino beam with an intrinsic electron neutrino component of the order of a percent, which comes from the decays of kaons and muons. The beam stability and direction are monitored by the Muon Monitor~\cite{Matsuoka2010591,Suzuki:2014jyd} which measures the muons of energies higher than about 5~GeV that are able to penetrate the beam dump, and also by INGRID, the on-axis near detector, which samples the neutrino beam 280~m from the proton beam target~\cite{Abe2012}.

\subsection{Off-axis near detector ND280}
\begin{figure}
\begin{center}
 \includegraphics[width=7cm]{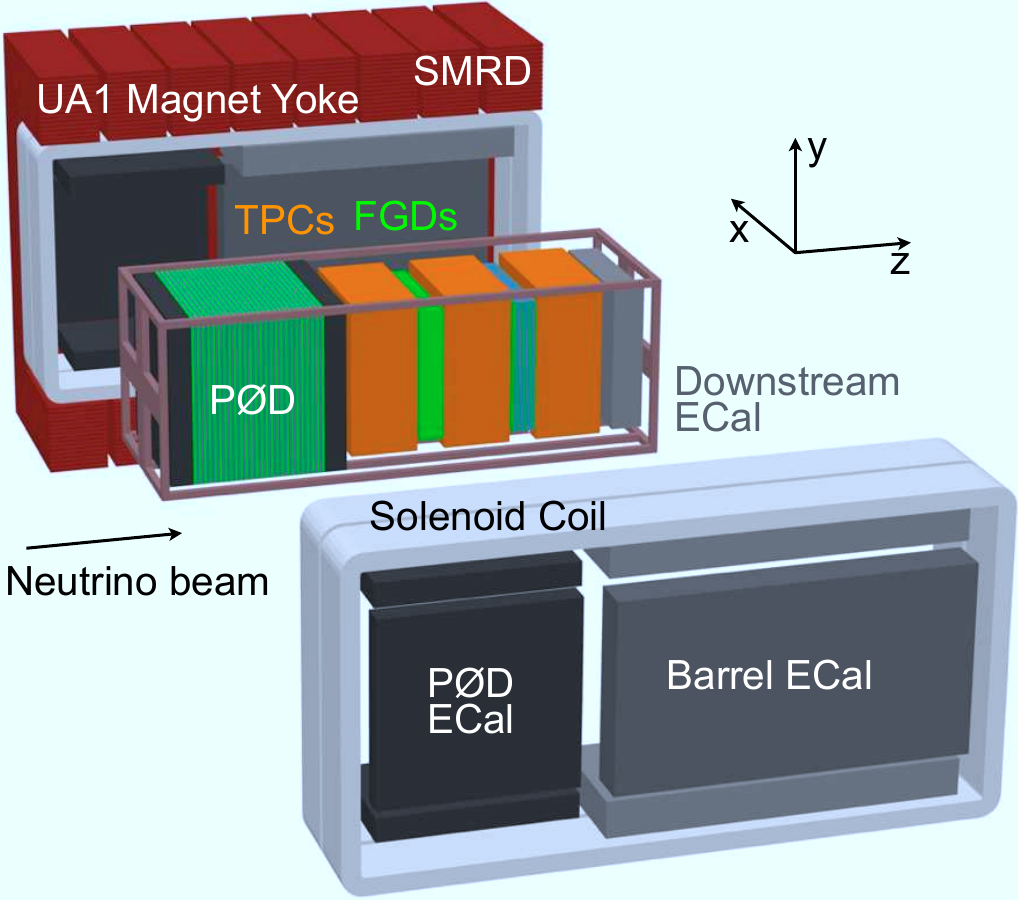}
\end{center}
\caption{Exploded view of the ND280 off-axis detector from Ref.~\cite{Abe2011106}. 
Only one half of the UA1 Magnet Yoke is shown in this figure.
}
\label{fig:ND280}
\end{figure}

The ND280 is composed of a series of sub-detectors located 280~m from the target in a direction making a 2.5$^{\circ}$ angle with the average neutrino beam direction and placed within the refurbished UA1/NOMAD magnet which generates a 0.2T magnetic field.
The neutrino beam first passes through the Pi-Zero Detector (P0D)~\cite{Assylbekov201248} and then the Tracker detector, which is used for the present measurement. The Tracker is made up of two Fine Grained Detectors (FGD)~\cite{Amaudruz:2012pe} and three Time Projection Chambers (TPC)~\cite{Abgrall:2010hi}.  Those detectors are surrounded by electromagnetic calorimeters (ECals)~\cite{Allan:2013ofa} and side muon range detectors (SMRDs)~\cite{Aoki:2012mf},
as can be seen in Fig.~\ref{fig:ND280}. 
The ND280 reference system is also shown in Fig.~\ref{fig:ND280}: the muon angle is defined as the polar angle ($\theta$)
between the muon momentum and the $z$ axis (which corresponds, to a good approximation, to the beam direction and thus to
the average neutrino direction). In general, most of the muons are expected to be `forward' ($\theta<90^\circ$), i.e. to move
in the same direction as the incoming neutrino beam, while events with `backward' muons ($\theta>90^\circ$) correspond typically to interactions
with high transferred $Q^2$.

In the Tracker, the target mass is provided by the FGDs. The first FGD (FGD1) is made only of scintillator bars, while the second FGD (FGD2) is made of alternating layers of scintillator bars and water. To measure cross-sections on carbon, neutrino interactions occurring in FGD1 are selected. 
The FGD1 fiducial volume has an elemental composition of 86.1\% carbon and 7.35\% hydrogen with remaining contributions 
from oxygen (3.70\%) and small quantities of other elements (Ti, Si, N)~\cite{Amaudruz:2012pe,Abe:2013jth}.  While the neutrino target for the present result is FGD1 with its complicated elemental composition, the measurement can be effectively considered as on scintillator (C$_8$H$_8$); the difference to the analysis from subtracting these small components is expected to be less than 1\%. The number of neutrons (nucleons) in the fiducial volume is $2.75 \times 10^{29}$ ($5.50 \times 10^{29}$).

\section{Analysis Strategy\label{sec:analysis}}

The measurement of the $\nu_\mu$ double-differential flux-integrated cross-section of the 
charged current process on carbon without pions in the final state (\cczeropi) 
is presented as a function of the muon momentum and angle. %
Two analyses have been performed which make use of different selections and
different cross-section extraction methods. The consistency between the results 
of the two analyses is an indication of the robustness of the measurement.
Particular care has been taken to perform a measurement that is highly model-independent:
\begin{itemize}
\item cross-sections are measured as a function of the kinematics of the outgoing muon, as opposed to reconstructed variables that relate 
to the neutrino, such as the neutrino energy or the transferred momentum ($Q^2$), which would depend on assumptions made
about the nuclear model; 
\item a flux-integrated cross-section is extracted, rather than a flux-averaged or flux-unfolded cross-section, thus avoiding $E_\nu$-dependent flux corrections;
\item the signal is defined in terms of the particles which exit the nucleus and can be observed in the detector. Compared to a signal defined in terms of interactions at the nucleon level, this removes the dependence on the modeling of the re-interactions of the final state particles in the nuclear medium. The definition used here includes CCQE interactions, but also events where, for example, one pion is produced at the interaction point and then reabsorbed in the nuclear environment;%
\item the cross-section measurement is designed to be robust to background-modeling uncertainties through the use of control
samples or a reduced phase space, thereby removing regions with a small signal-to-background ratio. 
\end{itemize}
The first analysis (Analysis I in the following) uses a dedicated selection for CCQE-like events where a single muon (with a proton above or below detection threshold) is required and no other tracks. The cross-section is extracted through a binned likelihood fit.
The second analysis (Analysis II in the following) follows the T2K oscillation analysis and MiniBooNE selection strategy, where CCQE-like interactions are identified by vetoing the presence of pions in the final state and Bayesian unfolding is used to correct for
detector effects.
\subsection{Event samples and simulation} 
\subsubsection{Data samples}\label{sec:realdatasets}
The analyses presented here use data from the three T2K run periods between November 2010 and May 2013,
where T2K was operating with a beam of mostly muon neutrinos.
Only data recorded with all detectors correctly working are used, corresponding to $5.734 \times 10^{20}$ protons on target (POT).

\subsubsection{Monte Carlo samples}\label{sec:mc}
\par In order to correct for the detector response, acceptance and efficiency, simulations have been produced which correspond to ten times the data POT used,
where the specific detector and beam configuration during each data run was modeled. 
The flux of neutrinos reaching the detectors---assuming the absence of oscillations---is predicted using simulations tuned to external measurements. Details of the simulation can be found in Ref.~\cite{PhysRevD.87.012001}. Interactions of protons in the graphite target and the resulting hadron production are simulated using the FLUKA 2008 package~\cite{Ferrari:2005zk,Battistoni:2007zzb}, weighted to match measurements of hadron production~\cite{PhysRevC.85.035210,eichten,allaby,PhysRevC.77.015209,Abgrall:2015hmv}. The propagation and decay of those hadrons is performed in a GEANT3~\cite{GEANT3} simulation which uses the GCALOR package~\cite{GCALOR} to model hadron re-interactions and decays outside the target. Uncertainties on the proton beam properties, horn current, hadron production model and alignment are taken into account to produce an energy-dependent systematic uncertainty on the neutrino flux. Flux tuning using NA61/SHINE data~\cite{PhysRevC.84.034604,PhysRevC.85.035210,Abgrall:2015hmv} reduces the uncertainty on the overall normalization of the integrated flux to 8.5\%.

\par Neutrinos are then propagated through the ND280 detector and interactions are simulated with the NEUT neutrino event generator.
NEUT 5.1.4.2 \cite{Hayato:2002sd, Hayato:2009} uses the Llewellyn-Smith CCQE neutrino-nucleon cross-section formalism~\cite{llewellyn-smith} with the relativistic Fermi gas (RFG) model by Smith and Moniz~\cite{smith-moniz} as the nuclear model. Dipole forms were used for both the axial-vector and vector form factors. From tuning to Super-Kamiokande atmospheric data and K2K data, the nominal axial mass $M_\text{A}^\text{QE}$ was set to 1.21 GeV. 
Neutrino-induced pion production is simulated based on the Rein Sehgal model~\cite{rein-sehgal} in NEUT with the axial mass $M_\text{A}^\text{RES} = 1.21$~GeV. The parton distribution function GRV98~\cite{Gluck:1998xa} with corrections by Bodek and Yang~\cite{Bodek:2003wd} is used for the deep inelastic scattering interactions.
Secondary interactions of pions inside the nucleus (so-called final state interactions (FSIs)) are simulated using an intranuclear cascade model based on the method of Oset~\cite{Salcedo:1987md}, tuned to external $\pi$-$^{12}$C data.
\par
The GENIE neutrino generator v2.6.4 \cite{Andreopoulos:2009rq} has been used as an alternative simulation to test the dependence of the analyses on the assumed signal and background models, with the primary difference to NEUT arising from different values of $M_\text{A}^\text{QE}$=0.99~GeV~\cite{Kuzmin:2007kr} and $M_\text{A}^\text{RES}$=1.12~GeV~\cite{Kuzmin:2006dh}. 

\subsubsection{Event pre-selection}\label{sec:preselection}
\par The FGD1 detector is used as the target for the neutrino interactions, and particles
are reconstructed in the FGD1 itself and in TPC2, which is situated immediately downstream from FGD1. 
Initially, a $\nu_{\mu}$ charged current selection is performed by looking for a muon candidate. The further event selection
depends on the analysis strategy and will be explained in the sections below.
The common pre-selection criteria are:

\begin{description}
  \item [Event quality] only good beam spills are used, and every ND280 sub-detector must have been functioning correctly at the time.
  \item [Bunching] tracks are identified as belonging to a specific beam bunch, based on hit timing.
  \item [TPC track quality] TPC tracks with good reconstruction quality are required. %
  \item [Muon candidate search] the muon candidate is identified as the highest-momentum negatively-charged track which passes the TPC track quality cut and starts in the FGD fiducial volume (FV). 
  \item [PID] the muon candidate is required to have a muon-like particle identification (PID) based on d$E$/d$x$ measurements in the TPC. 
  \item [Entering backgrounds cut] further cuts are applied to remove events where the interaction happens outside the FV
but the muon track has been mis-reconstructed as two tracks, one of which starts inside the FV.
 \end{description}

After the full \cczeropi selection is applied,
the background comes from events with one, or a number of, true
pions which are misidentified or not reconstructed (\cconepi and \ccother), neutral
current interactions (\nc) and interactions that occurred outside of the fiducial volume but were 
reconstructed inside (\oofv).

\subsubsection{Control samples for detector systematics}\label{sec:controlsamples}
The detector systematics are described in detail in~\cite{Abe:2015awa}.
The systematics on track efficiency in the FGD and TPC, particle identification,
charge identification and momentum scale and resolution were evaluated by dedicated
data and simulation comparisons using independent control samples (specially-selected event samples that are designed to be sensitive to specific sources of uncertainty). %
In addition to reconstruction-related detector uncertainties, we also 
estimated the uncertainties on the number of simultaneous events (pileup) and OOFV events.
Pions and nucleons from initial neutrino interactions
in FGD1 can re-interact and be absorbed further in the detector. While the intranuclear final
state interactions are simulated by the NEUT generator, the secondary interactions (SI) which follow these
from pions and nucleons are treated by Geant4.9.4 \cite{Geant4}. This additional uncertainty has also been evaluated
from a control sample and is one of the dominant detector systematics. 

\subsubsection{Uncertainties due to neutrino interaction model}
Uncertainties on the neutrino interaction model are described in detail in Ref.~\cite{Abe:2015awa}. 
A set of systematic parameters characterizes the uncertainties on the predictions of the NEUT generator and are propagated through the analyses to estimate the uncertainty on the background and signal modeling, as well as the effect of the final state interactions. A number of those parameters are normalization uncertainties for the different interaction modes simulated by NEUT (energy dependent for the dominant modes at the T2K neutrino energy spectrum). Other parameters describe uncertainties on the values of the axial mass (using separate parameters for CCQE and resonant interactions), of the binding energy and of the Fermi momentum. An additional systematic parameter covers the difference between the predictions obtained with the default relativistic Fermi gas model used by NEUT and a spectral function describing the momentum and energy of nucleons inside the nucleus \cite{Benhar:1995}.

Finally, for the analysis using information from the presence or absence of a reconstructed proton to separate events between different categories (Analysis I, described below), the effects of the re-interactions of produced protons in the nuclear medium are evaluated using the GENIE neutrino interaction generator \cite{Andreopoulos:2009rq} by varying the parameters of the intranuclear cascade model describing those final state interactions.

\subsection{Analysis I}
This analysis uses a binned likelihood fit performed simultaneously in four signal regions
and two control regions to constrain the backgrounds caused by resonant pion
production and deep inelastic scattering (DIS).

The signal includes events where the muon is reconstructed in the FGD but does not reach the TPC,
thus increasing the efficiency for muons that have small momentum or are emitted at high angle.
Signal events with and without a reconstructed proton are treated separately, thus allowing the proton FSI parameters to be constrained using data. This is a first step towards a future differential measurement as a function of the proton kinematics.

The normalization and the shape of the background are extracted from data:
the various background processes are parametrized in the same way as for the T2K oscillation analysis~\cite{Abe:2015awa} 
and the values of such parameters are constrained by simultaneously fitting the control regions.

A likelihood fit is performed to the number of \cczeropi events, in bins of muon momentum and angle.
Detector, flux and model uncertainties are included as nuisance parameters and
a penalty term is added to the likelihood to constrain detector and model uncertainties. 
All the systematics are evaluated with `toy' Monte Carlo experiments
sampling over the values of the nuisance parameters, as described in detail in Section~\ref{sec:systematicsI}.

\subsubsection{Event selection}

\begin{figure}
\begin{center}
 \includegraphics[width=7cm]{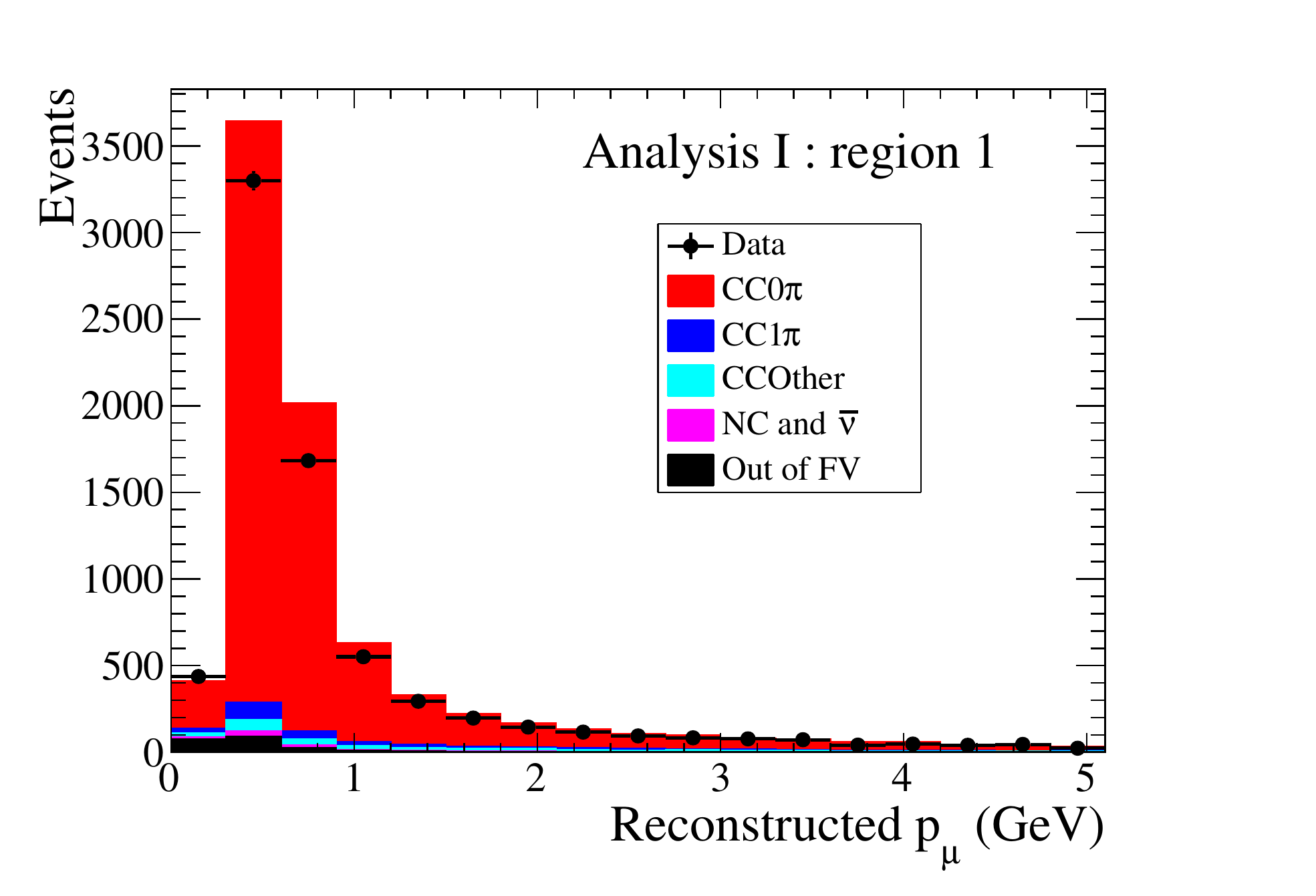}
 \includegraphics[width=7cm]{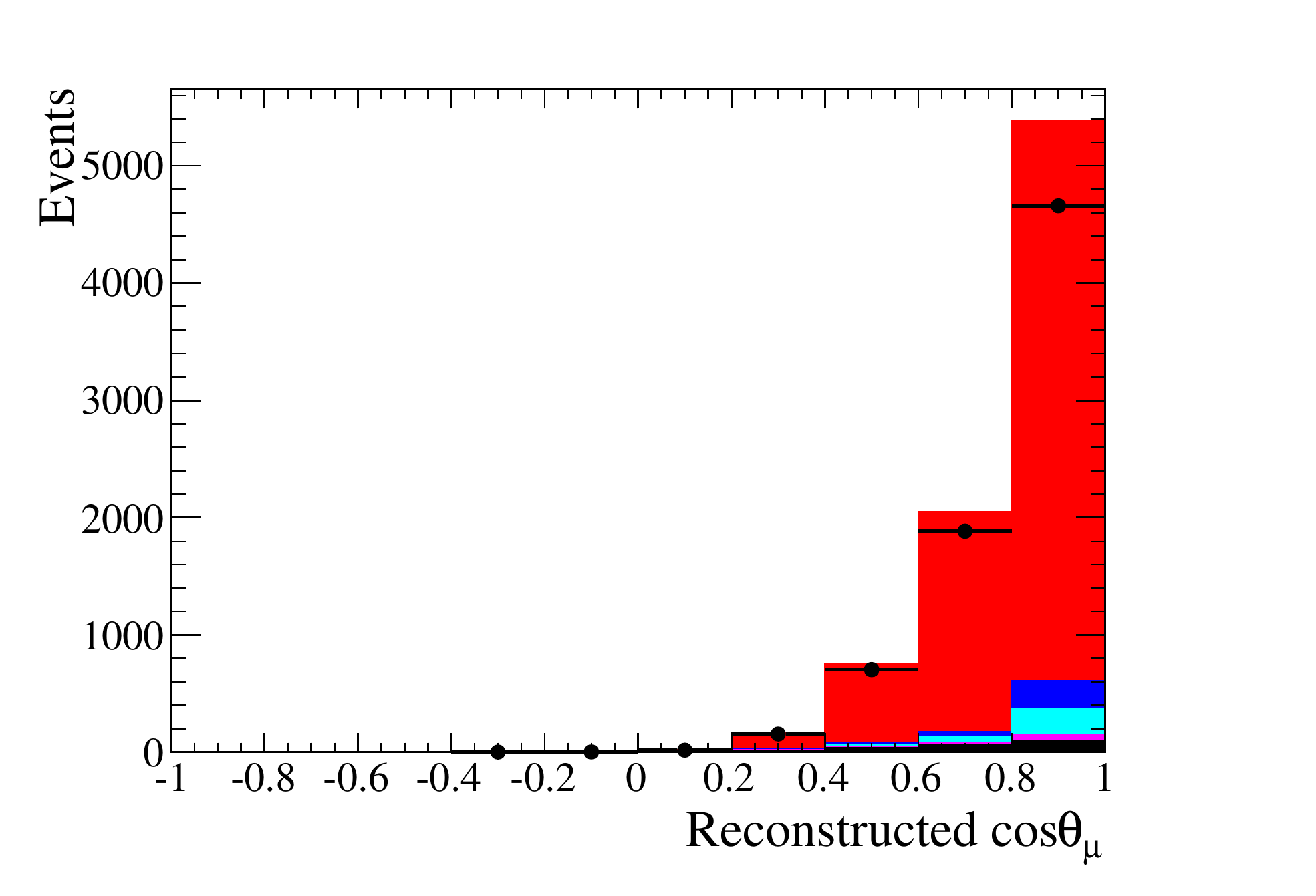}\\
 \includegraphics[width=7cm]{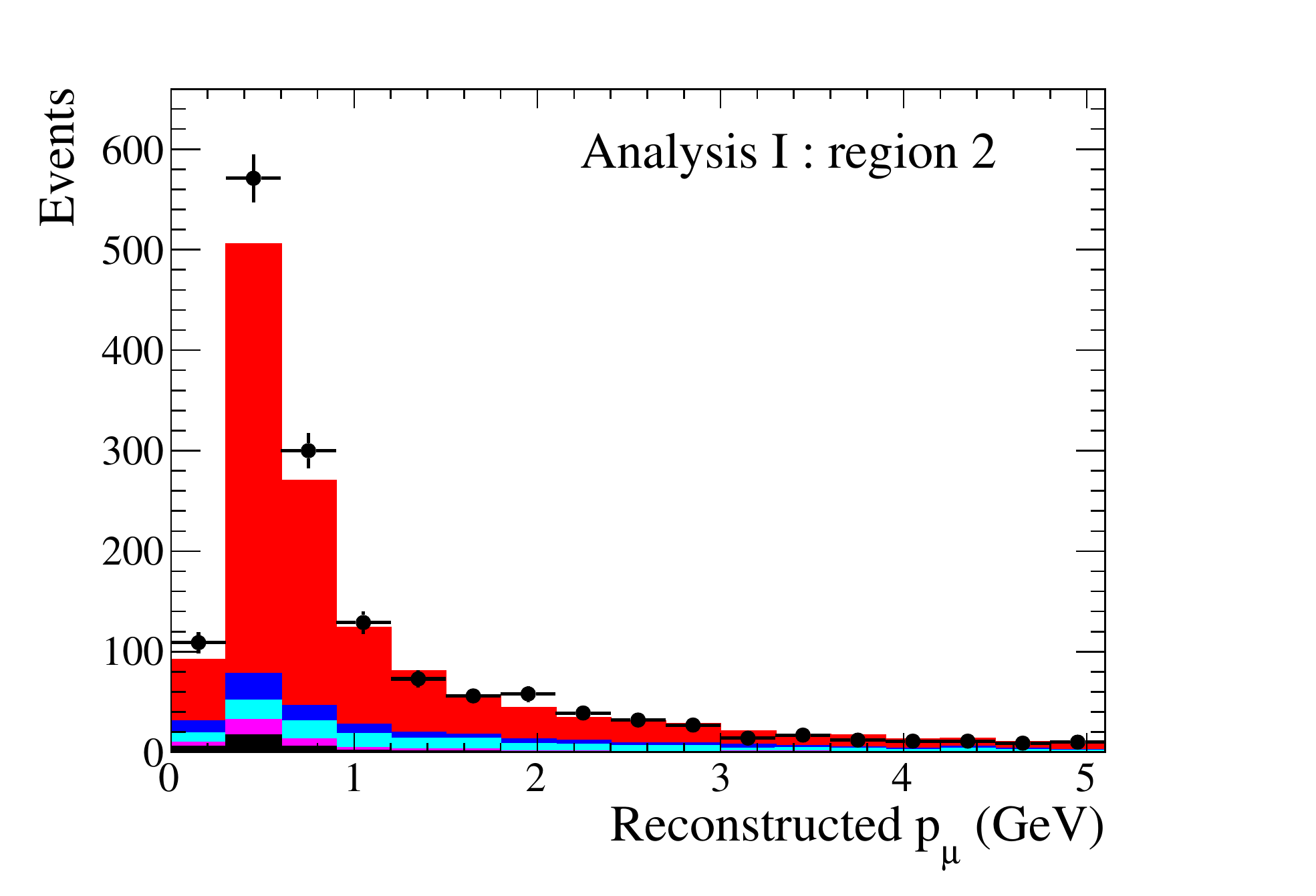}
 \includegraphics[width=7cm]{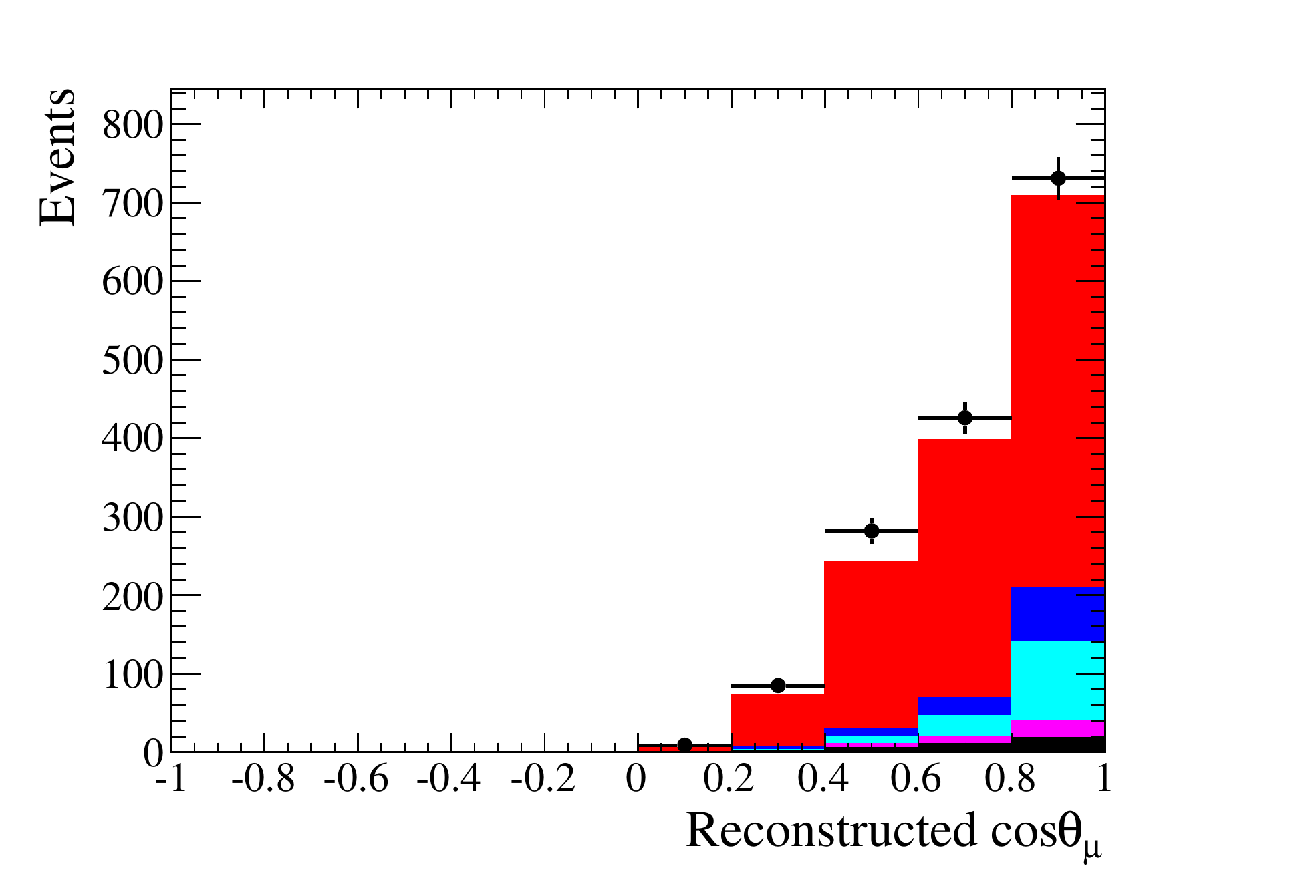}\\
 \includegraphics[width=7cm]{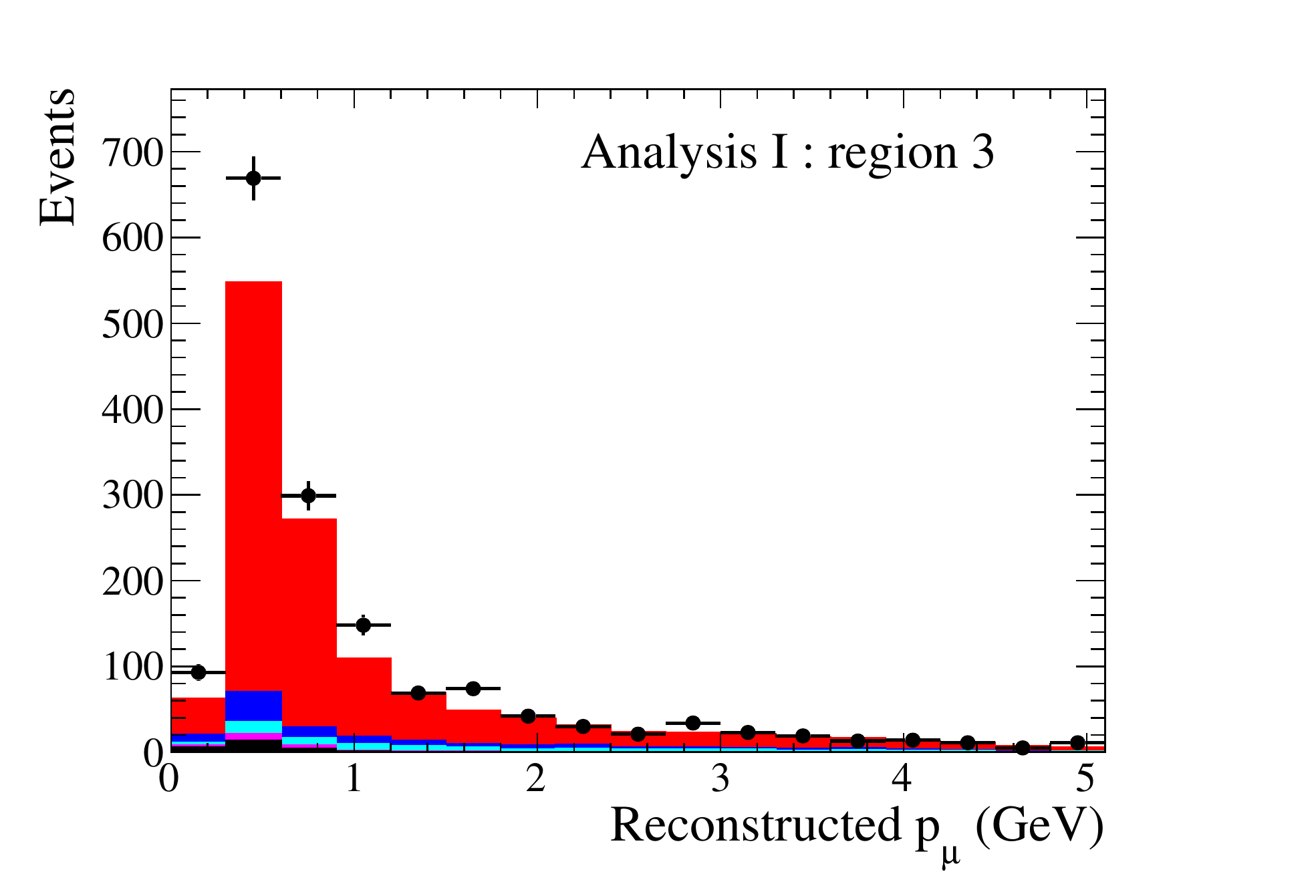}
 \includegraphics[width=7cm]{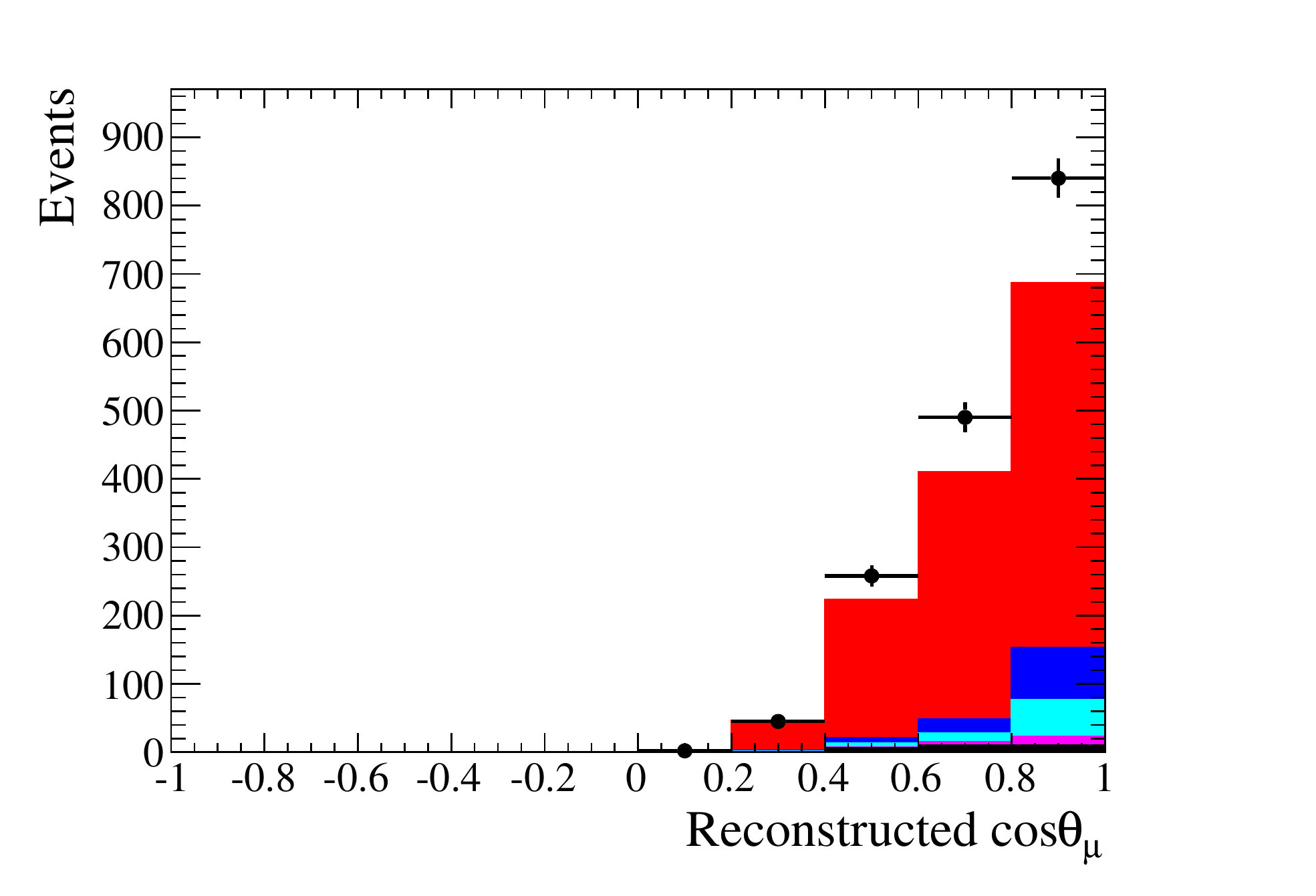}\\
 \includegraphics[width=7cm]{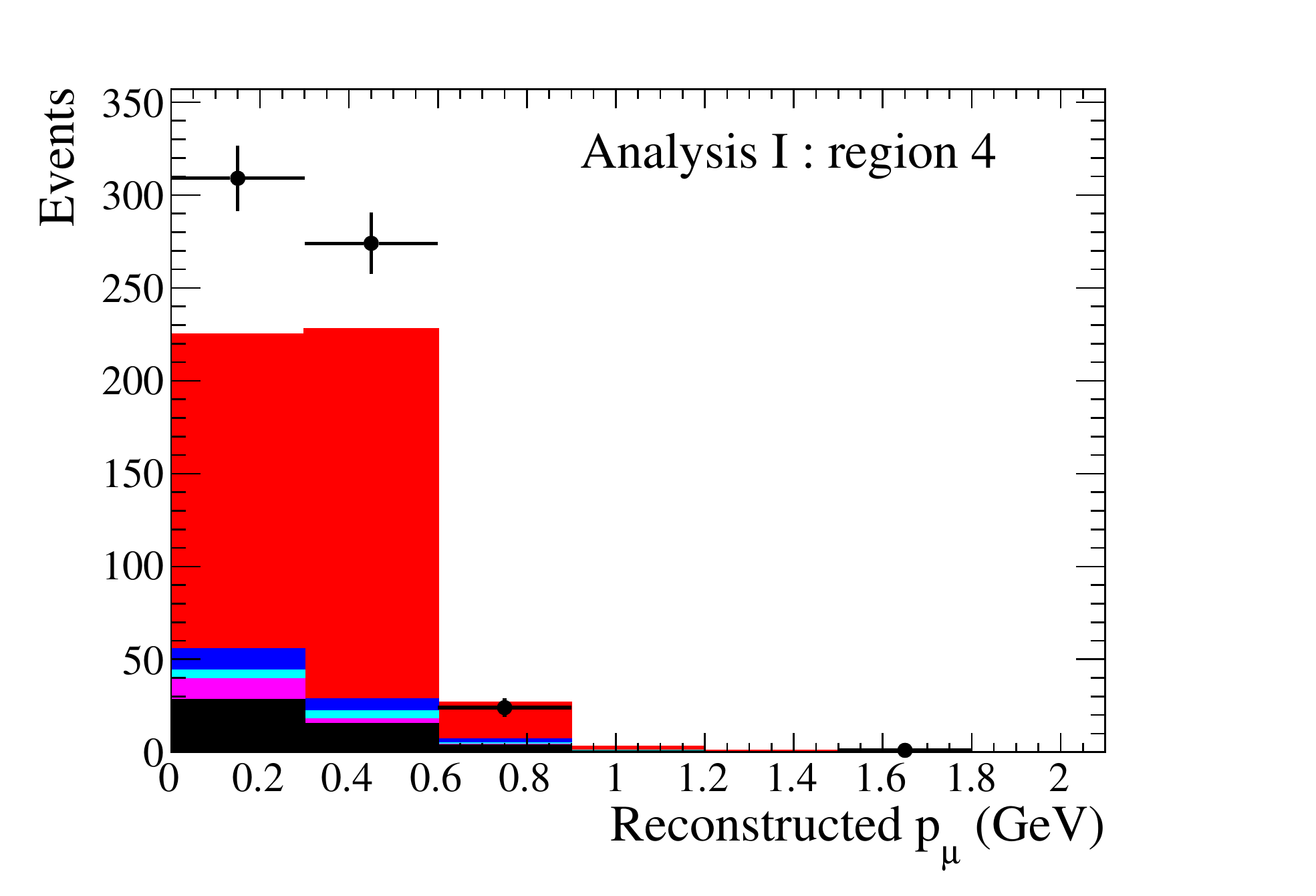}
 \includegraphics[width=7cm]{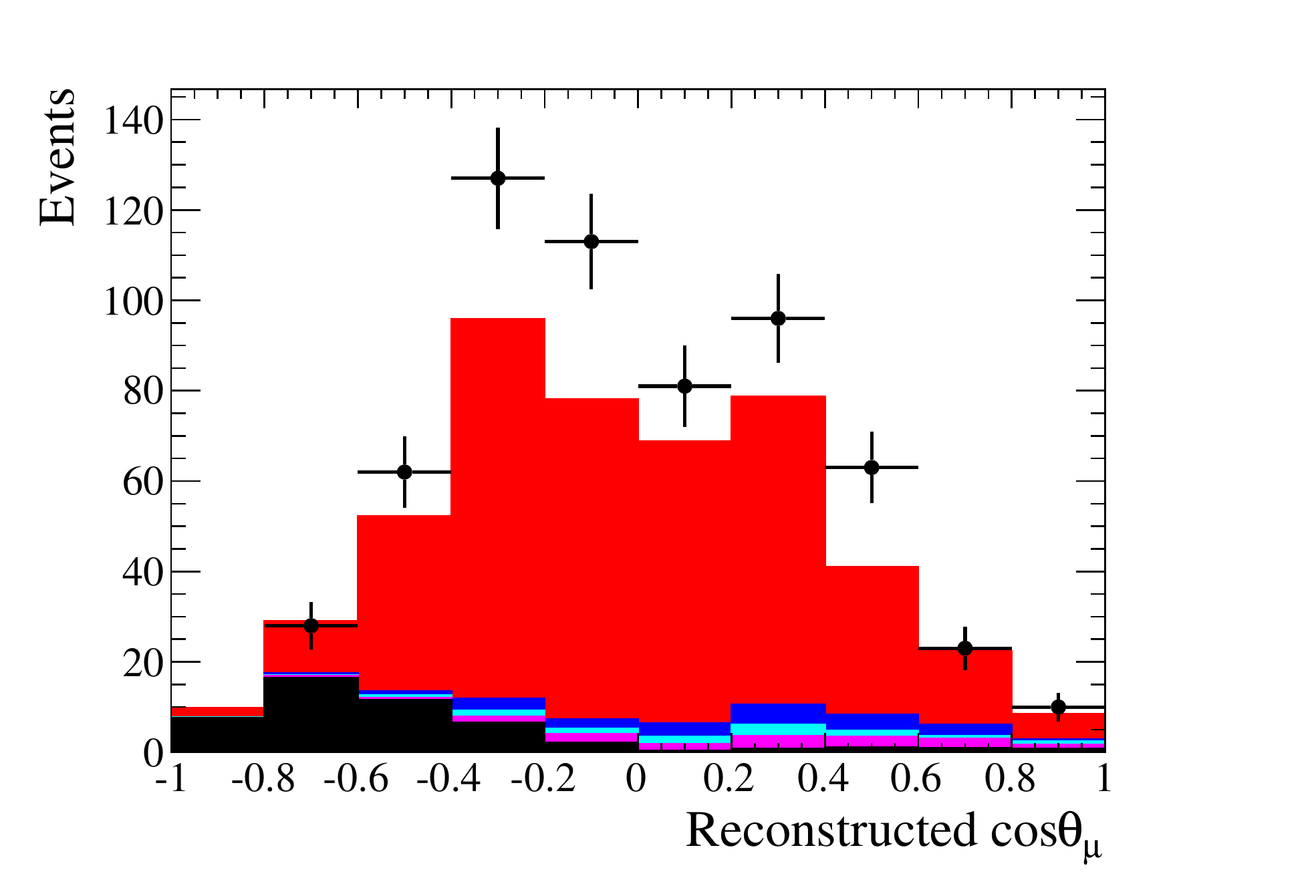}\\
\end{center}
\caption{Distribution of events in different regions for Analysis I.
Each row corresponds to a signal region from 1 (top) to 4 (bottom). Figures in the left column are plotted against the reconstructed
muon momentum and the right column against the reconstructed muon $\cos\theta$. Histograms are stacked.}
\label{fig:eventsDistributionsSig2}
\end{figure}

\begin{figure}
\begin{center}
 \includegraphics[width=7cm]{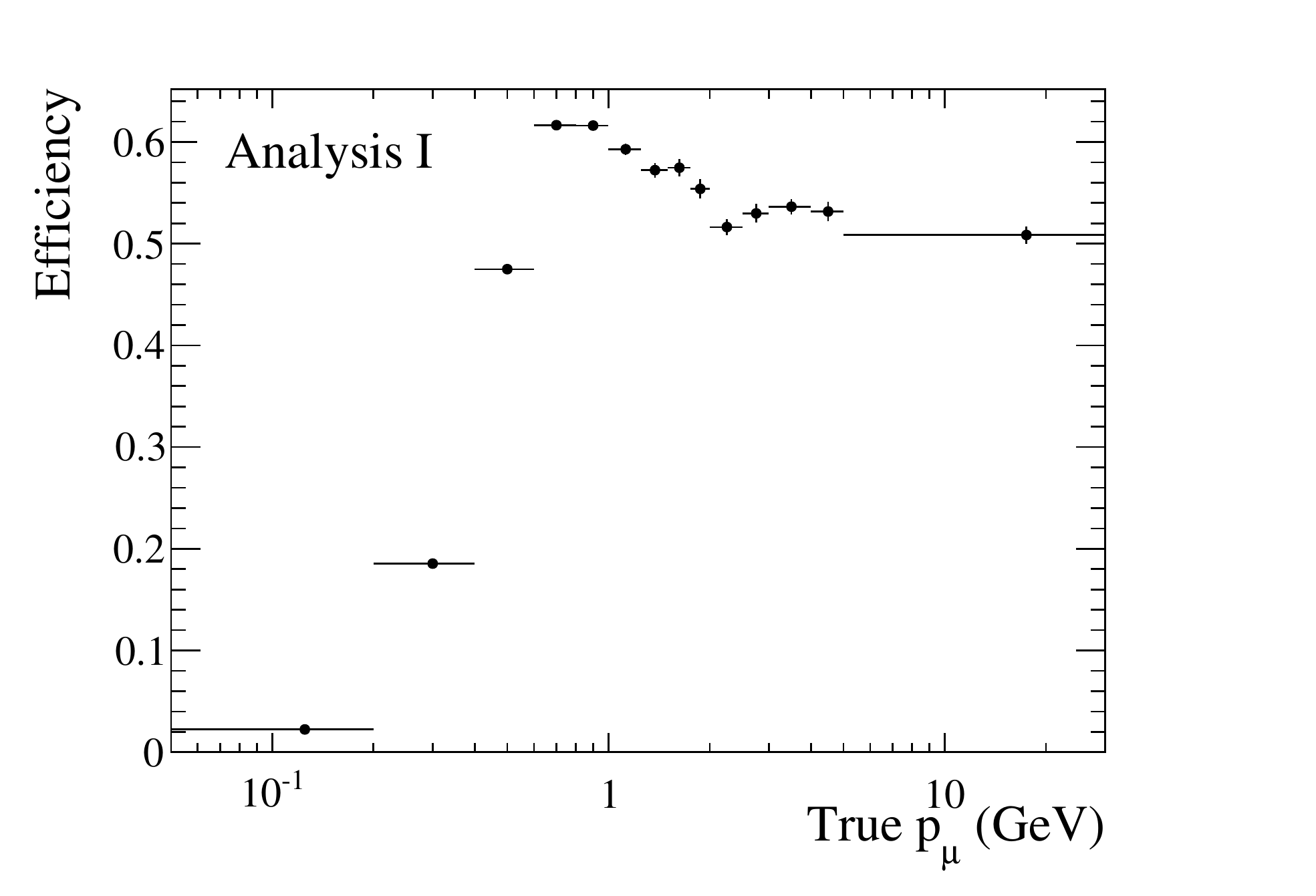}
 \includegraphics[width=7cm]{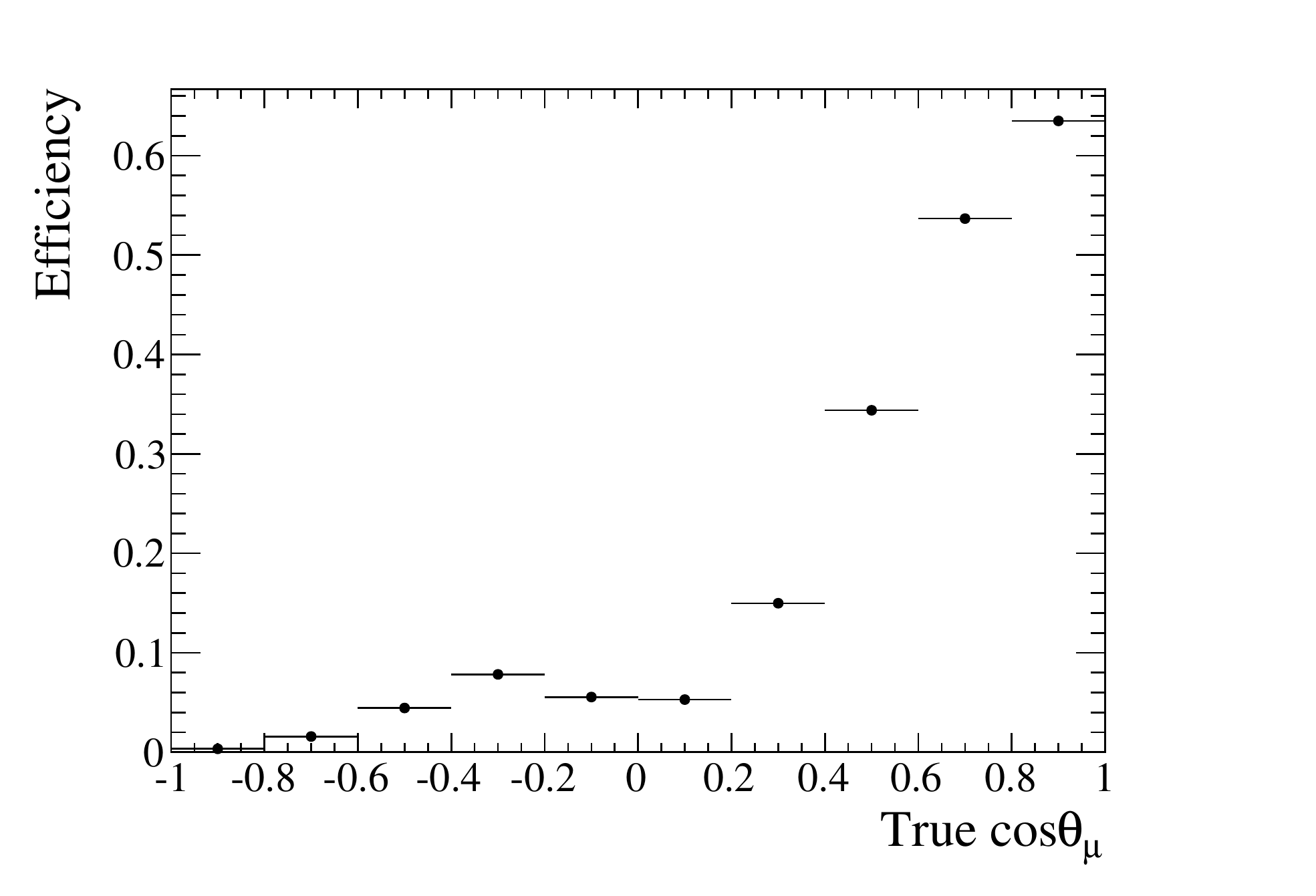}\\
\end{center}
\caption{Efficiency of reconstruction and selection of \cczeropi events for Analysis I.}
\label{fig:efficiency}
\end{figure}

The pre-selection described in Section~\ref{sec:preselection} is extended to also include 
muons which do not reach the TPC (i.e., are FGD-only or FGD plus the ECAL or the SMRD or both).
In this case the proton needs to be reconstructed as a positively-charged track in the TPC with a vertex 
in the FGD FV, where this track has to pass the TPC track quality cut and a proton-like
PID is required.
The pre-selected events are then divided in four signal regions:
\begin{description}
\item  [region 1] single-track events with a muon candidate in the TPC,
\item  [region 2] two-track events with a muon and a proton candidates in TPC, %
\item  [region 3] two-track events with a muon candidate in TPC and a proton candidate in FGD,
\item  [region 4] two-track events with a proton candidate in TPC and a muon candidate in FGD (possibly reaching the ECAL or the SMRD or both).
\end{description}
Muon and proton candidates are identified using the $\frac{\text{d}E}{\text{d}x}$ measurement in the TPC or the energy deposited in the FGD. 
The kinematics of the muon candidate in each selection region
for the CC0$\pi$ signal and the various backgrounds are shown in Fig.~\ref{fig:eventsDistributionsSig2}.
The selection is highly dominated by events with one reconstructed muon and no other tracks (region 1).
The signal regions where the muon is reconstructed in the TPC (regions 1,2,3) have very similar 
momentum distributions, although events with a reconstructed proton (regions 2,3) tend to have 
muons at slightly larger angles, while the region with the muon in the FGD
and the proton in the TPC (region 4) have muons with much smaller momenta and larger angles.
The overall selection efficiency for the CC0$\pi$ selection is 39\%; 
Fig.~\ref{fig:efficiency} shows the efficiency as a function of the muon candidate momentum and angle. 
The loss in efficiency at low momentum is due to the detector threshold for
muon tracking in FGD, while the loss for backward muons is due to limitations in the reconstruction algorithm.
This is the first T2K ND280 analysis with a non-zero efficiency and non-negligible event statistics for backwards-going
muons; future analyses will benefit from a new reconstruction algorithm with further improvements in backwards-track reconstruction.


\begin{figure}
\begin{center}
 \includegraphics[width=7cm]{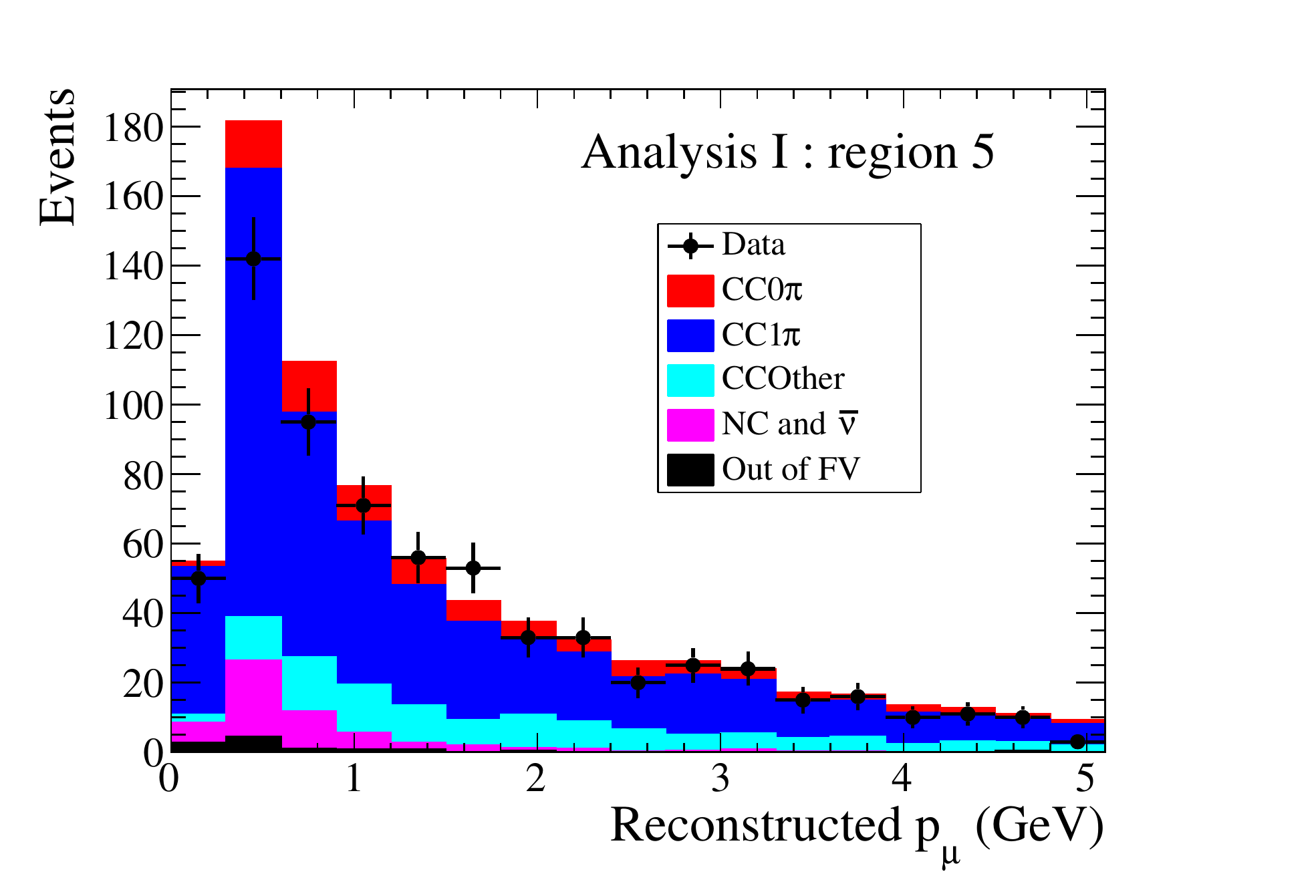}
 \includegraphics[width=7cm]{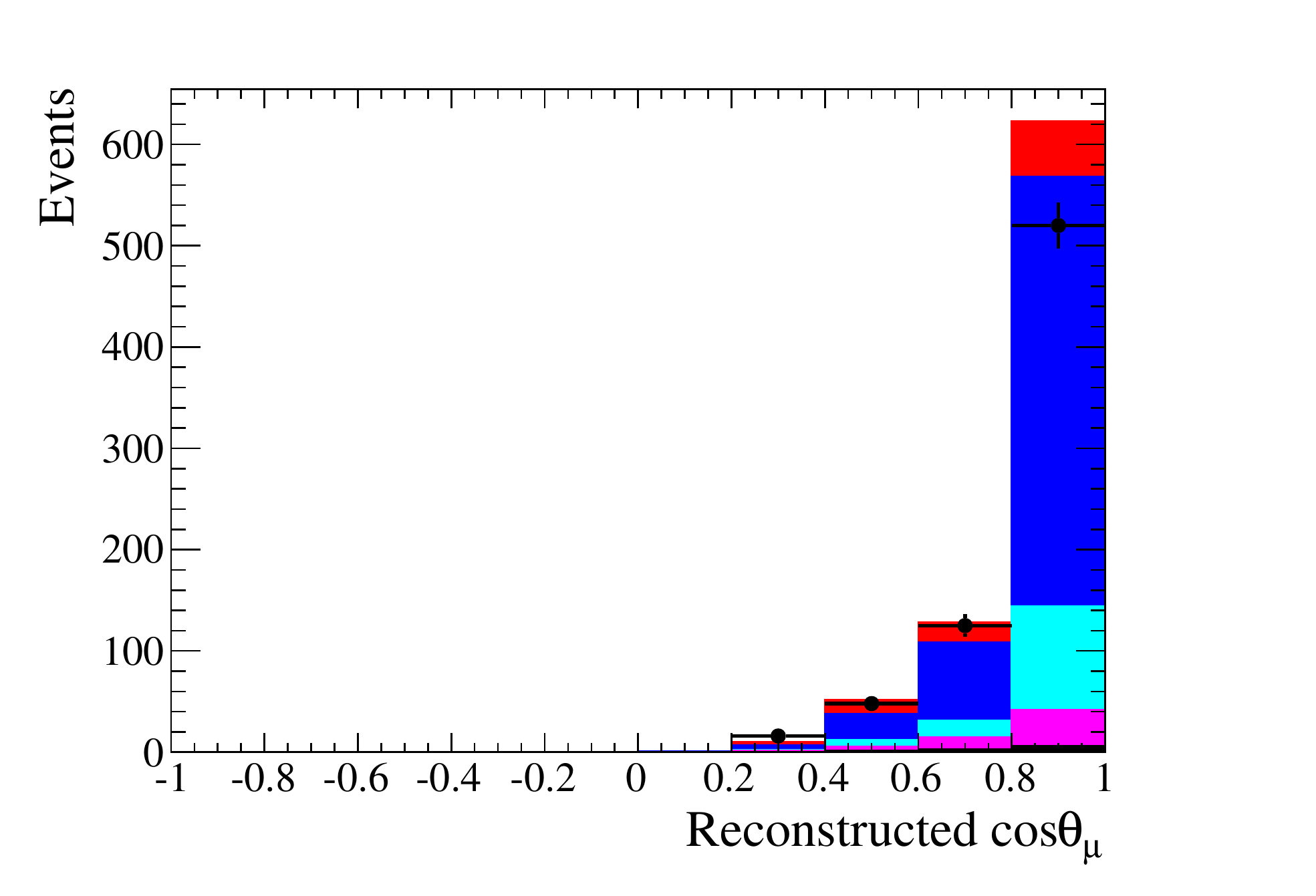}\\
 \includegraphics[width=7cm]{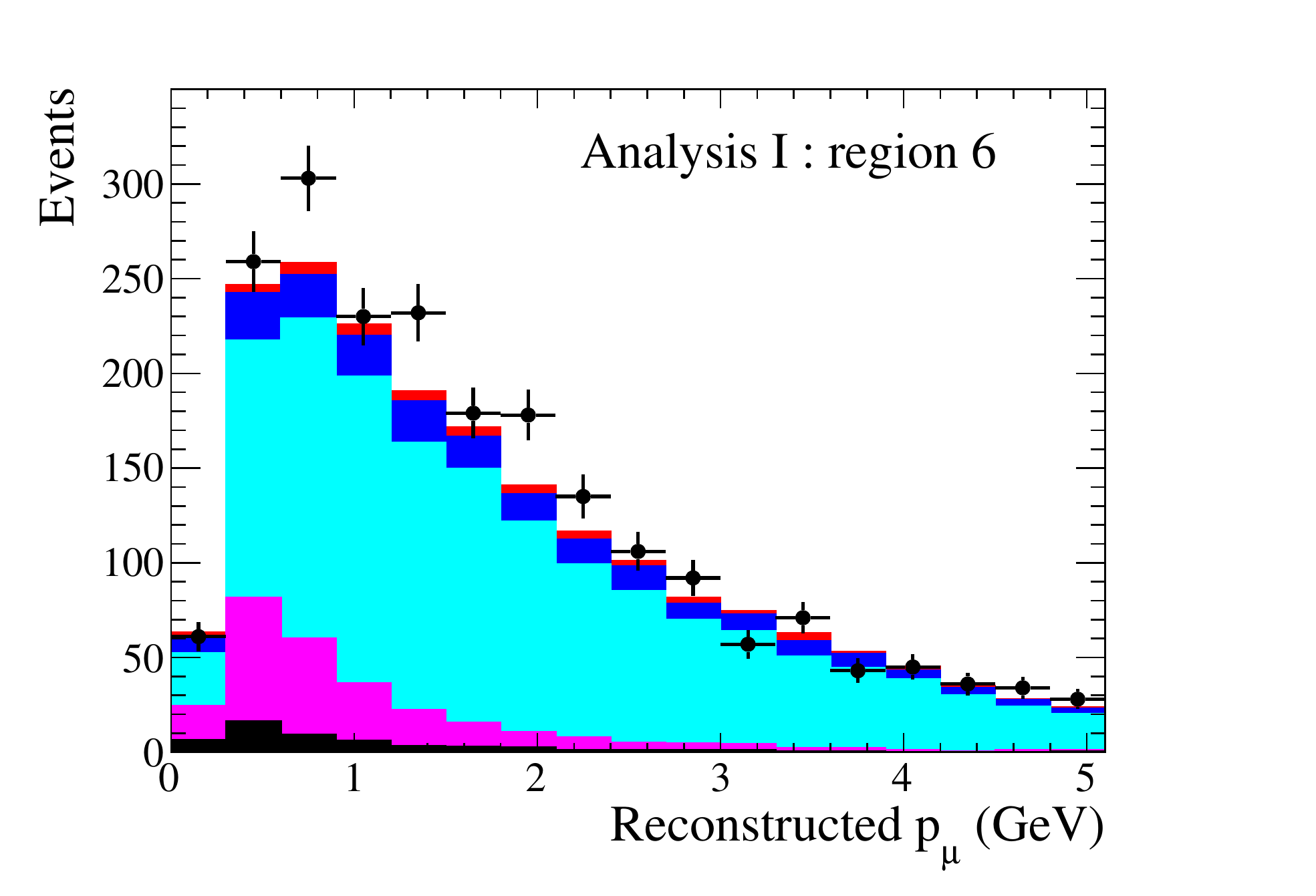}
 \includegraphics[width=7cm]{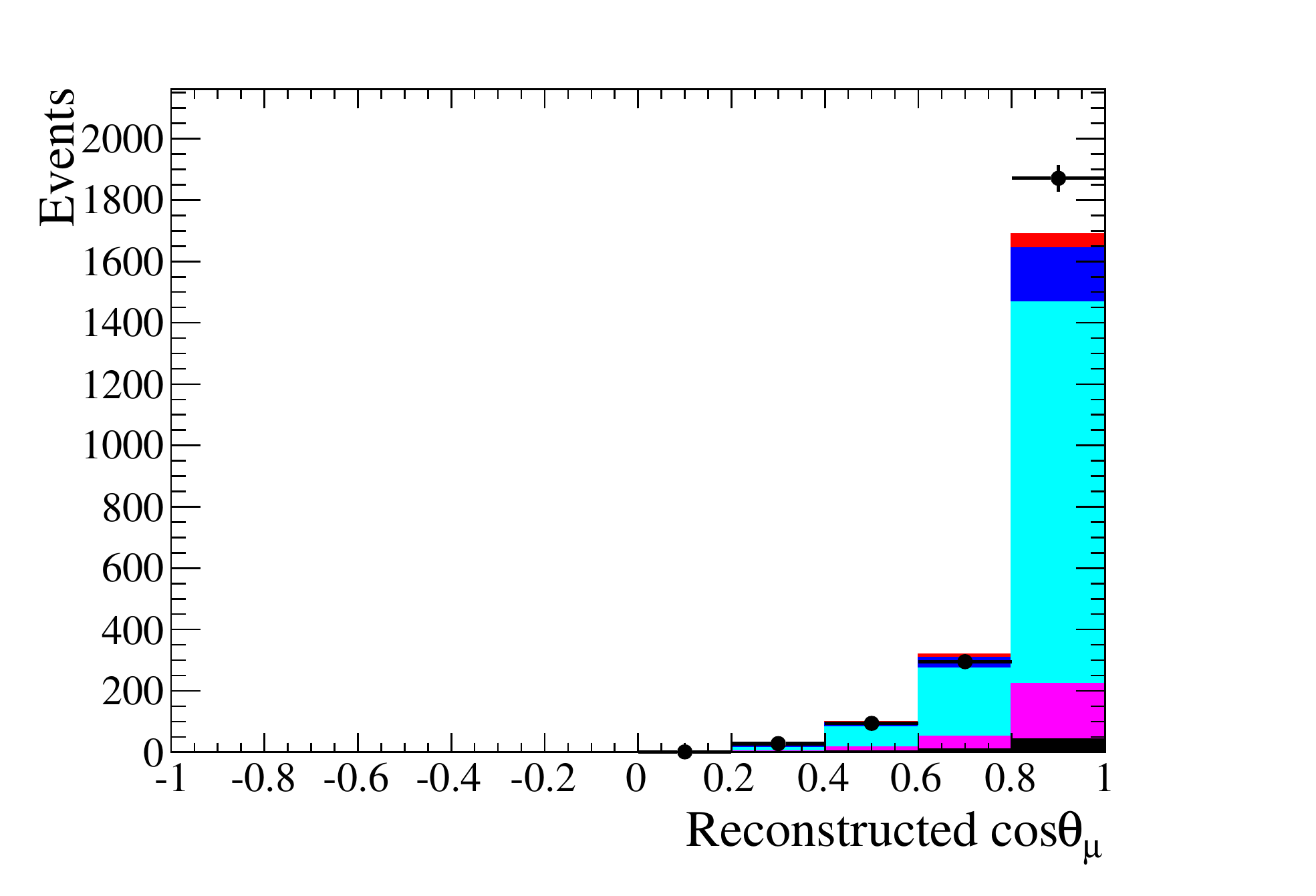}\\
 \end{center}
\caption{Distribution of events in different control regions for Analysis I.
The first row is the CC1$\pi$ control region (region 5) and the second row 
is the DIS control region (region 6). Figures in the left column are plotted against the reconstructed
muon momentum and the right column against the reconstructed muon $\cos\theta$. Histograms are stacked.}
\label{fig:eventsDistributionsBkg}
\end{figure}

Two additional control regions are selected to constrain charged current event rates
with single-pion and multiple-pion production. 
After pre-selection, a reconstructed negative track in TPC with muon-like PID and a positive track in TPC with pion-like 
PID are required.
Events with exactly two tracks are included in \textbf {region 5 (\cconepi control region)}
while events with more than two tracks are included in \textbf {region 6 (\ccother control region)}.
The background composition of the control regions as a function of muon momentum
is shown in Fig.~\ref{fig:eventsDistributionsBkg}.
The fraction of \cczeropi signal in the control regions is very low and the \cconepi (\ccother)
purity is quite good in region 5 (region 6), thus allowing unbiased
constraints to be put on the background shape and normalization.

\subsubsection{Cross-section extraction}
We perform a fit to the number of selected events as a function of the
muon kinematic variables ($p_\mu$ and $\cos\theta_\mu$), simultaneously in the four signal and two control regions.
The detector-related and theoretical systematic uncertainties are parametrized and
included in the fit through nuisance parameters.
The number of selected events in each signal region and in each bin of reconstructed kinematics $j$ is computed as
\begin{equation}
  \label{eq:systematics} %
N_j =  \sum_i^\text{true bins}\left[c_i \left(N^\text{MC \cczeropi}_i  \prod_a^\text{model} w(a)_{ij}^\text{\cczeropi} \right) + \sum_k^\text{bkg reactions} N^\text{MC bkg $k$}_{i} \prod_a^\text{model} w(a)_{ij}^{k} \right]t_{ij}^\text{det} r_j^\text{det},
\end{equation}
where $i$ runs over the bins of the `true' muon kinematics prior to detector smearing effects, 
$k$ runs over the background reactions (\cconepi, \ccother etc.),
$c_i$ are the parameters of interest which adjust the Monte Carlo \cczeropi cross-section to match with the observation in data,
$t_{ij}^\text{det}$ is the transfer matrix from the true ($i$) to the reconstructed ($j$) muon kinematics bins,
and $r_j^\text{det}$ represents the free nuisance parameters in the fit describing the detector systematics
and which are constrained by a prior covariance matrix. 
The product $\prod_a^\text{model} $ runs over the
systematics related to the theoretical modeling of signal and background. Each $w(a)^{k}_{ij}$ term
is a weighting function describing how the generated and reconstructed muon kinematics change
(in bins $i$, $j$ and for each signal and background process) as a function of the value of a particular theoretical parameter $a$.
All the parameters $a$ are free nuisance parameters in the fit, constrained by a prior covariance matrix. 

For simplicity, we use the same binning for the fit to the reconstructed $p_\mu$, $\cos\theta_\mu$ distribution
and for the extraction of the data/Monte Carlo cross-section corrections $c_i$. 
A non-rectangular $p_\mu$, $\cos\theta_\mu$ binning (different $p_\mu$ binning for each $\cos\theta_\mu$ bin) is chosen on 
the basis of the available event numbers (which are much smaller in the high angle and backward regions), 
the signal-over-background ratio (which is much smaller in the high momentum region) 
and of the detector resolution (to avoid large migrations of events between nearby bins).

As the parameters of interest for the fit rescale the overall number of \cczeropi events
in the four signal regions and two control regions together, the resulting cross-section is 
extracted inclusively for all the regions simultaneously. A future analysis will measure separate cross-sections, with and
without reconstructed protons. 

A binned likelihood fit is performed %
\begin{equation}
\label{eq:chi2}
\chi^2 = \chi^2_\text{stat} + \chi^2_\text{syst} = \sum_j^\text{reco bins} 2(N_j-N_j^\text{obs}+N_j^\text{obs} \ln\frac{N_j^\text{obs}}{N_j}) +  \chi^2_\text{syst},
\end{equation}
where $\chi^2_\text{syst}$ is a penalty term for the systematics:
\begin{eqnarray}
\label{eq:chi2syst}
\chi^2_\text{syst} &=& (\vec{r}^\text{ det}-\vec{r}^\text{ det}_\text{prior})(V^\text{det}_\text{cov})^{-1}(\vec{r}^\text{ det}-\vec{r}^\text{ det}_\text{prior})   \nonumber \\
&+& (\vec{a}^\text{ model}-\vec{a}^\text{ model}_\text{prior})(V^\text{model}_\text{cov})^{-1}(\vec{a}^\text{ model}-\vec{a}^\text{ model}_\text{prior})  
\end{eqnarray}
where $\vec{r}^\text{ det}$ and $\vec{a}^\text{ model}$ are the parameters for detector and theory systematics 
running over the reconstructed bins $j$,
and $\vec{r}^\text{ det}_\text{prior}$, $\vec{a}^\text{ model}_\text{prior}$ indicate our initial knowledge of the detector response and theory parameters
and $V^\text{det}_\text{cov}$, $V^\text{model}_\text{cov}$ are the corresponding covariance matrices.

The number of selected events as a function of the `true' kinematics extracted from the fit:
\begin{equation}
\label{eq:finalObservable}
N^{\cczeropi}_i = \sum_t^\text{regions}\sum_j^\text{reco bins}c_i  N^\text{MC\:\cczeropi}_{jt} t_{tij}^\text{det} r_{tj}^\text{det} 
\prod_a^\text{model syst} w(a)_{ij}^\text{\cczeropi} 
\end{equation}
is corrected by the selection efficiency in each bin $\epsilon_i$ and
divided by the overall integrated flux and the number of neutrons in the fiducial volume
to extract a flux-integrated cross-section:
\begin{equation}
\label{eq:xsec}
\frac{\text{d}\sigma}{\text{d}x_i}= \frac{N^\text{\cczeropi}_i}{\epsilon_i \Phi N^\text{FV}_\text{neutrons}} \times \frac{1}{\Delta x_i}
\end{equation}

\subsubsection{Treatment of systematic uncertainties}\label{sec:systematicsI}
The detector systematics are stored in a covariance matrix ($V_{jk}^\text{det}$) as uncertainties on the total number of reconstructed events in bins of reconstructed muon momentum and angle for each signal and control region.

The flux systematic uncertainties affect the measured cross-section in two ways: 
they affect the fit, by varying the signal differently in each bin and varying the signal over background ratio bin by bin,
and thus changing the shape of the measured cross-section,
and they also affect the overall cross-section normalization.
These two contributions are treated separately in the analysis.
While the fit to the control regions has the power to constrain the flux systematic uncertainties,
the flux is not included as a nuisance parameter in the fit; this is to reduce the impact of model-dependent assumptions
when the signal-over-background distribution is extrapolated from the control regions to the signal regions.

The systematics due to signal and background modeling are based on the parametrization discussed 
in Section~\ref{sec:mc}.
The systematics due to background modeling and pion and proton FSIs
are included as nuisance parameters in the fit. The fit to the control regions
reduces the background modeling and pion FSI systematics by about a factor of four.
Systematics related to signal cross-section modeling, on the other hand, are not constrained from data
because including them in the fit with a specific parametrization (e.g: $M_\text{A}^\text{QE}$)
would introduce a model-dependent bias to the result.
The effect of signal modeling on the estimation of the efficiency 
in Eqn.~\ref{eq:xsec} is therefore described by a large systematic uncertainty, without trying to constrain
it from the fit to the data. The small uncertainty on the efficiency that arises from proton and pion FSIs is also included.

In summary, all the systematic uncertainties are included in the fit as nuisance parameters,
except for signal modeling and flux uncertainties.
Finally the effect from the statistical uncertainty on the Monte Carlo samples is included in the bin-by-bin efficiency in Eqn.~\ref{eq:xsec}.

To evaluate all the systematic and statistical uncertainties, we produce a large number of toy experiments.
To asses the statistical uncertainty, the number of reconstructed events in each bin 
is fluctuated according to the Poisson distribution in each of the toy experiments. 
To evaluate the systematics, the values of the parameters governing the various systematic uncertainties are varied
in each toy dataset according to a Gaussian distribution, following the prior covariance matrices. 

\begin{figure}
\begin{center}
 \includegraphics[width=6cm]{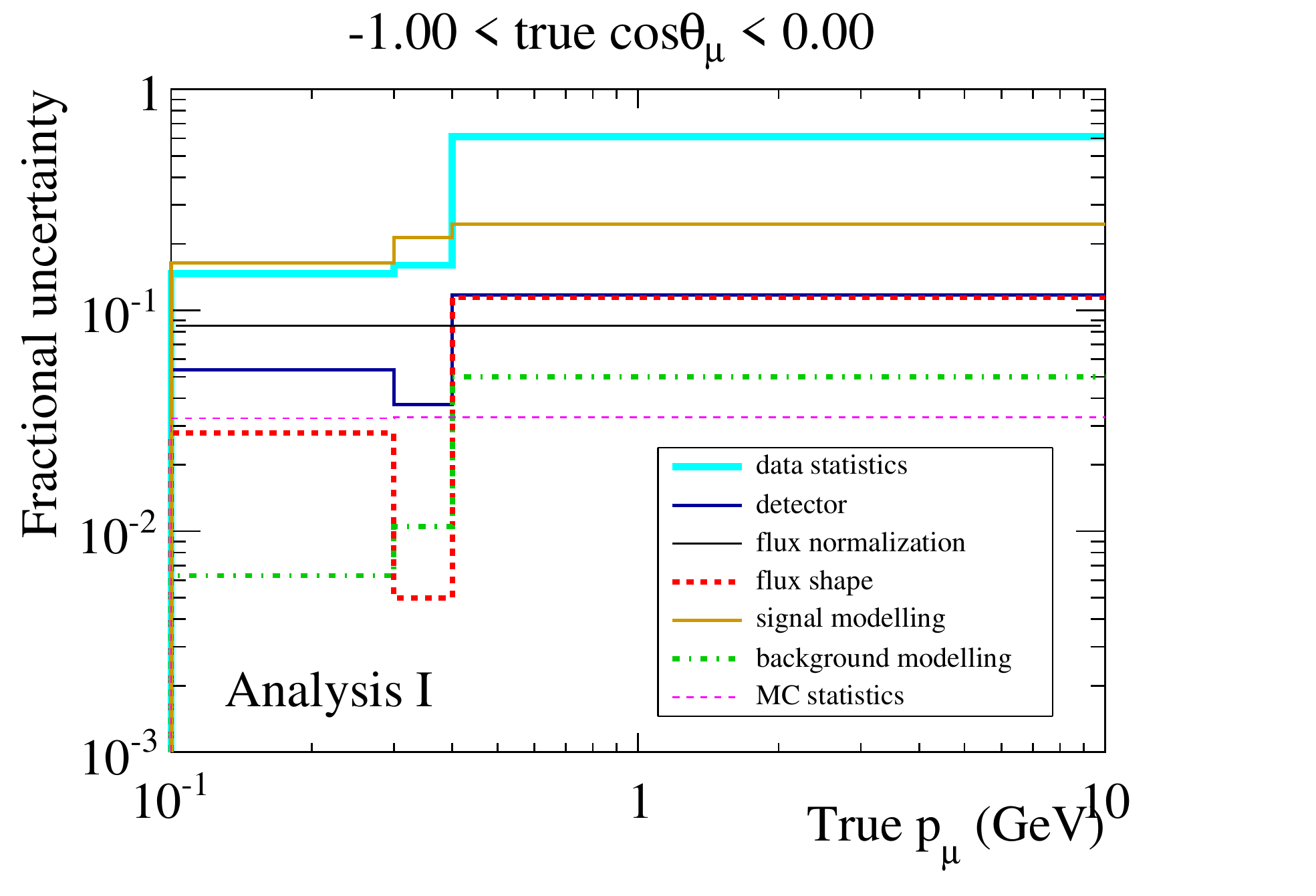}
 \includegraphics[width=6cm]{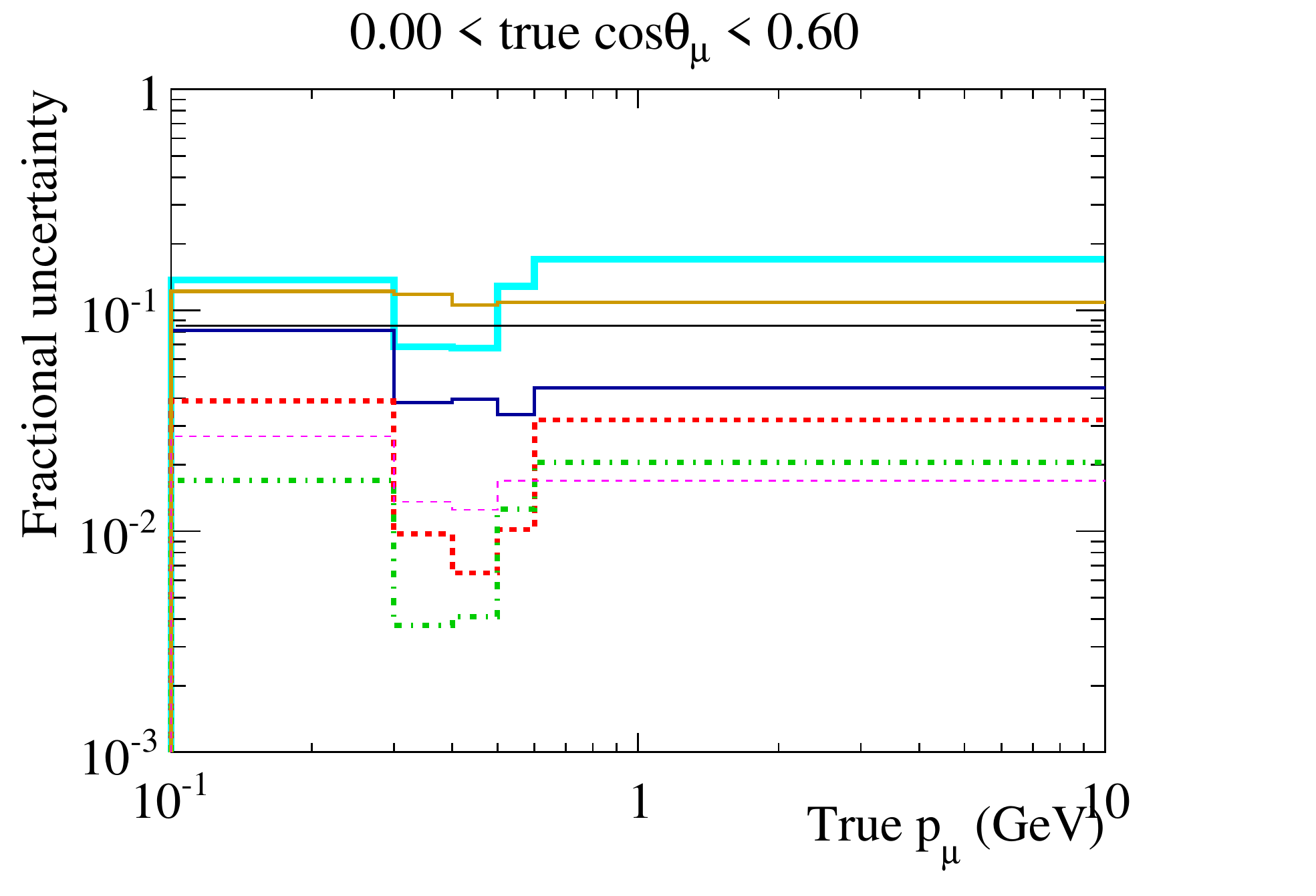}\\
 \includegraphics[width=6cm]{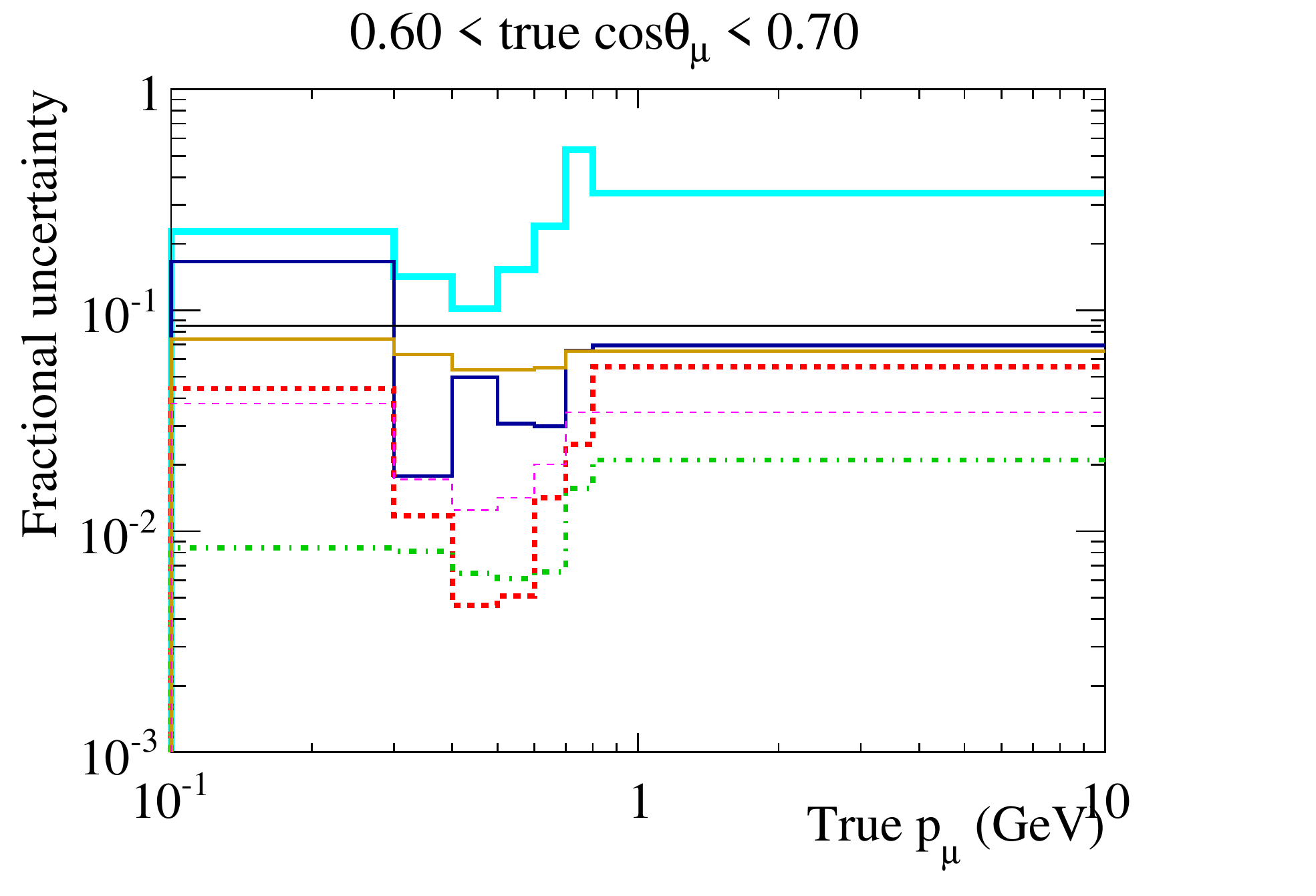}
 \includegraphics[width=6cm]{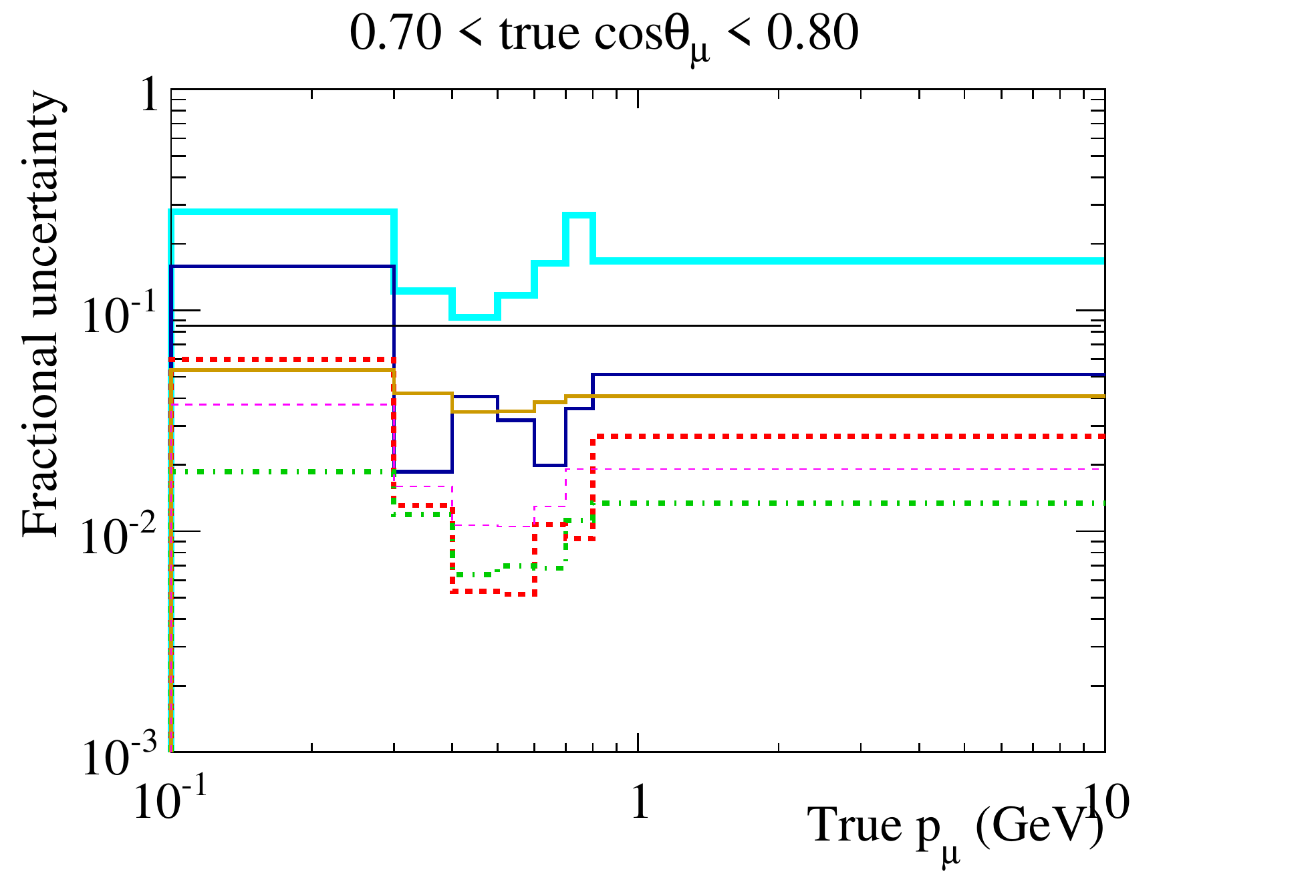}\\
 \includegraphics[width=6cm]{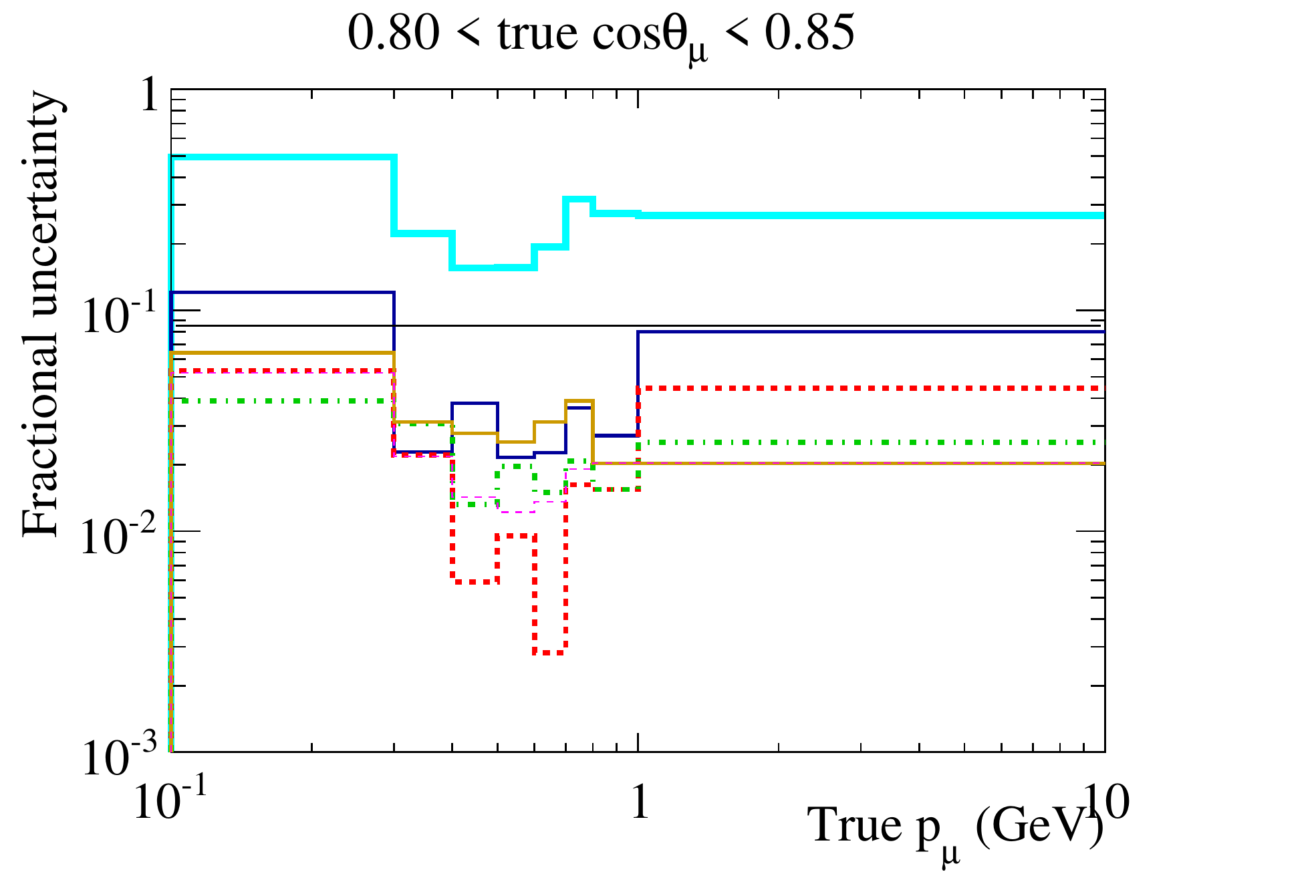}
 \includegraphics[width=6cm]{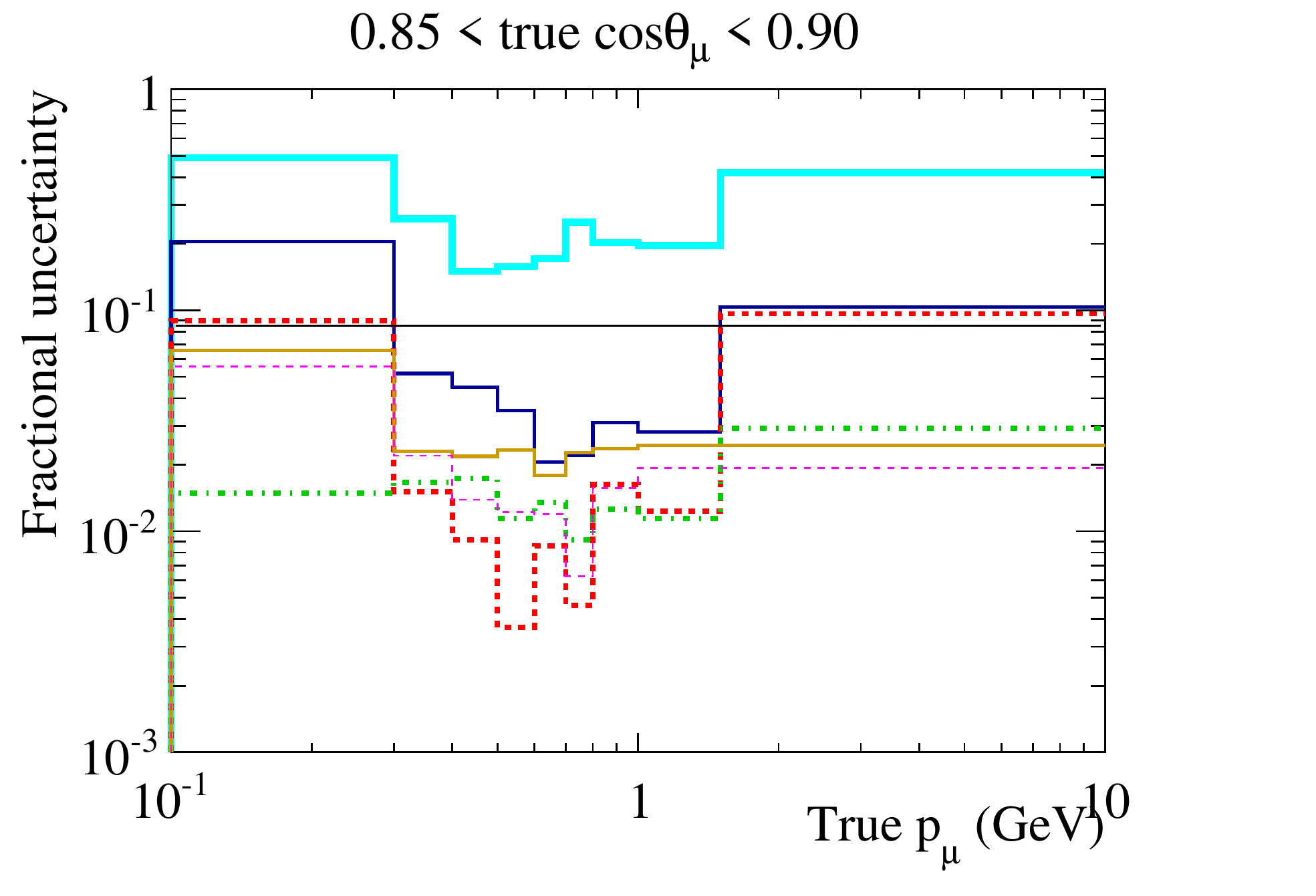}\\
 \includegraphics[width=6cm]{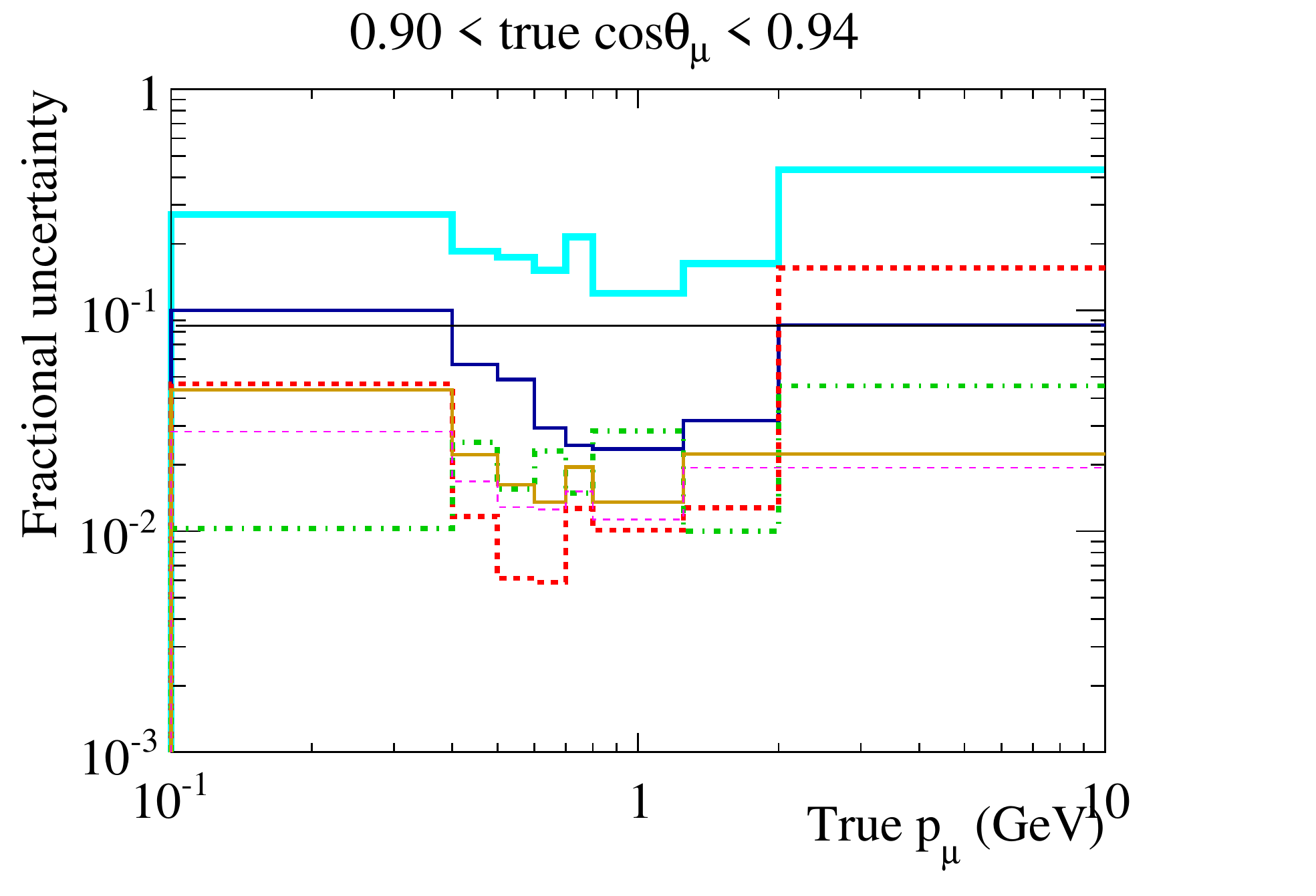}
 \includegraphics[width=6cm]{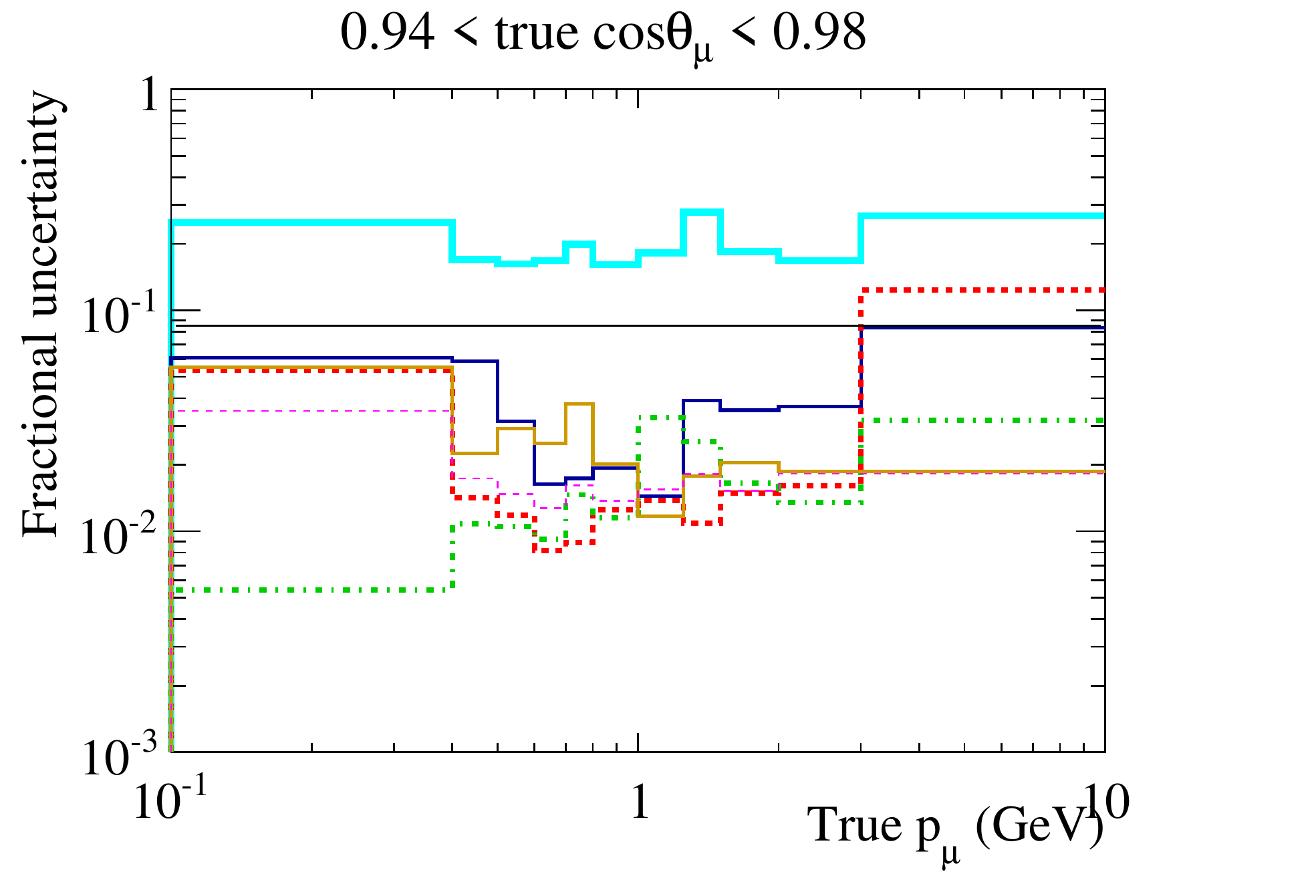}\\
 \includegraphics[width=6cm]{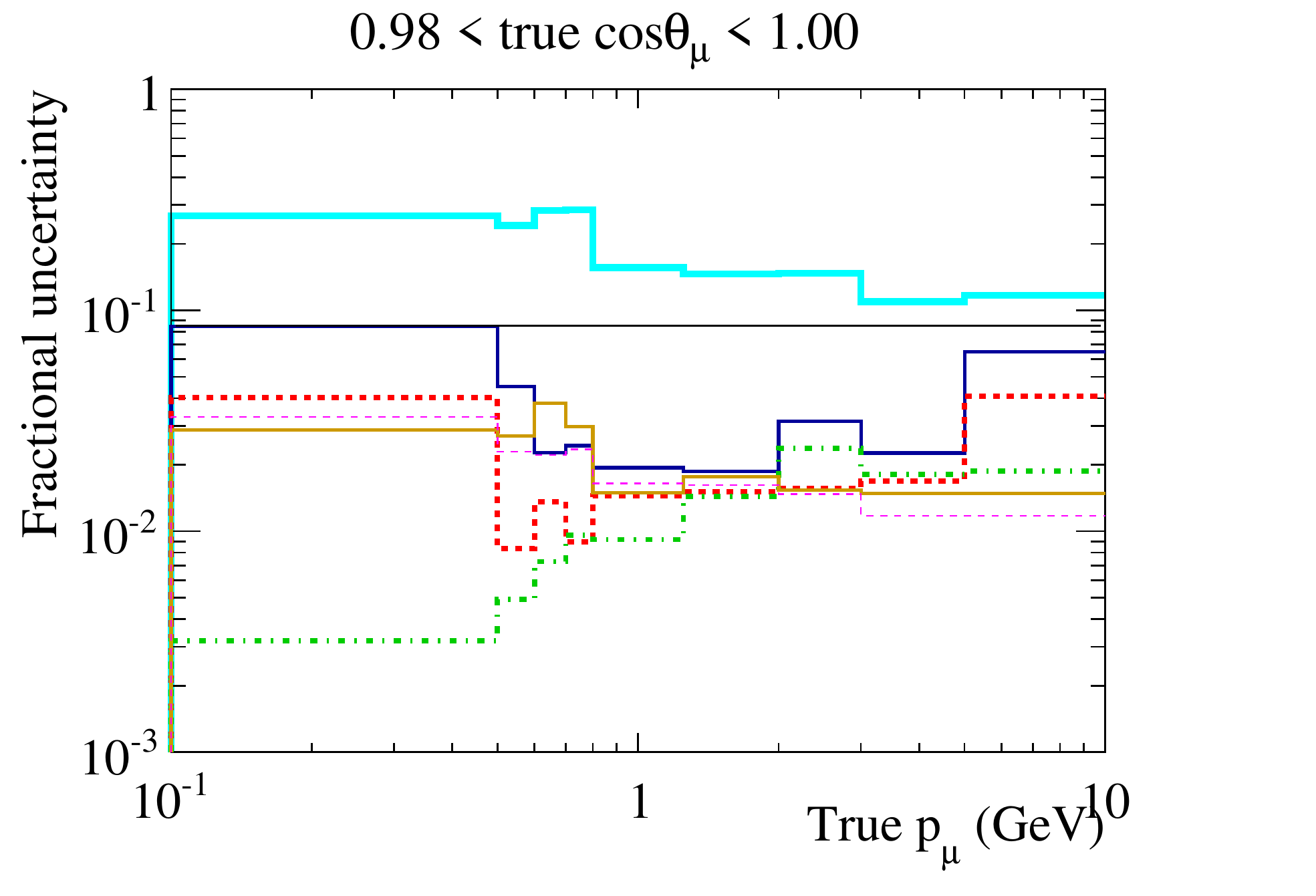}
\end{center}
\caption{All uncertainties in bins of true muon kinematics for Analysis I.}
\label{fig:allErrors}
\end{figure}

A summary of statistical and all the systematic errors is shown in Fig.~\ref{fig:allErrors}.
Theoretical uncertainties for the background cross-section, pion FSIs
and proton FSIs are varied together, while separate toy experiments are made
for the signal-modeling systematics, the flux systematics and the detector systematics. 
With the chosen binning, the statistical uncertainty is dominant.
The largest uncertainty is from the flux normalization (8.5\%) but, being
fully correlated between all bins, it does not affect the cross-section shape.
The effect of the flux uncertainty on the cross-section shape is small ($\leq$1\%) in the
region relevant for the signal ($p_\mu \simeq 0.3$--$1$ GeV) while it reaches
5--10\% at low and high momenta, where the magnitude of the effect is similar to that from the detector systematics. 
The systematics due to the model of background cross-sections and pion FSIs are larger in the forward and high momentum regions where
most of the background is located, but even in these regions they remain below 2\%, due to
the constraint of nuisance parameters from the control regions.  
The Monte Carlo statistical uncertainty on the efficiency is about $2\%$ or lower
in most of the bins, but it is as much as $4\%$ in the lowest and highest momentum bins, 
where the statistical uncertainties from data are also large.
The systematics on the efficiency due to signal modeling
are typically about a few percent, except in the high-angle region where the efficiency is lower
and therefore we depend more on the simulation to extrapolate to the full phase space.
Finally the detector systematics are of the order of a few percent and become larger (up to 10\%) in the low and high momentum regions,
where the detector resolution and efficiency is less well-known.
The detector systematics are the dominant shape uncertainties 
in most of the phase space, except at very high angles where the uncertainty on the signal modeling is larger.
For the final results, the systematics uncertainty is evaluated on a separate set of toy experiments by varying
all the theory nuisance parameters and the flux parameters at once. 
The uncertainties due to detector systematics and the systematics on the efficiency due 
to Monte Carlo statistics and signal modeling are added in quadrature.

\subsubsection{Results}
The total signal cross-section per nucleon integrated over all the muon kinematics phase space is:
\begin{equation}
\label{eq:xsecInt}
\sigma = (0.417 \pm 0.047 \text{(syst)} \pm 0.005 \text{(stat)})  \times 10^{-38} \text{cm}^2 \text{nucleon}^{-1}
\end{equation}
to be compared with the NEUT prediction of $0.444 \times 10^{-38} \text{cm}^2 \text{nucleon}^{-1}$.%
The uncertainty is dominated by the flux normalization
systematics (8.5\%), while other sources of systematic uncertainty are a few percent or less.

\begin{figure}
\begin{center}
  \includegraphics[width=6cm]{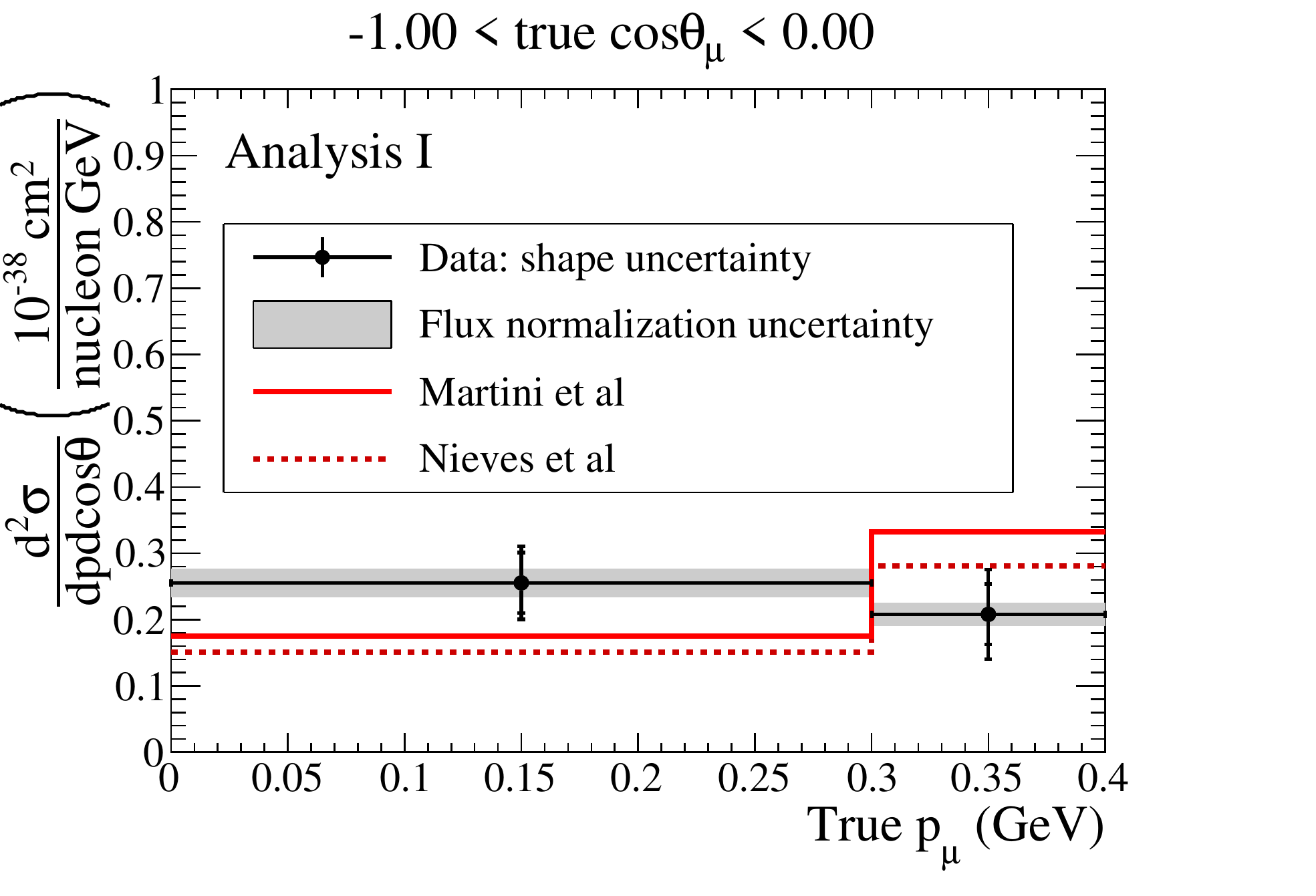}
 \includegraphics[width=6cm]{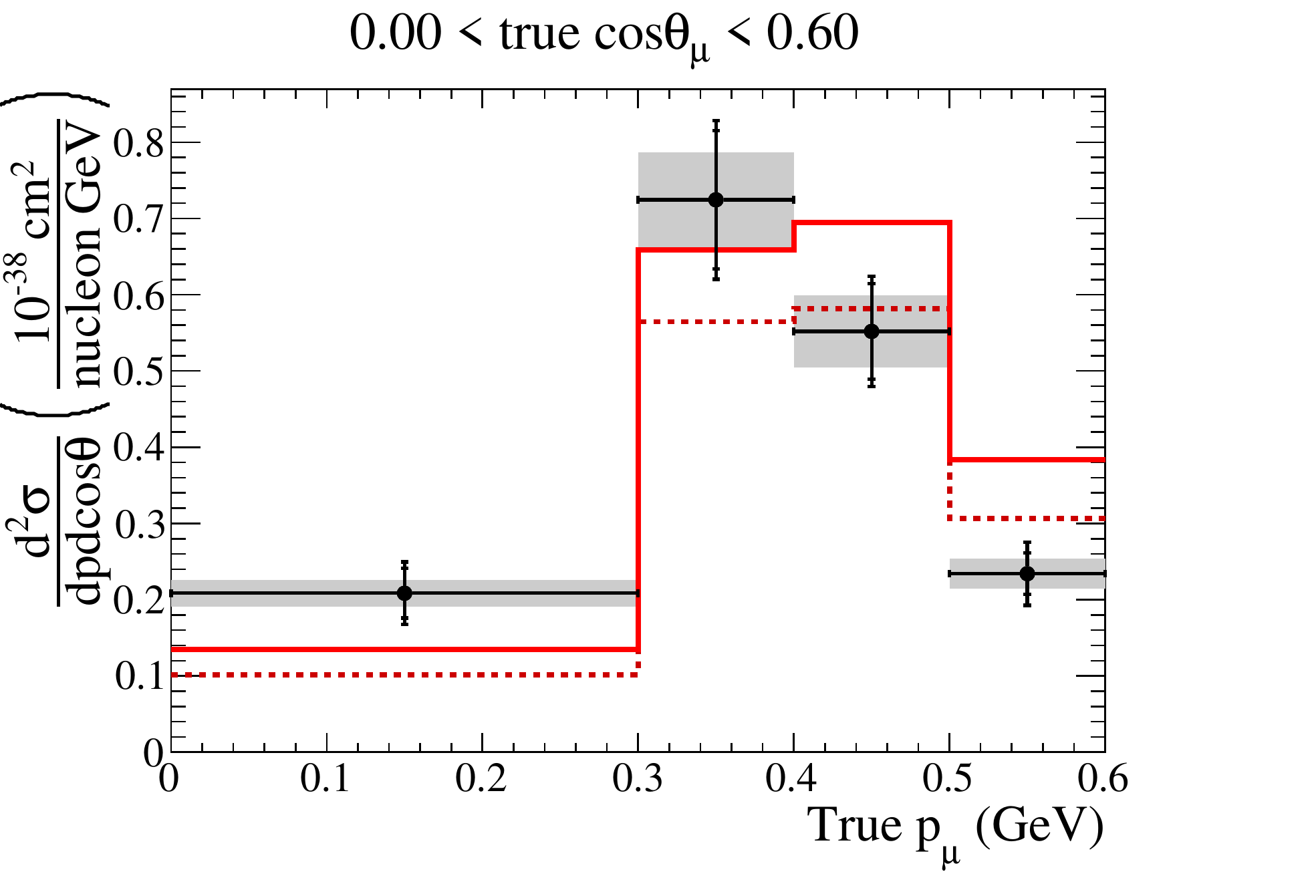}\\
 \includegraphics[width=6cm]{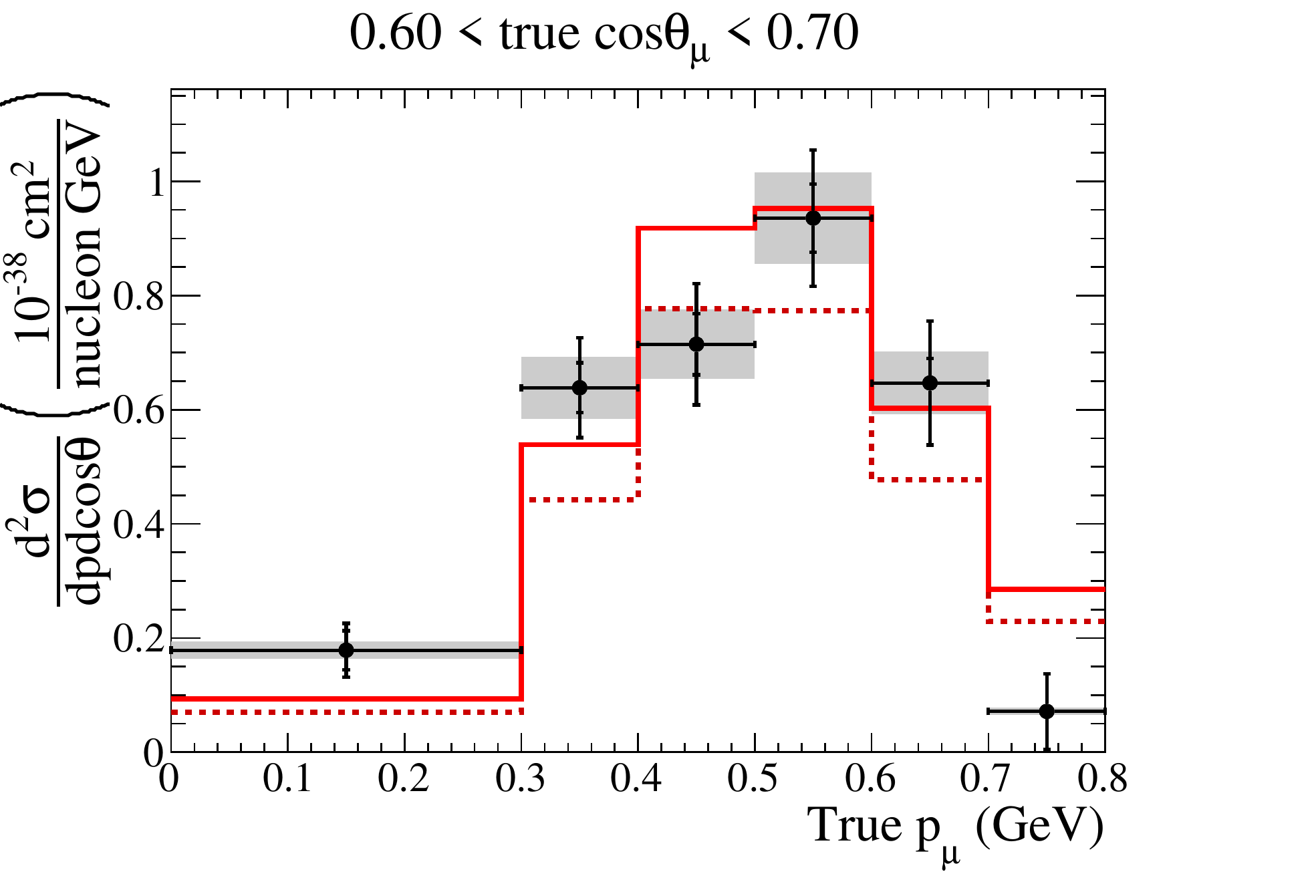}
 \includegraphics[width=6cm]{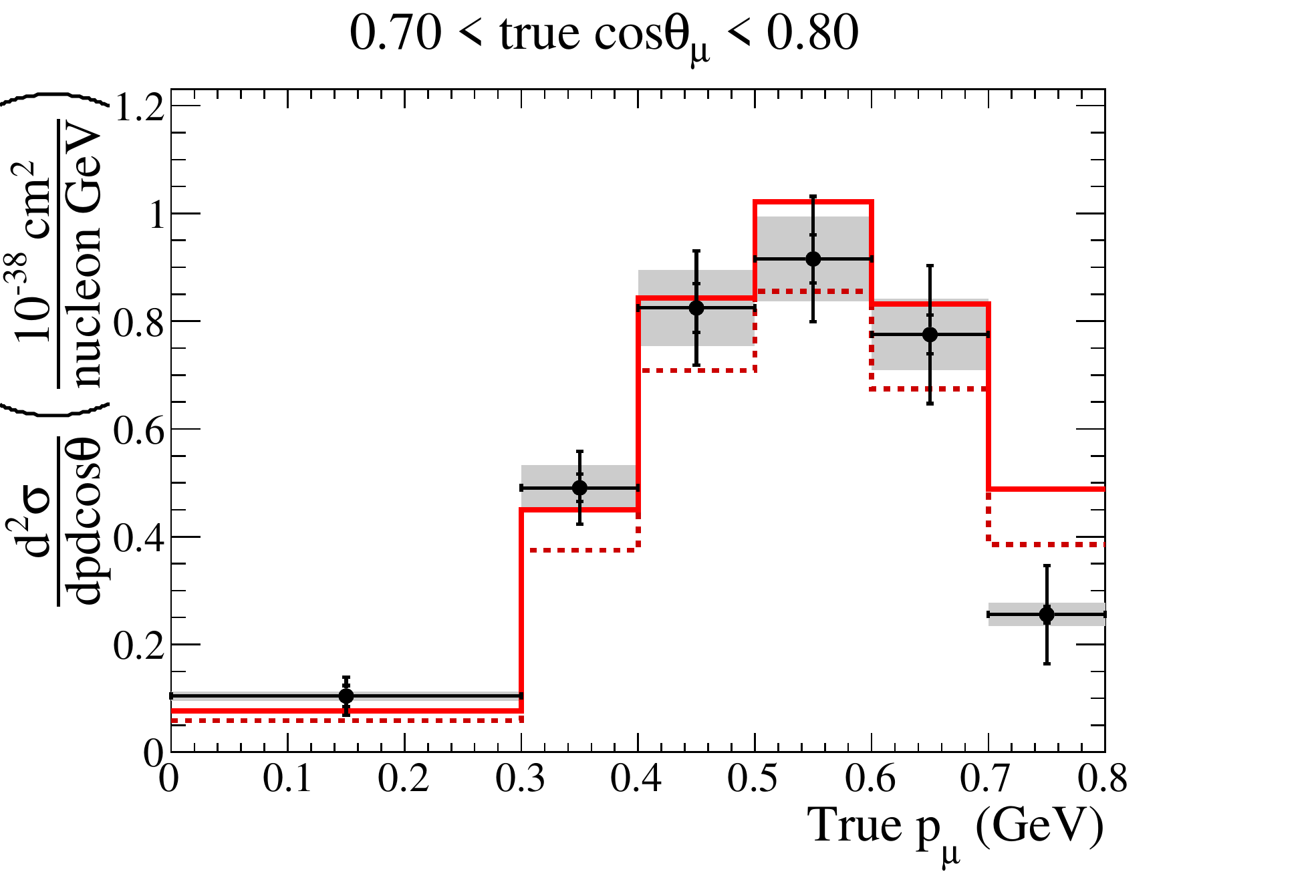}\\
 \includegraphics[width=6cm]{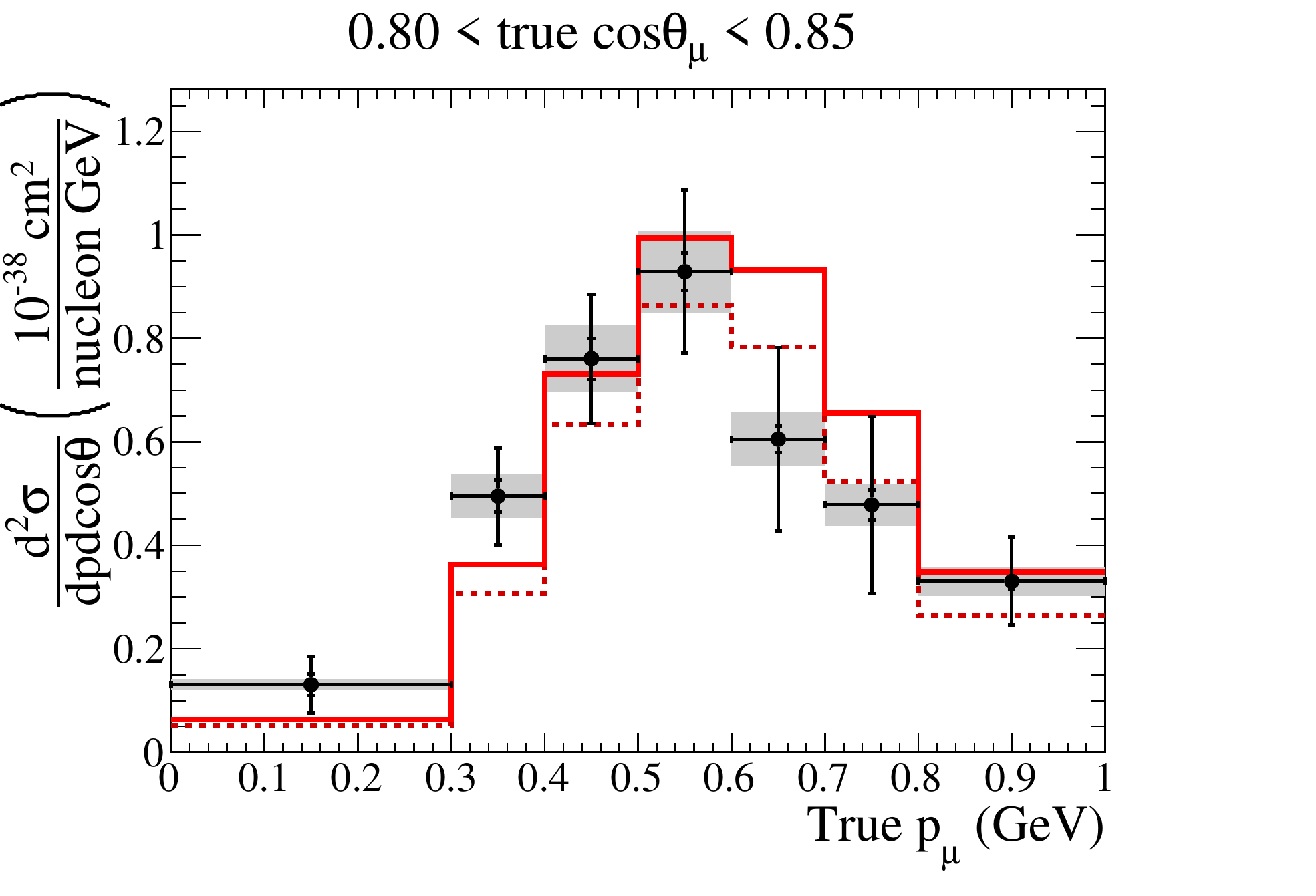}
 \includegraphics[width=6cm]{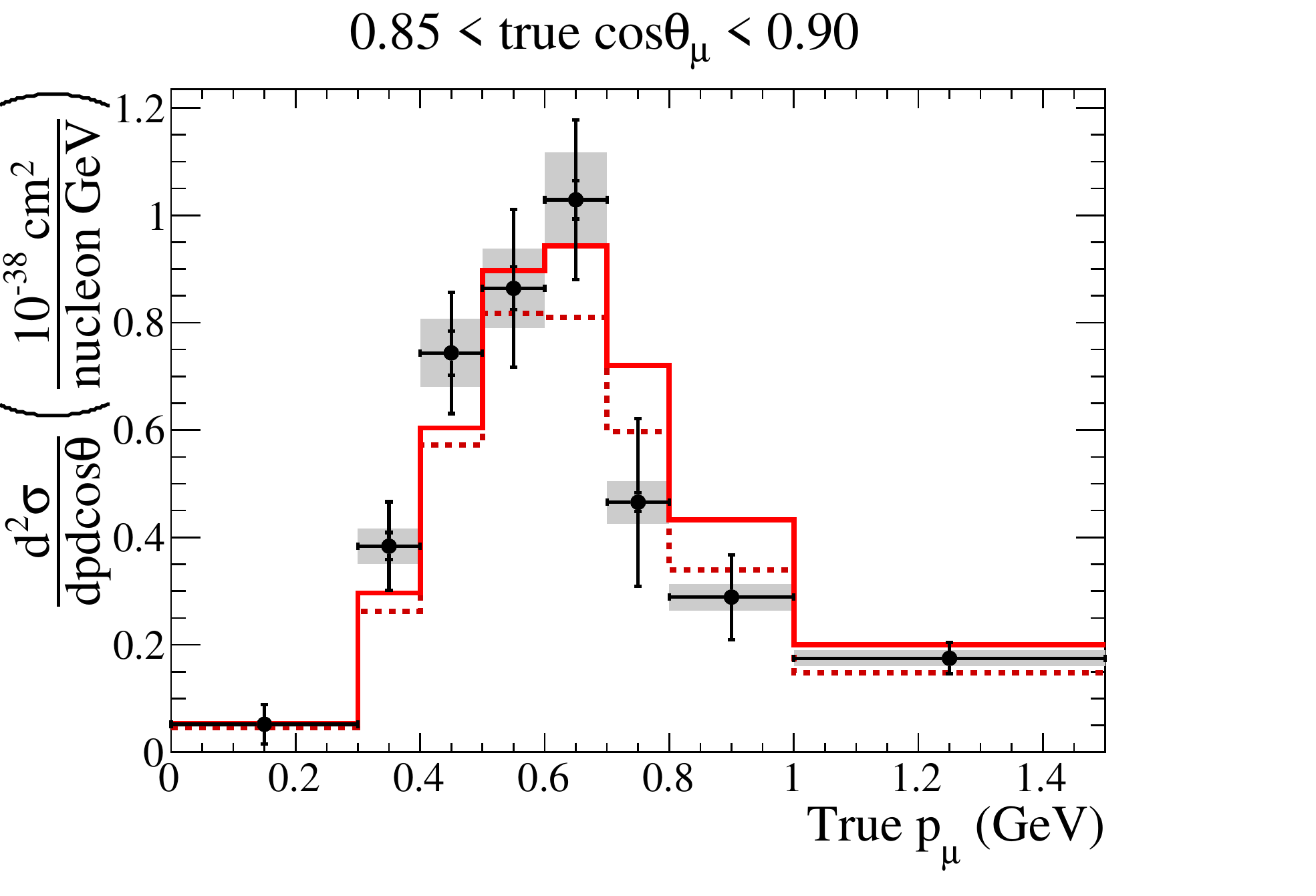}\\
 \includegraphics[width=6cm]{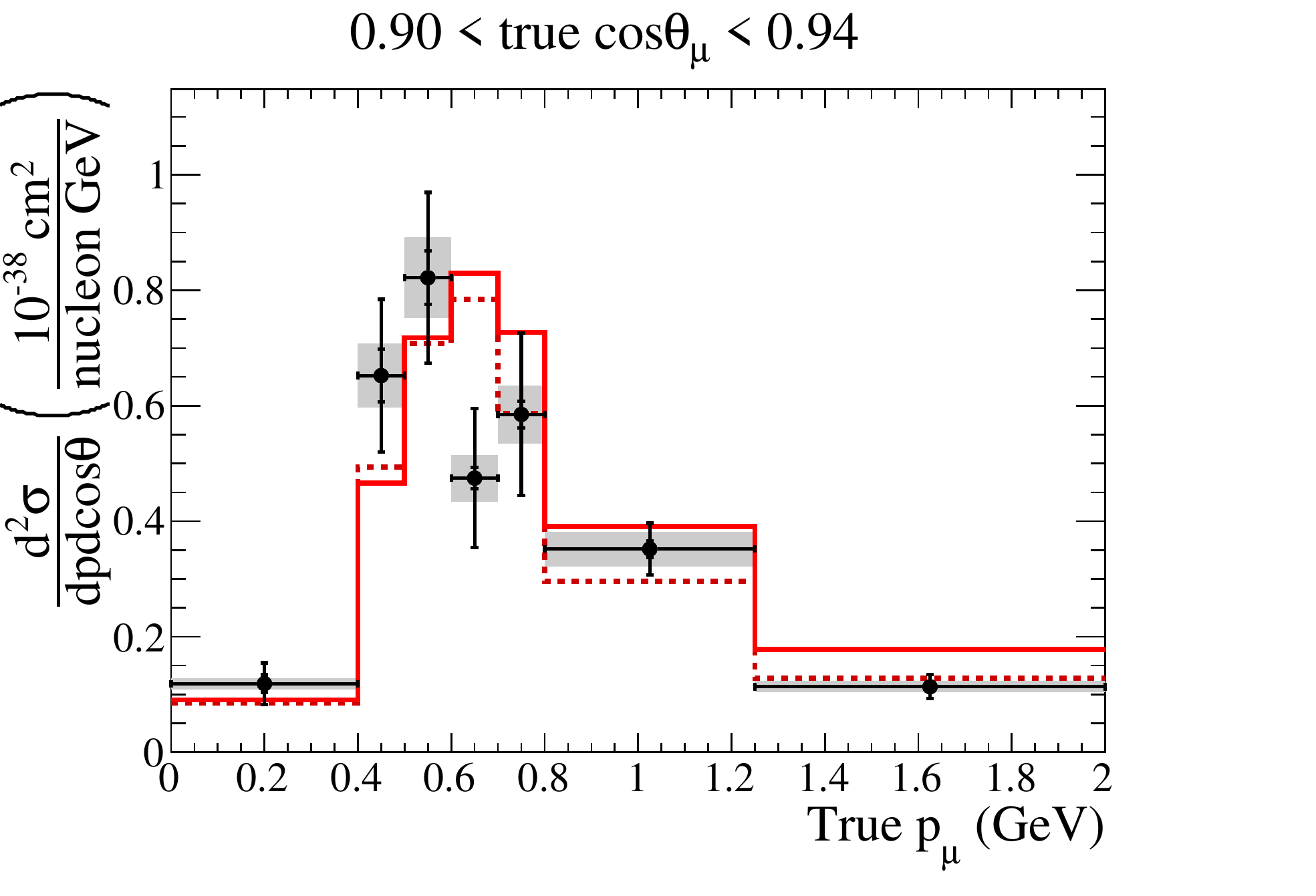}
 \includegraphics[width=6cm]{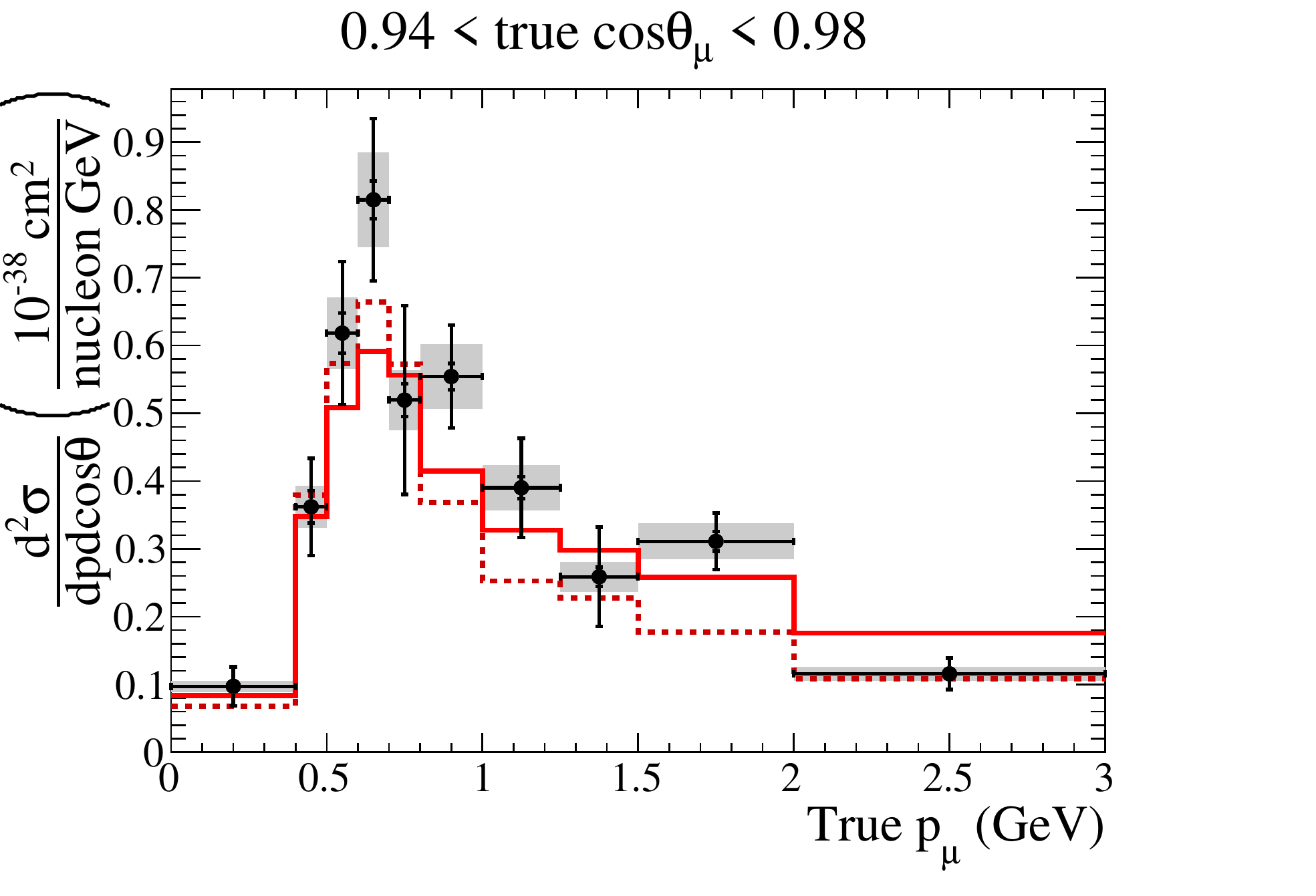}\\
 \includegraphics[width=6cm]{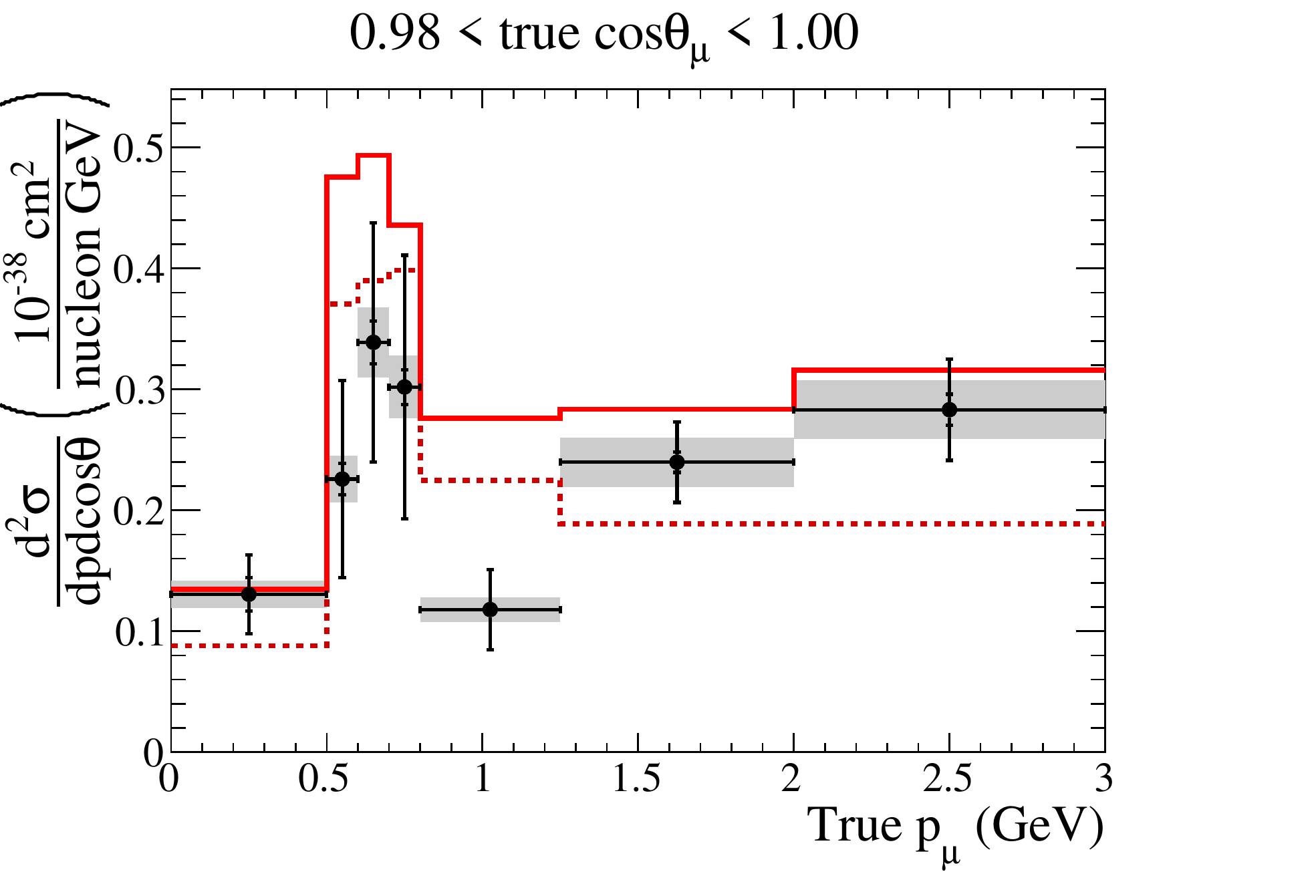}
\end{center}
\caption{Measured cross-section with shape uncertainties (error bars: internal systematics, external statistical) 
and fully correlated normalization uncertainty (gray band).
The results from fit to the data are compared to the predictions from Nieves {\it et al} (red dashed line), 
and from Martini {\it et al} (red solid line).}
\label{fig:xsecResultsLin2}
\end{figure}

The double-differential flux-integrated cross-section is shown in Fig.~\ref{fig:xsecResultsLin2}.
Here the systematic uncertainties and the data statistical uncertainties are
summed in quadrature and shown as error bars.  
The uncertainty related to the flux normalization is given as a gray band.
The results are compared
to the model of Nieves {\it et al}~\cite{Nieves:2012,Nieves:2012yz} (with a cut on the three-momentum transfer of $q_3<1.2$ GeV) 
and to the model of Martini {\it et al}~\cite{Martini:2009,Martini:2010}.
These models include corrections to the interaction for collective nuclear effects
calculated with Random Phase Approximation, as well as 2p2h contributions, i.e. neutrino interactions with 
nucleon-nucleon correlated pairs and Meson Exchange Currents (MEC).
These models do not include the contribution of \cconepi with pion re-absorption due to FSIs,
but they do include the production of a $\Delta$ resonance followed by pion-less decay in the MEC. 
Moreover the region of very small transferred $Q^2$ (most forward muon angles and higher muon momentum)
could be sensitive to the shell structure of the nucleus~\cite{pandey_2014}. 
Therefore the comparison to data has been limited
to muon momenta below 3 GeV. 
In the Appendix
the results are compared to the same models with and without the 2p2h contribution 
(Figs.~\ref{fig:xsecResultsLinMM} and \ref{fig:xsecResultsLinN})  and
to NEUT and GENIE Monte Carlo (Figs.~\ref{fig:xsecResultsLin} and \ref{fig:xsecResultsLog}).

\subsection{Analysis II}
Analysis II makes use of a Bayesian unfolding procedure to extract the CC0$\pi$ differential cross-section from a single selection which is designed around the vetoing of pions.
The selection used is the same as that used for the near detector fits in recent T2K oscillation analyses.
In this analysis there is no direct constraint on the background. The background uncertainties are taken from fits to external 
data from MiniBooNE performed by T2K, which are also used as priors in the fit of Analysis I.
\subsubsection{Event selection}
After the pre-selection described in Section~\ref{sec:preselection}, the selection is sub-divided based on the observed number of pions as done in recent T2K oscillation analyses~\cite{Abe:2015awa}.
Charged pions are tagged by searching for either a pion-like track in the TPC, a pion-like track in the FGD, or a Michel electron 
from muon decay in the FGD.  
Neutral pions are tagged by searching for electron-like tracks in the TPC. Events with no evidence for pions, charged or neutral, are kept and placed in the so-called \cczeropi 
signal sample. This event selection is found to be 72\% pure according to our NEUT Monte Carlo prediction of background events. %

\begin{figure}
\begin{center}
 \includegraphics[width=7cm]{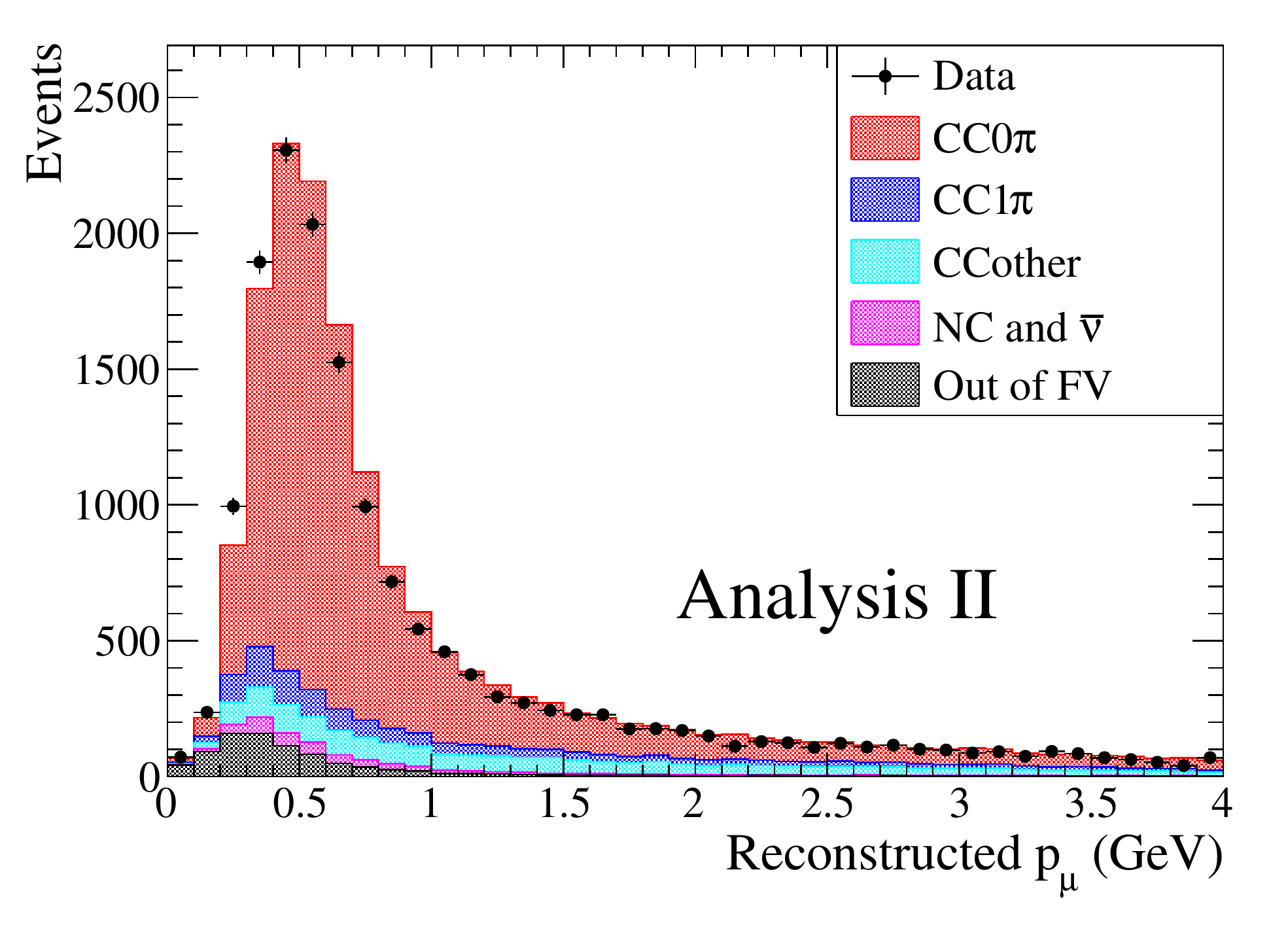}
 \includegraphics[width=7cm]{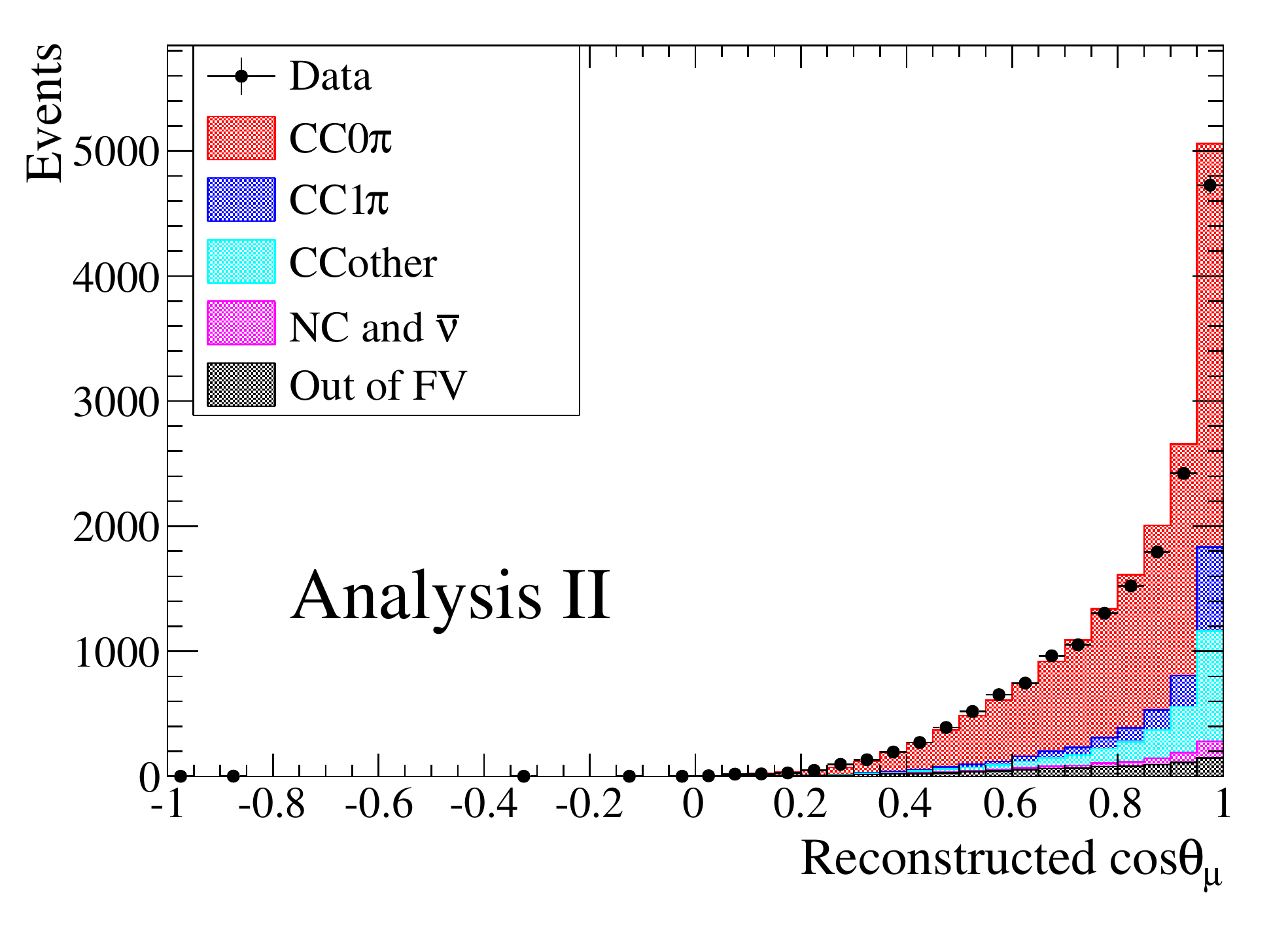}\\
\end{center}
\caption{Distribution of events in selection, separated by true final state topology. It is clear that in this selection there is negligible coverage at angles above 90 degrees.}
\label{fig:analysis2:selection_by_topology}
\end{figure}

\begin{figure}[h]
\begin{center}
 \includegraphics[width=7cm]{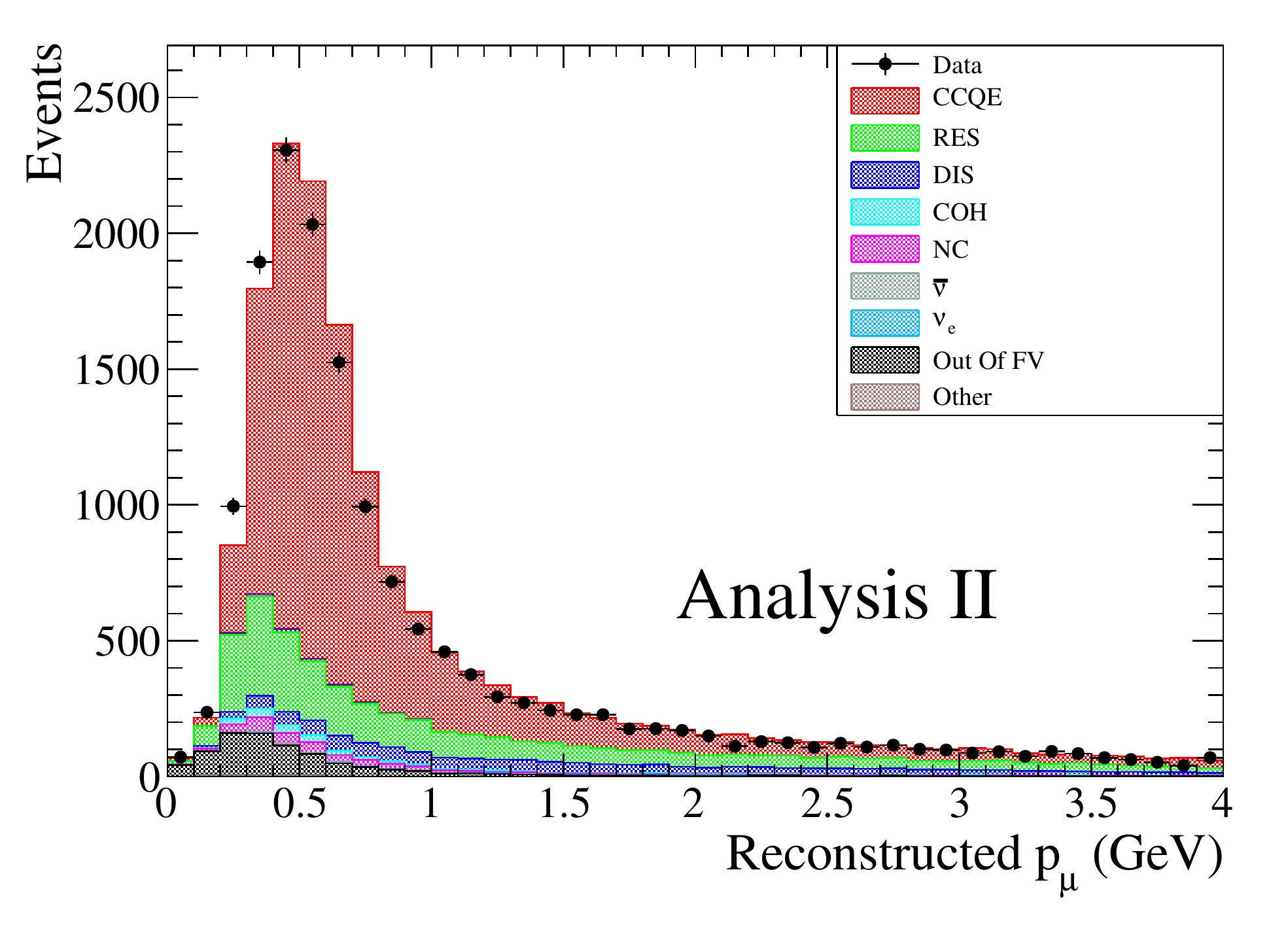}
 \includegraphics[width=7cm]{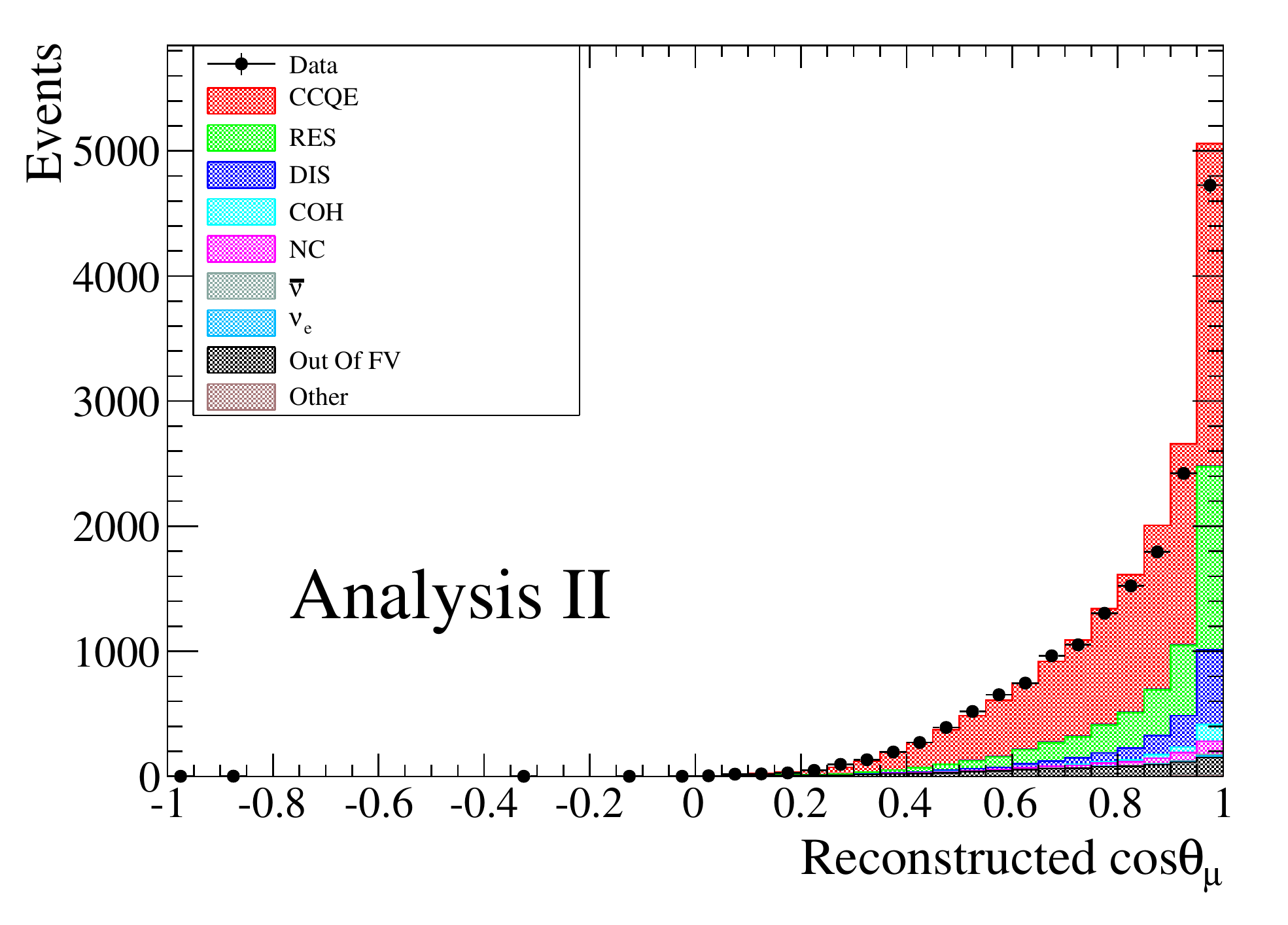}\\
\end{center}
\caption{Distribution of events in selection, separated by true reaction type.}
\label{fig:analysis2:selection_by_reaction}
\end{figure}

Fig.~\ref{fig:analysis2:selection_by_topology} shows the kinematics of the events selected as simulated by our NEUT prediction, and measured in data.
The Monte Carlo is shown divided by true final-state topology.
In Fig.~\ref{fig:analysis2:selection_by_reaction}, the same is shown but the simulation is divided by true reaction type.

\subsubsection{Cross-section extraction}
The Bayesian unfolding procedure as described by D'Agostini~\cite{D'Agostini1995487} is used to convert from reconstructed variables to true variables.

The inputs to the Bayesian unfolding are:

\begin{itemize}
  \item $P(t_i) = N(t_i) / \sum_{\alpha}^{\text{true bins}}N_{t_{\alpha}}$, the prior probability of finding an event in true bin $i$.
  \item $P(r_j|t_i)$, the probability of an event being reconstructed in bin $j$, given it originated in true bin $i$.
  \item $B_{r_j}$, the predicted background in each bin.
\end{itemize}

Using these inputs, it is possible to define the efficiency of selecting events in true bin $i$,
\begin{equation}
\epsilon_i = \sum_{\alpha}^{\text{reco bins}} P(r_{\alpha}|t_i)
\end{equation}
and the prior probability of being reconstructed in bin $j$,
\begin{align}
P(r_j) &= \sum_{\alpha}^{\text{true bins}} P(r_j|t_{\alpha}) P(t_{\alpha}) \\
       &= \frac{N_{r_j}}{\sum_{\alpha}^{\text{true bins}}N_{t_{\alpha}}}
\end{align}

Applying Bayes' theorem to the probabilities we have, results in the ``unsmearing'' matrix, which gives the probability for an event originated in true bin, $i$, given it was reconstructed in bin $j$,
\begin{equation}
P(t_i|r_j) = \frac{P(r_j|t_i) P(t_i)}{P(r_j)}
\end{equation}
and we can then use this unsmearing matrix, along with the efficiency, to obtain the unfolded estimate of the true event distribution.
\begin{equation}
N_{t_i}^{\text{\cczeropi}} = \frac{1}{\epsilon_i}\sum_{j}P(t_i|r_j)(N_{r_j} - B_{r_j})
\end{equation}

The cross-section is then calculated in the same manner as Analysis I---by scaling the unfolded true number of events in each bin by the flux, number of targets, efficiency, and bin-width, as in Eqn.~\ref{eq:xsec}.

It is possible to iterate this procedure by feeding the unfolded true distribution in the start as an updated prior.
Fake data studies showed that the first iteration is sufficient to correct for detector effects, even when the prior and fake data were generated according to models chosen to have exaggerated differences between each other.
This is because the reconstruction resolution is very good, with over 60\% of events being reconstructed in their true bin, and under 5\% of events being reconstructed more than one bin away. %

There are limitations to this approach.
We note that while the technique outlined in Ref.~\cite{D'Agostini1995487} should in principle unfold both background and signal, for this analysis, the unfolding procedure was applied subsequent to background subtraction. %
Because the background is subtracted before unfolding, this method can yield negative cross-sections in some bins that contain large backgrounds.
This cannot occur in Analysis I, because the background contributions are fit simultaneously.

\subsubsection{Treatment of systematic uncertainties}
All systematic uncertainties were propagated using a sample of toy experiments that were generated assuming different underlying parameters.
Each toy experiment was unfolded using the same algorithm, and the results were used to calculate a covariance matrix defined as:
\begin{equation}
V_{ij} = \frac{1}{N} \sum_{s_n = 1}^{N} \left ( \sigma^{(s_n)}_i - \sigma^\text{nominal}_i \right ) \left ( \sigma^{(s_n)}_j - \sigma^\text{nominal}_j \right ),
\label{eqn:analysis2:covarianceMatrix}
\end{equation}
where, for each source of uncertainty, labeled by $s$, $N$ pseudo experiments are performed, giving a new differential cross-section $\sigma^{(s_n)}$ each time, and the nominal cross-section in bin $i$ is given by $\sigma^\text{nominal}_i$.

\subsubsection{Region of reported results}
The region below 0.2 GeV in muon momentum contains a significant amount of external backgrounds, and suffer from 
a very low efficiency due to reconstruction difficulties.
For this reason, no result is reported below 0.2 GeV.
This should not be interpreted as measuring zero cross-section in this region, rather that the cross-section has not been measured in this region.

\subsubsection{Results}
\begin{figure}
\begin{center}
  \includegraphics[width=6cm]{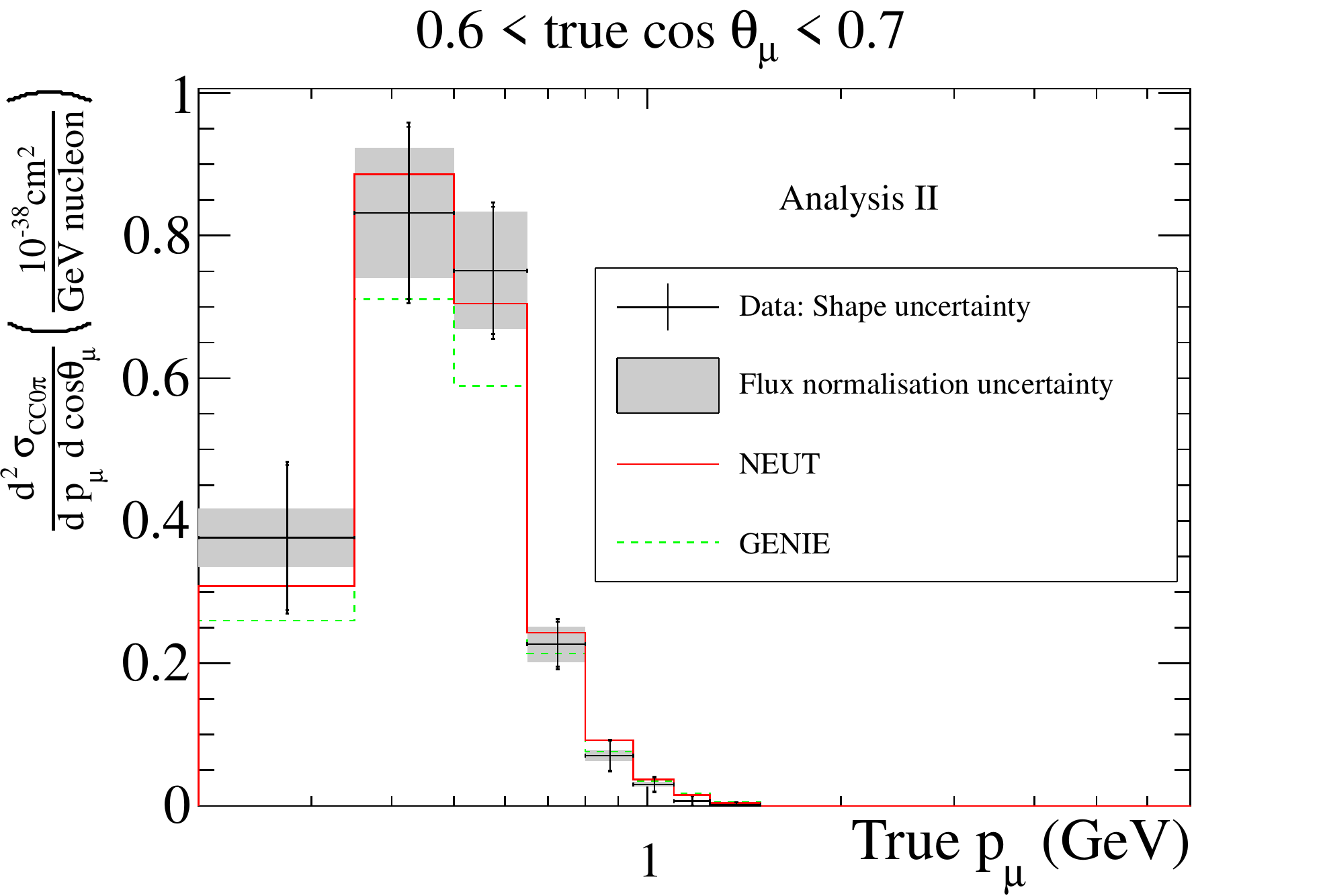}
  \includegraphics[width=6cm]{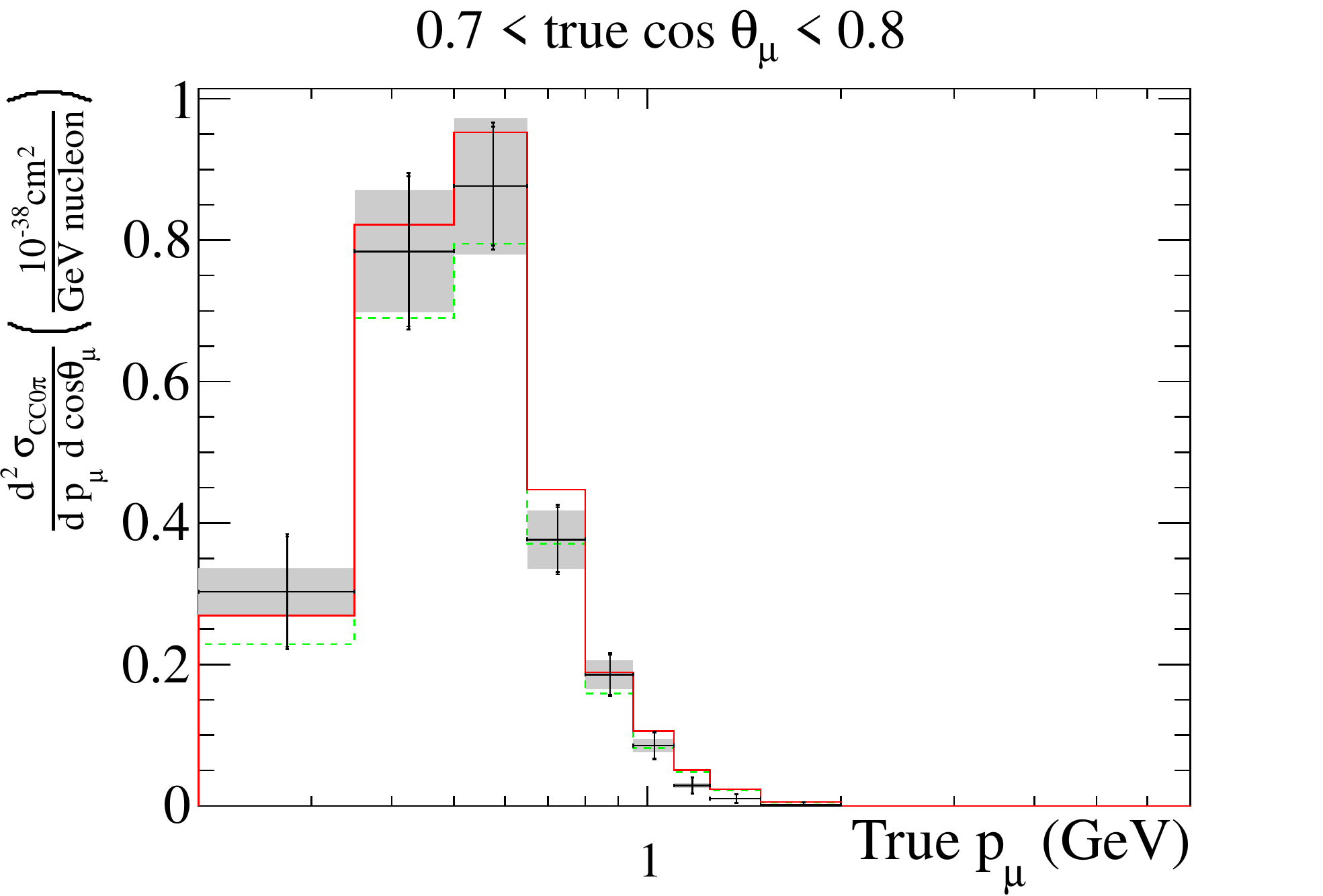}\\
  \includegraphics[width=6cm]{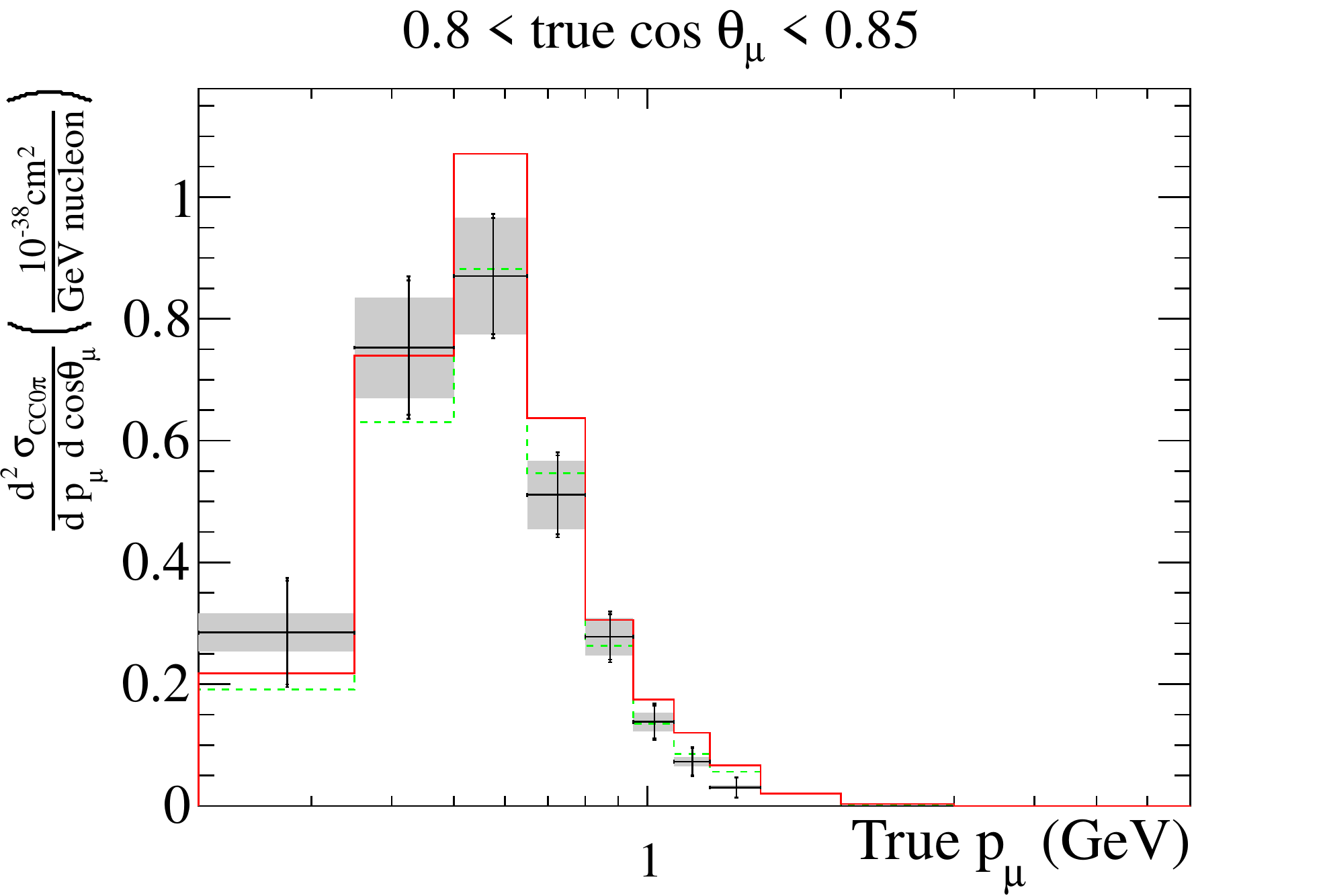}
  \includegraphics[width=6cm]{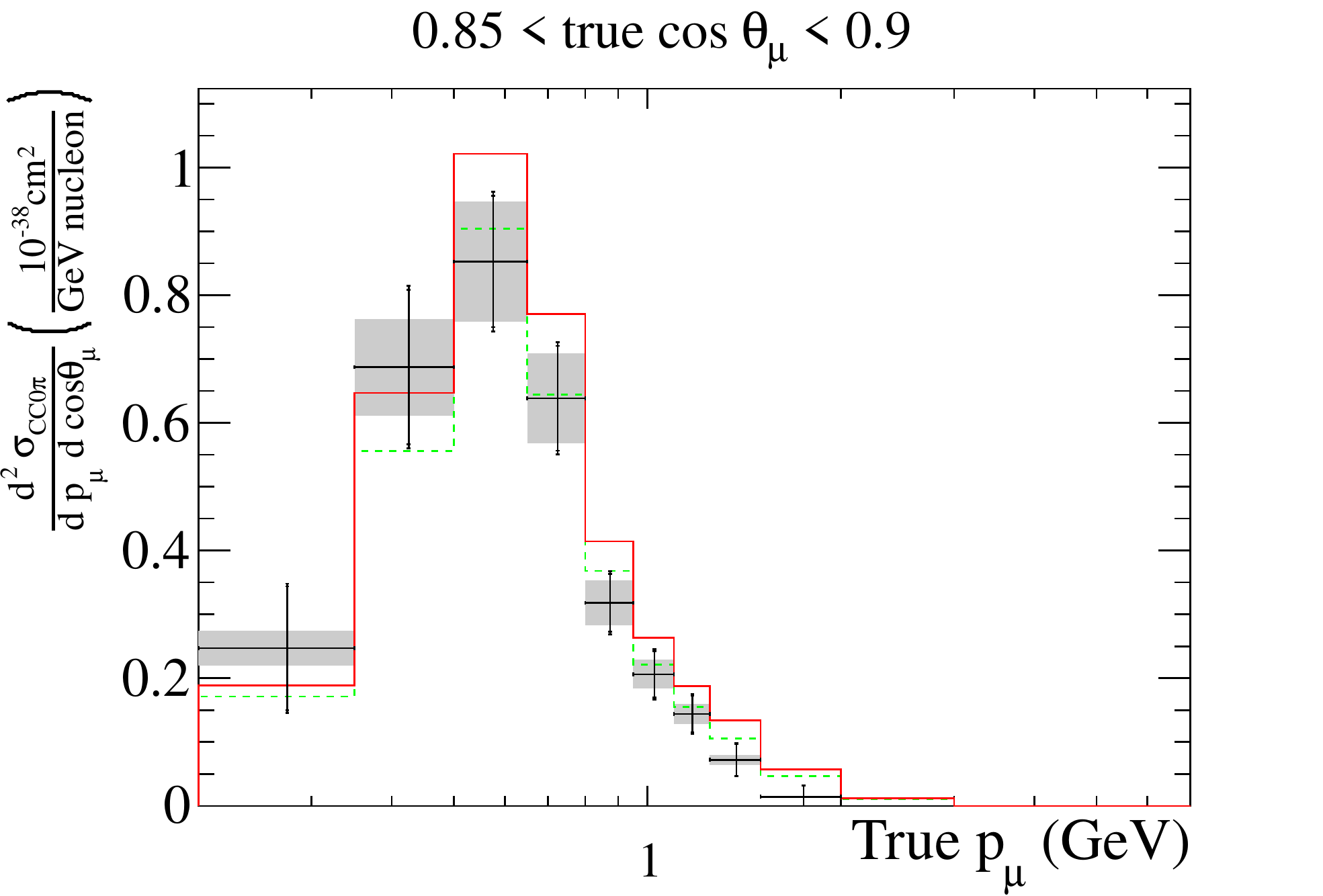}\\
  \includegraphics[width=6cm]{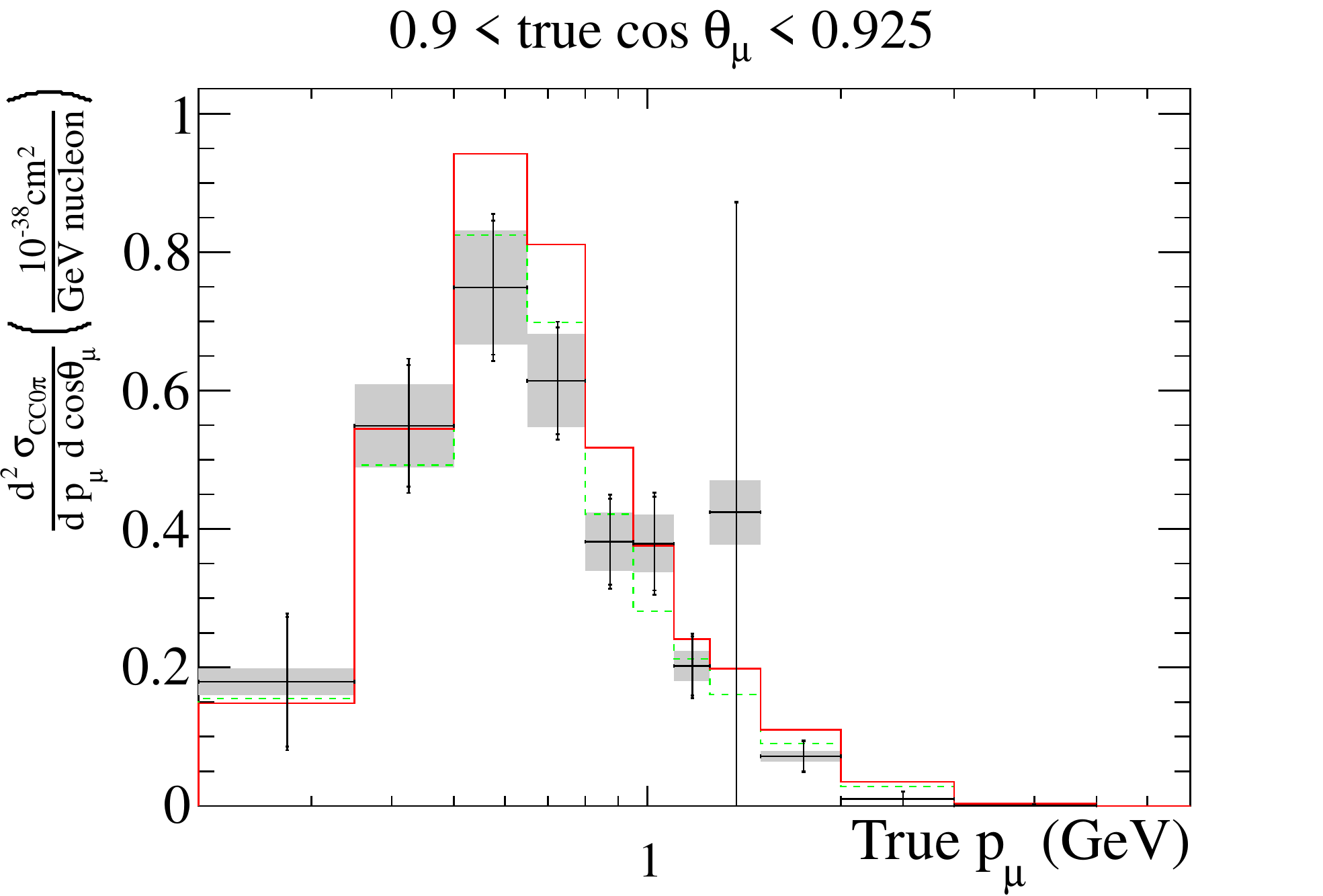}
  \includegraphics[width=6cm]{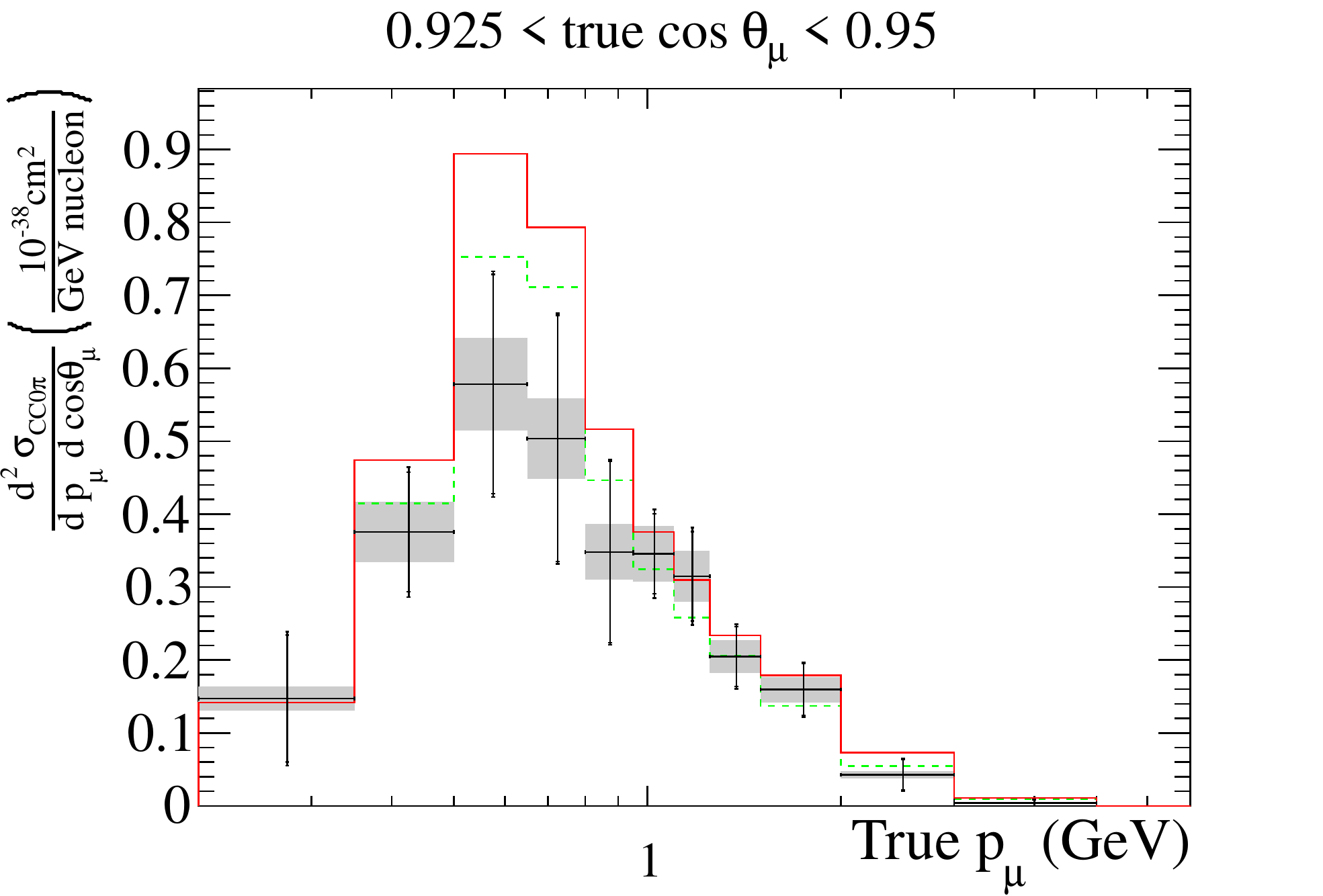}\\
  \includegraphics[width=6cm]{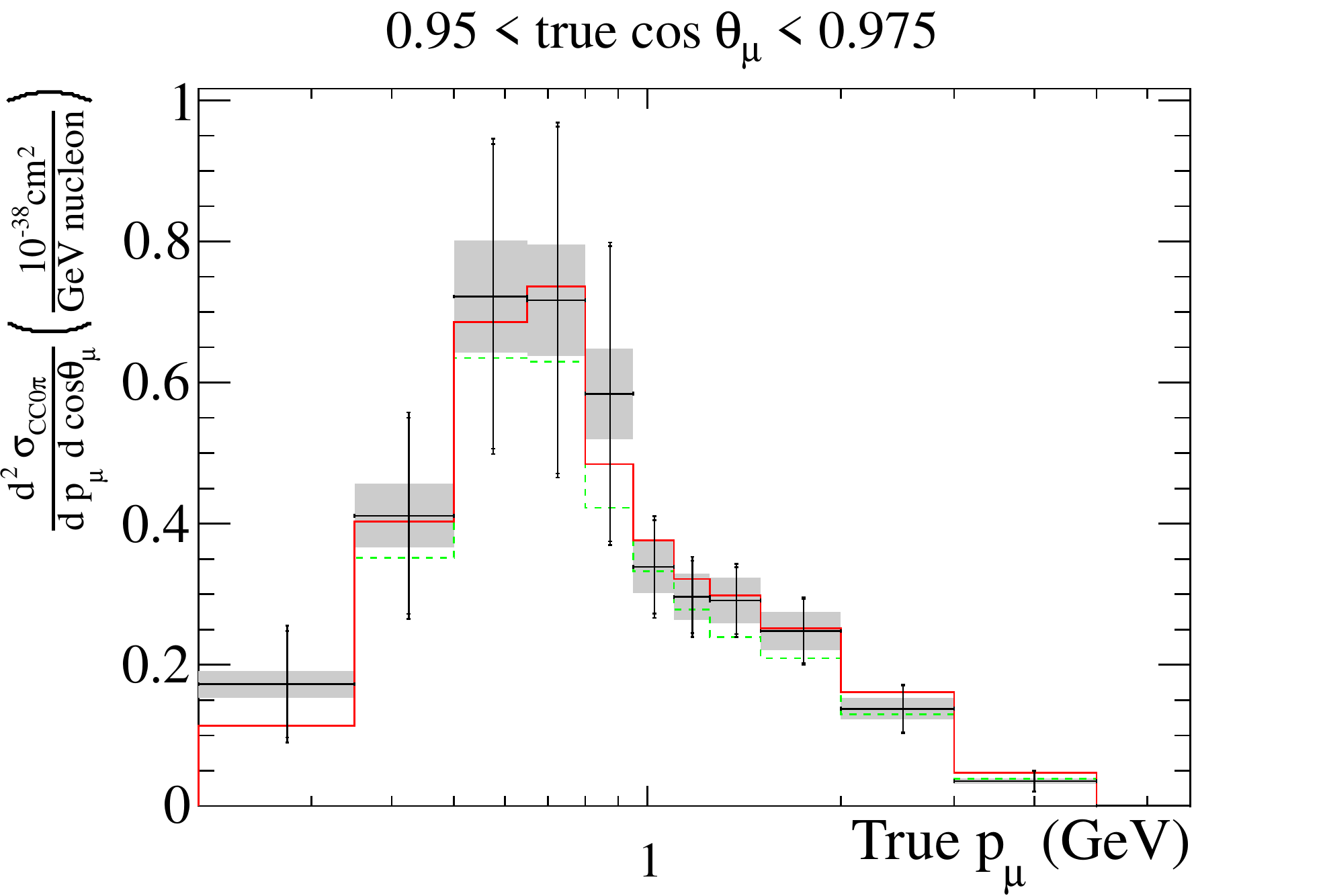}
  \includegraphics[width=6cm]{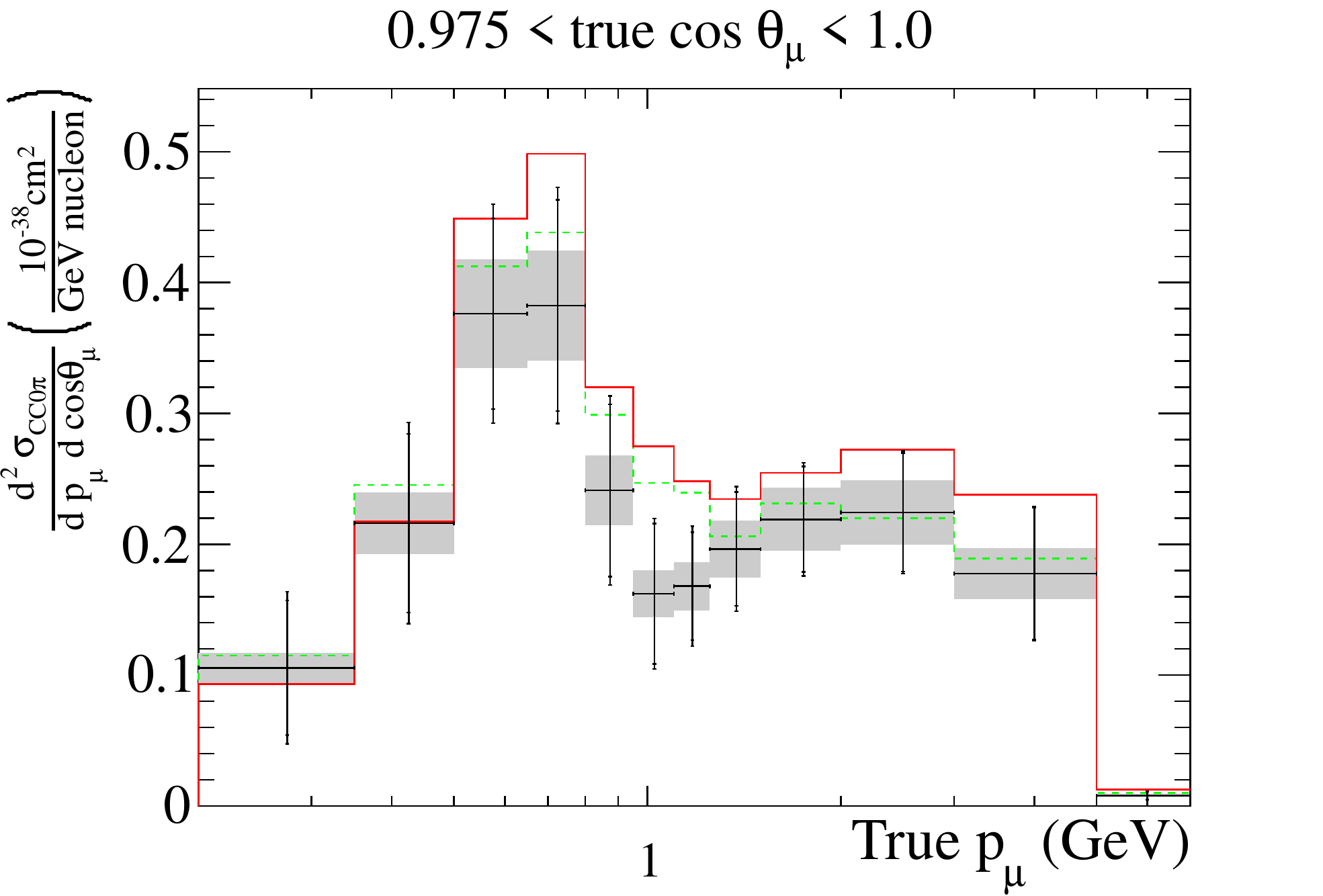}\\
\end{center}
\caption{Measured cross-section with shape uncertainties (error bars: internal systematics, external statistical) and fully correlated normalization
uncertainty (gray band) as a function of muon momentum for different angular bins.}
\label{fig:analysis2:doubleDiffXsec}
\end{figure}

The total signal cross-section per nucleon integrated over the restricted muon kinematics phase space ($p_{\mu} > 0.2$ GeV and $\cos\theta_{\mu} > 0.6$) is:
\begin{equation}
\label{eq:xsecInt_analysis2}
\sigma = (0.202 \pm 0.0359 \text{(syst)} \pm 0.0026 \text{(stat)}) \times 10^{-38} \text{cm}^2 \text{nucleon}^{-1}
\end{equation}
to be compared with the NEUT prediction: $0.232 \times 10^{-38} \text{cm}^2 \text{nucleon}^{-1}$. The uncertainty is fully dominated by the flux normalization.
When considering the full phase space the results agree very well with those in Analysis I, however they suffer from large uncertainties that arise from extrapolating beyond the visible phase space.
Fig.~\ref{fig:analysis2:doubleDiffXsec} shows the cross-section as a function of momentum for different angular bins.

\section{Discussion\label{sec:discuss}}

The results are only presented in the form of a double-differential cross-section, and a total flux integrated cross-section per nucleon.
Since Analysis II has no efficiency in the low-momentum and high-angle region, the phase space has been reduced in that analysis.%

Cross-sections are often calculated as a function of $Q^2$, providing 1-dimensional distributions which are useful for comparisons between models;
however, because of the limited acceptance of ND280, this would make the result very model-dependent.
Indeed, each $Q^2$ bin would contain contributions from events with different muon kinematics thus requiring very different efficiency corrections.
For instance the efficiency for forward muons in ND280 is very good, while it is lower for backward tracks (zero in Analysis II).
Therefore the efficiency correction in each bin of $Q^2$ depends strongly on the relative number of events with forward or backward muons,
which depends in turn on the particular model assumption.
Moreover, in fake data studies for Analysis II, it was shown that changing the assumed (prior) distribution in $(p_{\mu}$, $\theta_{\mu})$
space could have a drastic effect on the shape and normalization in $Q^2$, or $p_{\mu}$, and $\theta_{\mu}$ separately.
The efficiency corrections in bins of $(p_{\mu}$, $\theta_{\mu})$ are instead mostly model-independent since these are the actual
variables measured in the detector.

The cross-section is reported for all events without any pions in the final state. Thus the signal includes contributions from
\cconepi events where the pion has been reabsorbed through FSIs. Such contributions can only be estimated from Monte Carlo,
and are accompanied by very large uncertainties. The track counting-based selection of Analysis I has a smaller contamination
from these events (i.e., a larger CCQE purity) with respect to Analysis II; 
the contribution of \cconepi production with pion absorption is of the order of a few percent, 
except in the forward bin (up to 15\%) and in the first momentum bin of each angular bin (up to 50\%).

\begin{figure}
\begin{center}
 \includegraphics[width=6cm]{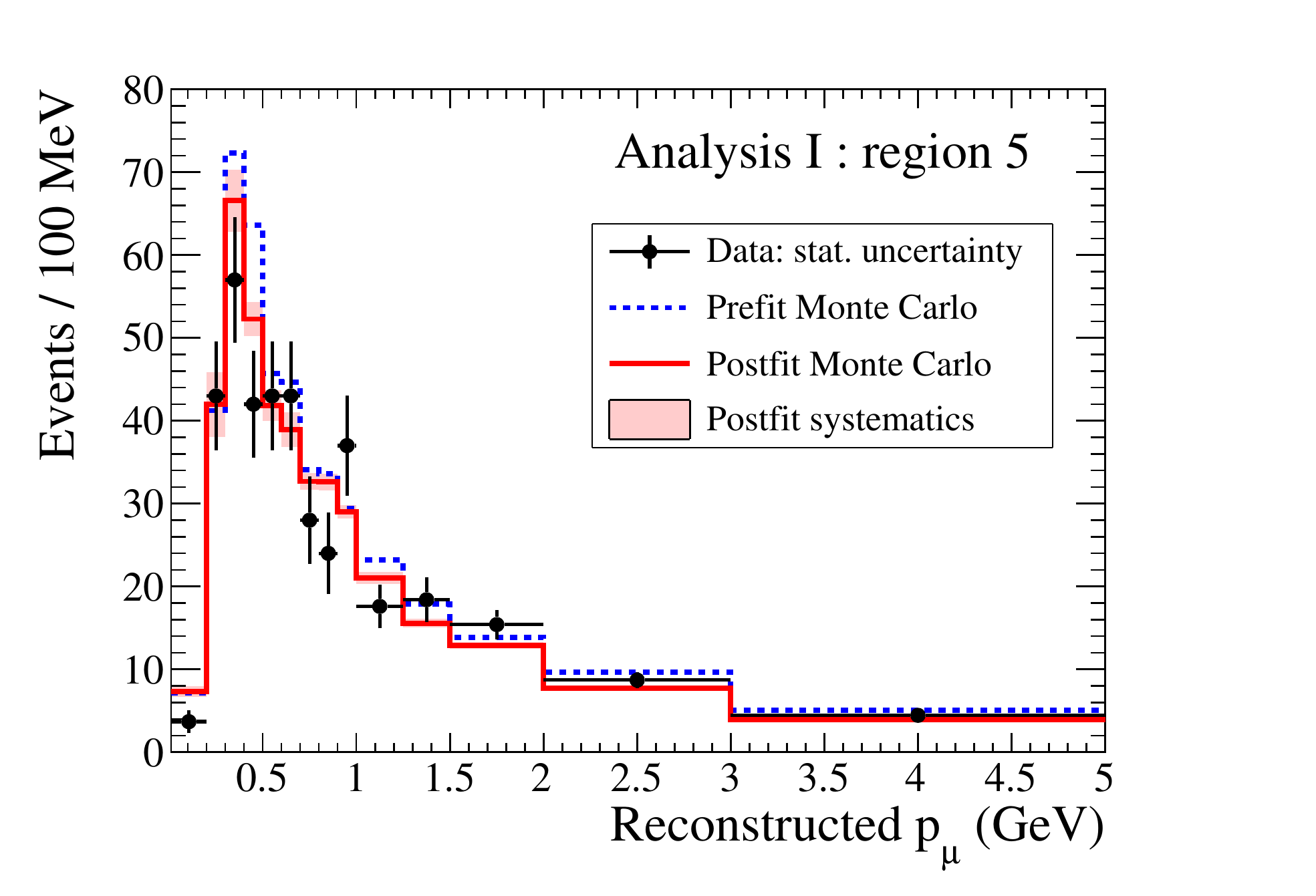}
 \includegraphics[width=6cm]{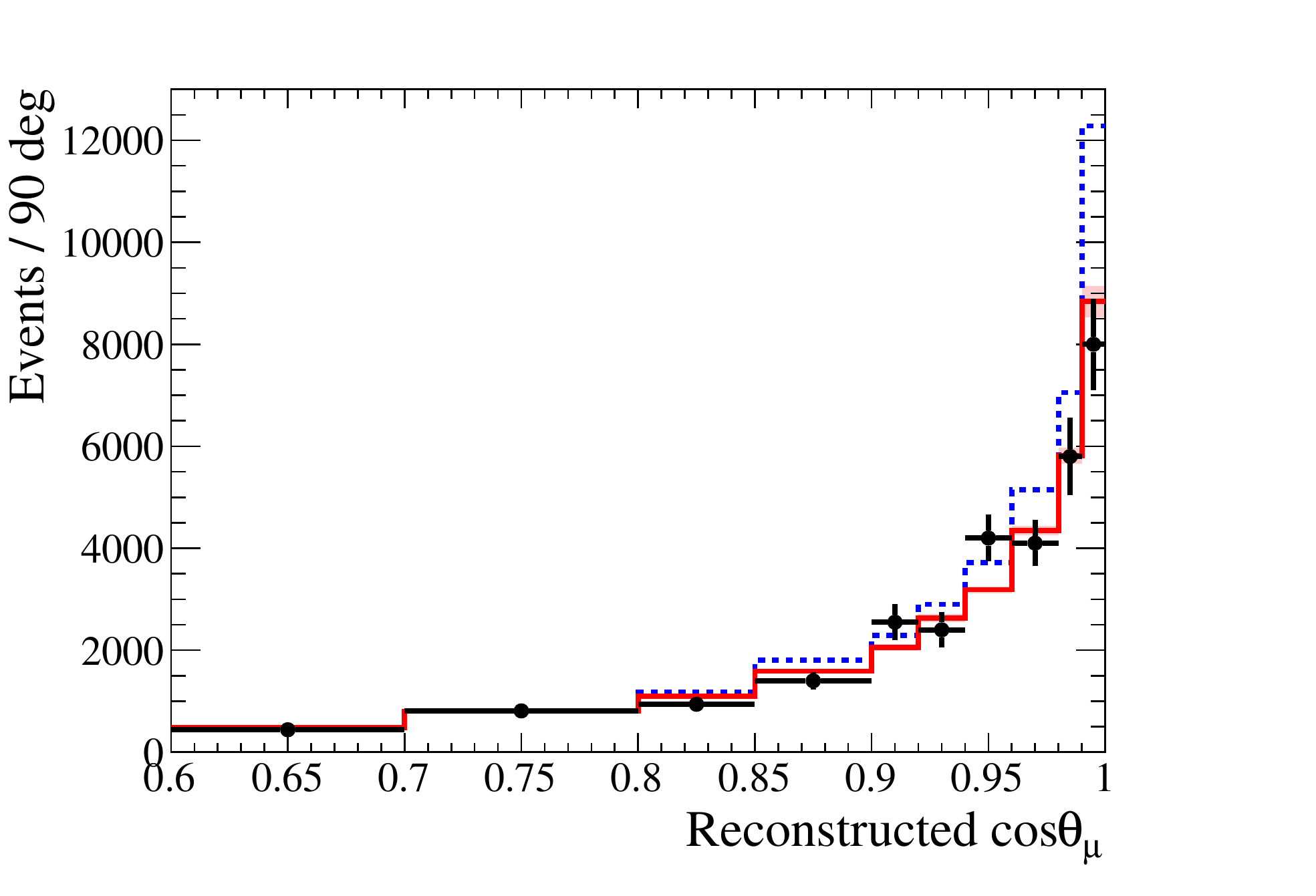}\\
 \includegraphics[width=6cm]{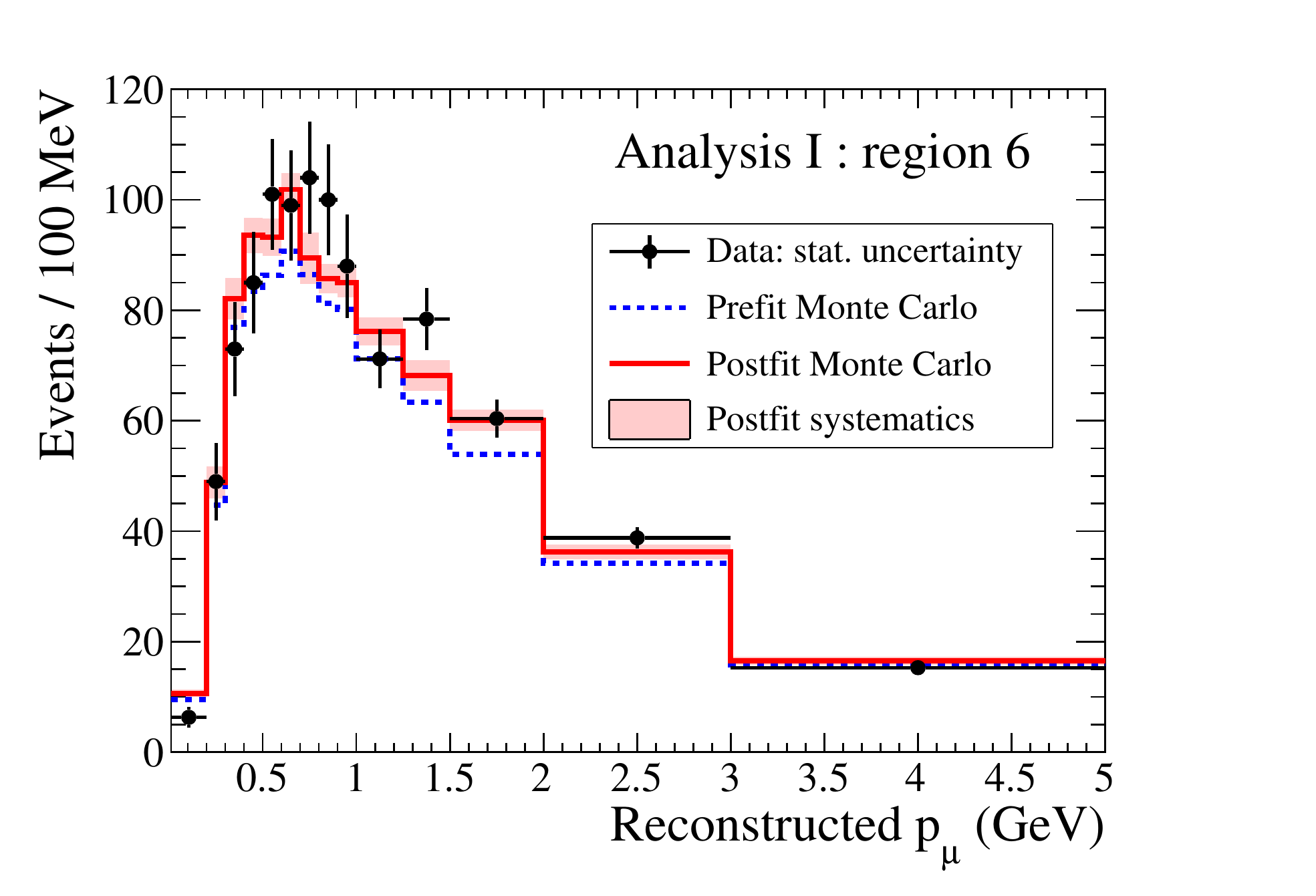}
 \includegraphics[width=6cm]{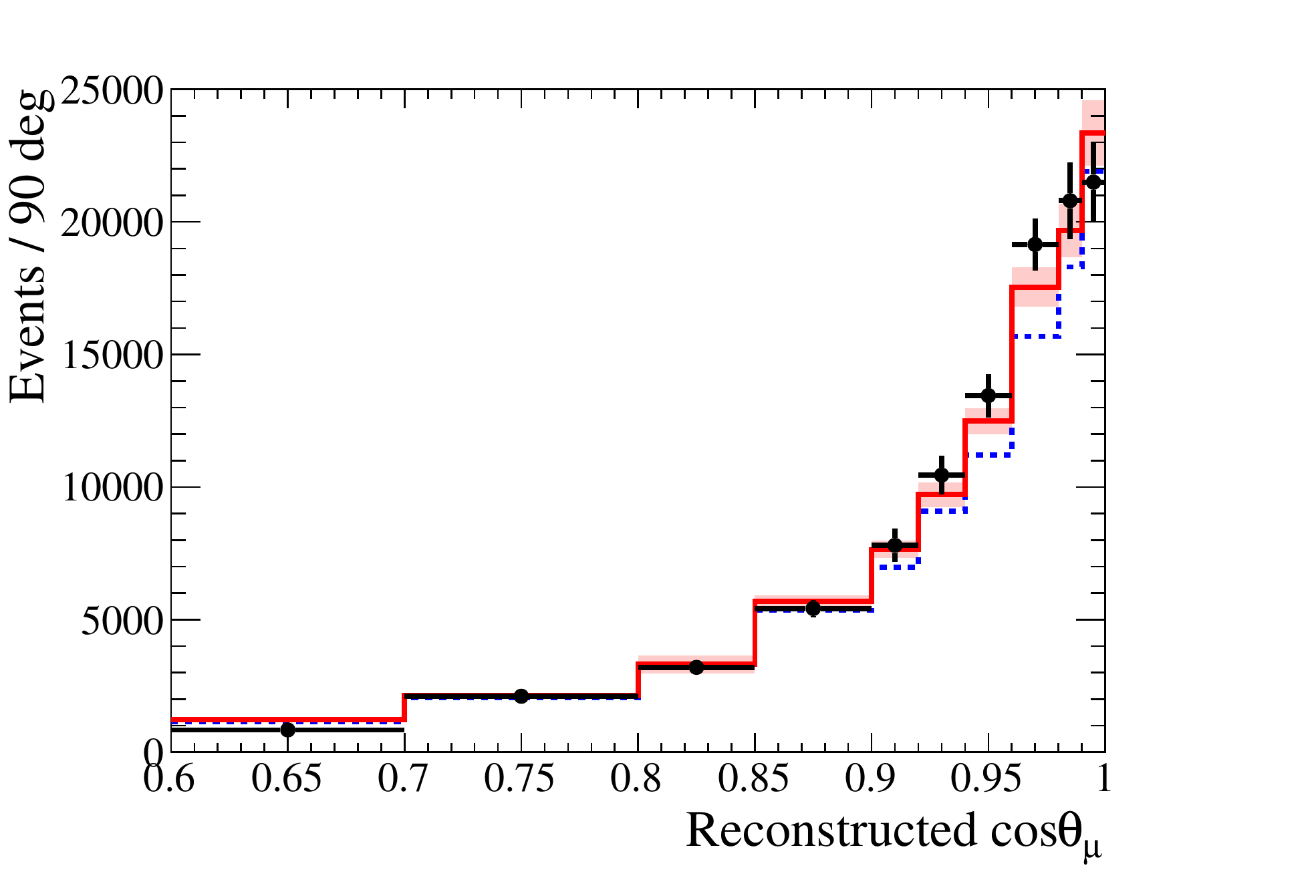}\\
\end{center}
\caption{Results of the fit to the control regions: the distribution of selected events in the \cconepi control region
(region 5, top) and in the DIS control region (region 6, bottom), as a function of muon momentum (left) and
muon $\cos\theta$ (right). The data are shown in black (with statistical errors), Monte Carlo predictions before
the fit are shown by the dotted blue line, and those after the fit are in solid red (with systematics errors indicated by the pink band).}
\label{fig:CRResults}
\end{figure}

The two analyses differ, most substantially, in the methods used for the estimation of backgrounds, signal selection, and cross-section extraction.
Analysis I uses two control regions to constrain the background parametrization, while
Analysis II makes use of Monte Carlo predictions to estimate the background.
Given the discrepancy between the present modeling of the \cconepi process and the available measurements 
in MiniBooNE~\cite{AguilarArevalo:2010zc} and MINER$\nu$A~\cite{Walton:2014esl,Fiorentini:2013ezn},
there are large uncertainties on this process and it is important to constrain them from T2K data.
In Fig.~\ref{fig:CRResults} the results of the fit to the control regions in Analysis I are shown;
the most visible effect is a reduction of the events with very forward muons in
the \cconepi-dominated region. The overall normalization of the DIS-dominated region is also slightly increased.
Analysis II includes large uncertainties in the backgrounds that are subtracted, to cover possible 
Monte Carlo mis-modeling of this type, and it has limited the phase space to above 0.2 GeV to avoid the region
where the signal-over-background ratio is low. 

The event selection for Analysis I, based on track counting and explicit proton identification,
results in a better \cczeropi purity (87\%), at the expense of a smaller efficiency (39\%),
with respect to the selection used in Analysis II (purity 73\%, efficiency 48\%). 
Analysis II is instead based on pion counting, and thus
is required to take into account uncertainties due to possible mis-modeling of pions in the detector.
Because of the detector geometry, pion rejection is worse at lower momenta and higher angles.
Analysis I includes events where the muon does not reach the TPC (region 4) thus
increasing the coverage of the phase space for high-angle muons, but in this case a proton
has to be reconstructed in the TPC to improve the purity. 
On the other hand, the explicit reconstruction and identification of protons in Analysis I can be affected by
cross-section modeling and proton FSI uncertainties. 
The selection of Analysis II is not subject to uncertainties related to the proton kinematics, as it does not attempt to measure the protons in the event.
In particular, events which have two protons, where both are energetic enough to
result in a reconstructible track in the detector, are included in Analysis II while they are excluded
in Analysis I. These events can be due to nucleon-nucleon correlations or
to meson exchange current processes or can be due to proton production through FSIs.  
Studies with generators which include these effects suggest that in this dataset 
we would expect to see around 70 events with two visible protons 
(to be compared with a number of signal selected events of about 10000).
This category of events will be studied in dedicated analyses in the future. %
It should be noted that although the event selections are different, there is an overlap of approximately 80\% between the two samples.

For the differential cross-section extraction, %
Analysis II uses an unfolding method to correct for detector effects, as with previous cross-section analyses by T2K~\cite{Abe:2013jth,Abe:2014agb}, MiniBooNE~\cite{AguilarArevalo:2010zc} and MINER$\nu$A~\cite{AguilarArevalo:2010zc,Fiorentini:2013ezn,Walton:2014esl}.
Analysis I uses a likelihood fit instead that is similar to the one used for T2K oscillation analyses~\cite{Abe:2015awa}.
The likelihood fit in Analysis I allows the fit parameters describing the theory systematics and the detector systematics to be kept separate,
in contrast to the unfolding procedure where the final result is a convolution of the detector and theory parameters.
Given the present poor knowledge of the modeling of signal and background, the likelihood fit
allows one to check that the systematics theory parameters converge to meaningful values.
The unfolding procedure in Analysis II has the feature that the statistical and systematics error estimates
depend on the amount of regularization (or number of iterations), which needs to be considered.
On the other hand, given the complexity of the fit in Analysis I and the large number of nuisance parameters it includes,
it is important to compare the results using an independent and simpler method, as represented by the unfolding used
in Analysis II.

\begin{figure}
  \centering
      \includegraphics[scale=0.4]{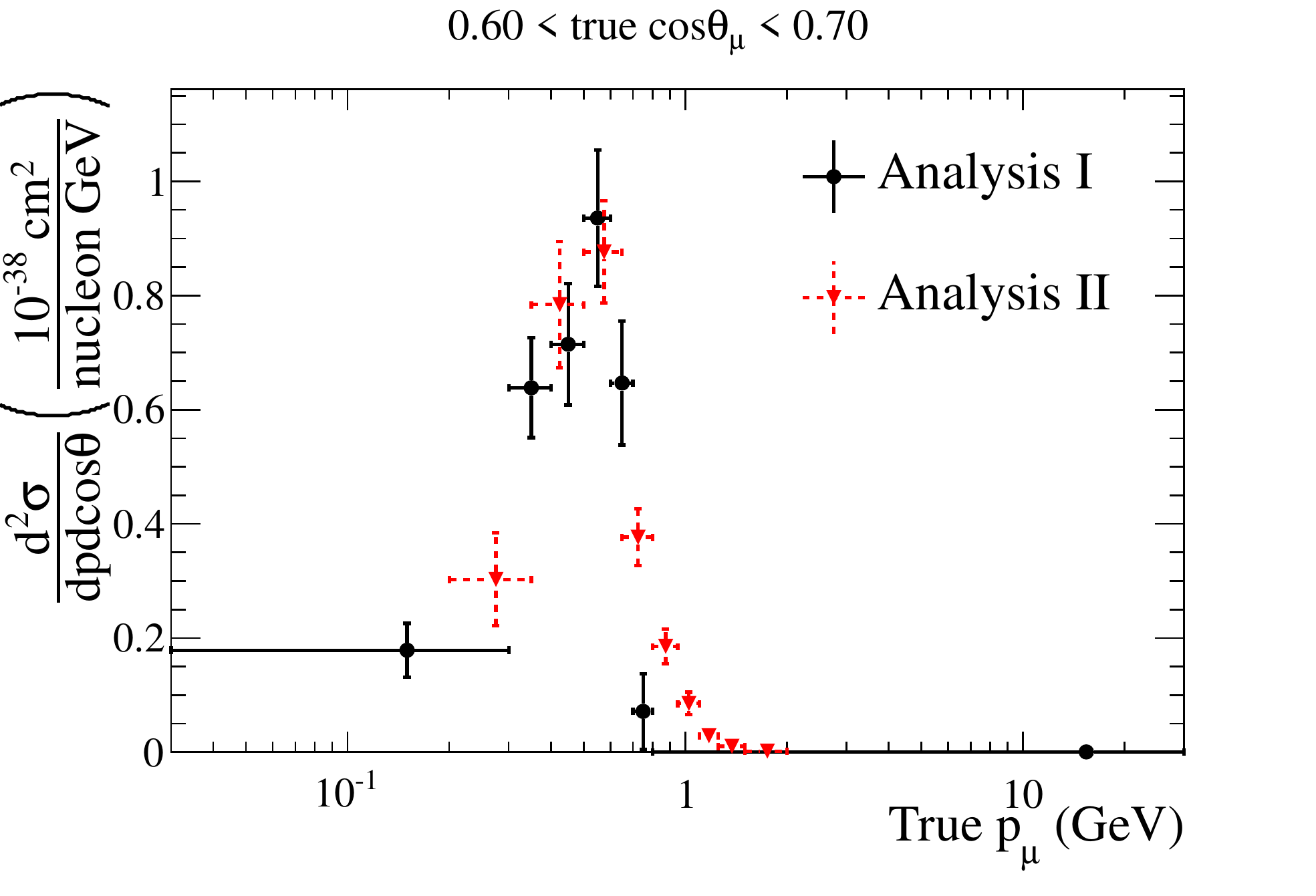}
      \includegraphics[scale=0.4]{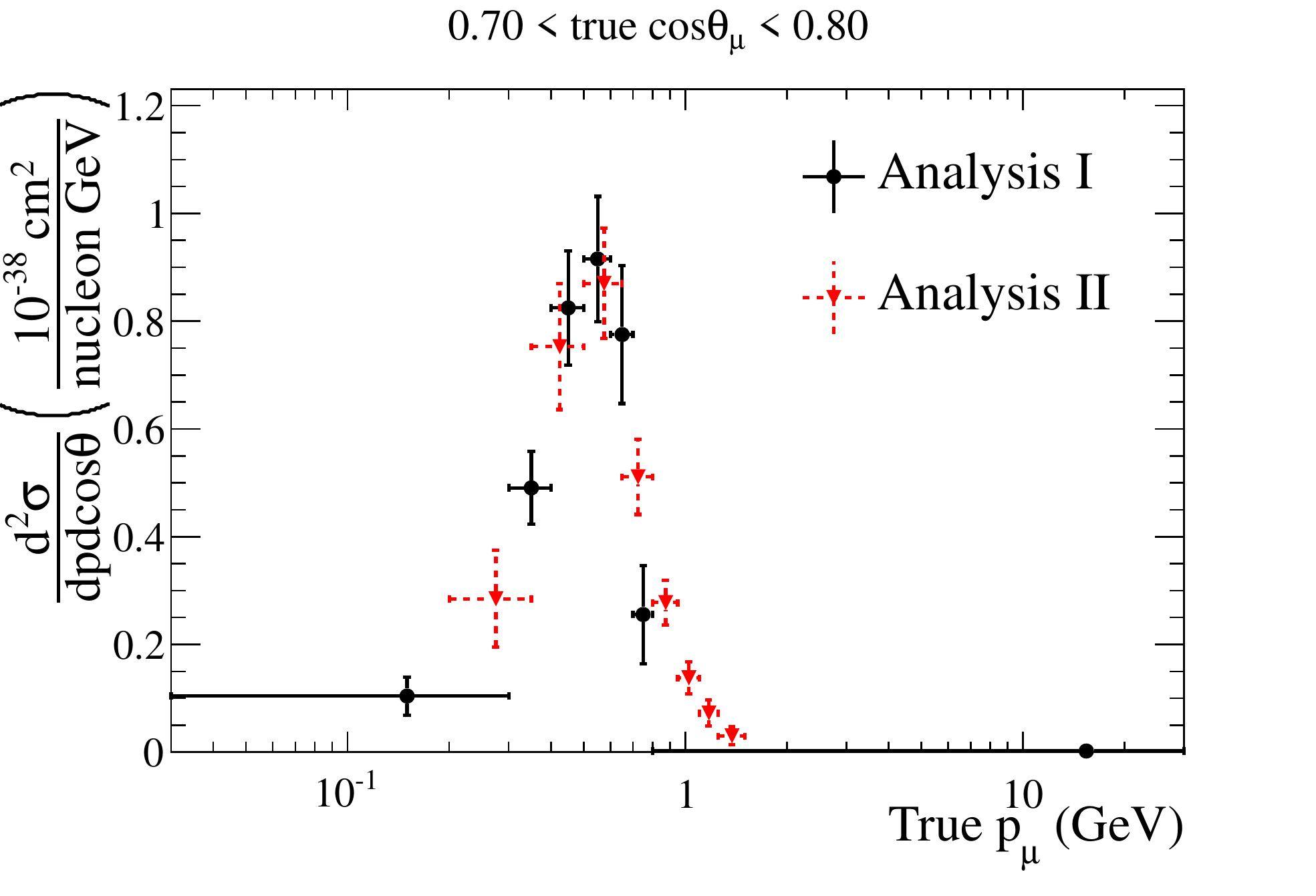}\\
      \includegraphics[scale=0.4]{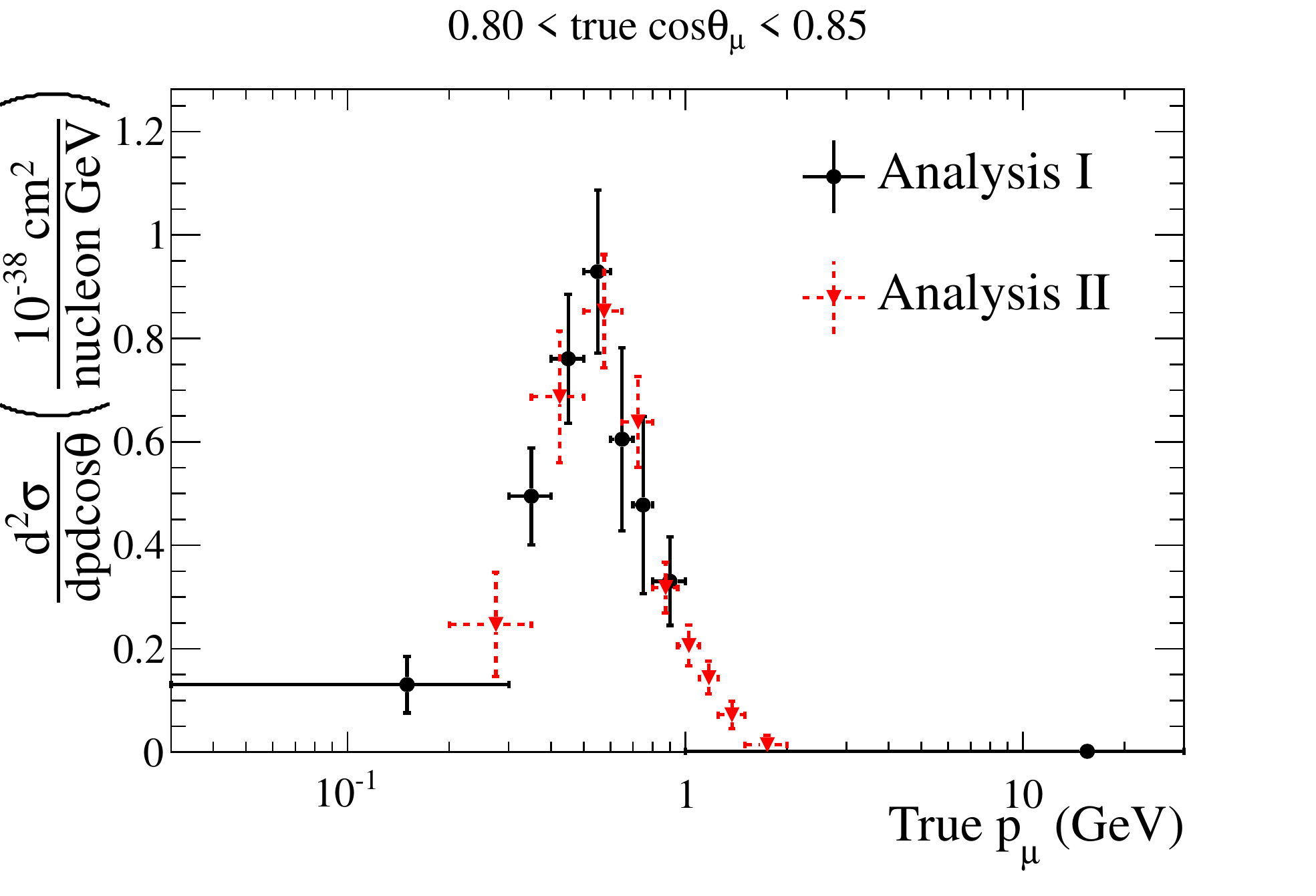}
      \includegraphics[scale=0.4]{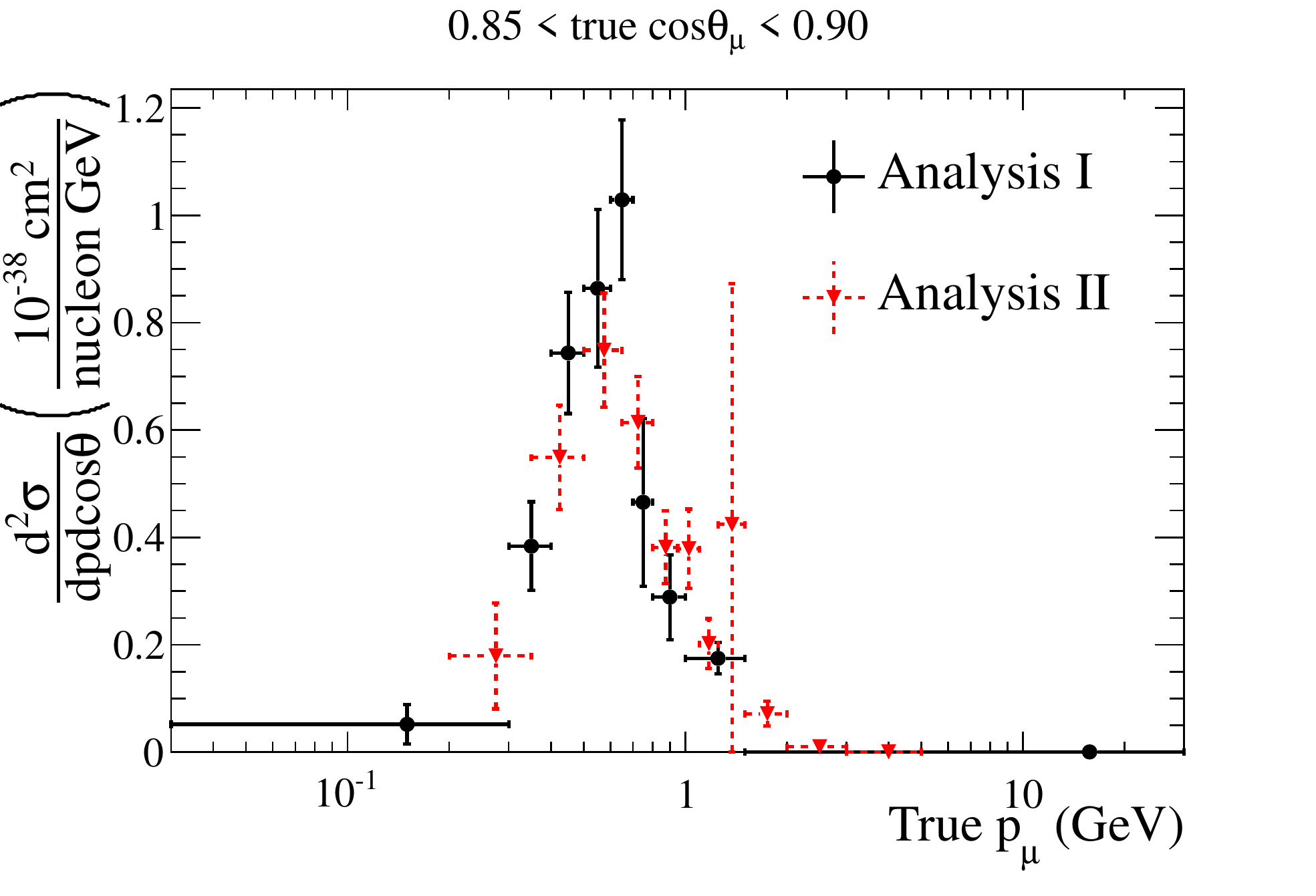}
  \caption{Comparison between the two analyses performed for the regions where the angular binning is the same (the measurement from Analysis II is reported only above
$p_{\mu}>200$~MeV).  A good level of agreement is found. Error bars include all uncertainties except for the flux normalization uncertainty.}
  \label{fig:analysisComparison}
\end{figure}

Despite the aforementioned differences between the two strategies,
the results from the two analyses are in good agreement (see Fig.~\ref{fig:analysisComparison}).
This is a strong demonstration of the model-independence of the results.
Moreover, both analyses were tested using different model assumptions, to understand the dependence of the result on the signal
and background model.
Different generators were used for these tests, such as GENIE~\cite{Andreopoulos:2009rq} and NuWro~\cite{Golan:2012wx}, as well as reweightings of the nominal NEUT model as a function of $E_{\nu}$, $Q^2$, or ($p_{\mu}$,$\cos\theta_{\mu}$). 
The results were found in good agreement in all the tests, and to be within the estimated uncertainties.

\section{Conclusions\label{sec:conc}}

In modern experiments, which use relatively heavy nuclear targets, CCQE cross-section 
measurements, previously considered to be well-understood, have been found to contain potentially significant contributions 
from nuclear effects that are not well known and difficult to disentangle experimentally.
The narrow-band T2K off-axis beam, which has a peak energy of 0.6~GeV, provides a powerful probe to study these CCQE interactions.
In this paper the measurements of interactions on carbon with the production of a muon 
with no associated pions is presented in the form of a double-differential, flux-integrated cross-section.

The results are compared to two sets of models. In more recent models from Martini {\it et al}~\cite{Martini:2009,Martini:2010} and Nieves {\it et al}~\cite{Nieves:2012,Nieves:2012yz}
the CCQE parameters are tuned to deuterium scattering data and very low momentum data and the nuclear effects are explicitly implemented
in the form of long- and short-range nucleon-nucleon correlations. %
We find that predictions from these new models agree with the data; in particular,
the data suggest the presence of 2p2h with respect to pure CCQE predictions with the Random Phase Approximation (RPA). 
On the other hand, the data also agree, inside current uncertainties, with the NEUT and GENIE simulations. In NEUT, the value of the $M_\text{A}^\text{QE}$ is higher than it is in GENIE to achieve agreement with recent datasets. Explicit implementation of nuclear effects (RPA and 2p2h) is included in more recent NEUT versions that will be used in future analyses.%

Quantifying the agreement between the data and the various models is not straightforward.
The correlations between uncertainties in different bins must be considered, but the experimental measurements are affected
by theoretical systematics, which are especially large in the backward region.
On the other hand, the models have known limitations (the lack of FSIs and large uncertainties related to nuclear effects, in particular for the
very forward region) which should be considered in the comparison to the data. For all these reasons, we do not attempt
a quantitative comparison of the experimental results with the various models. Such phenomenological studies
will be pursued further in the T2K collaboration, possibly combining measurements for different channels and targets.

\begin{figure}
  \centering
  \includegraphics[scale=0.8]{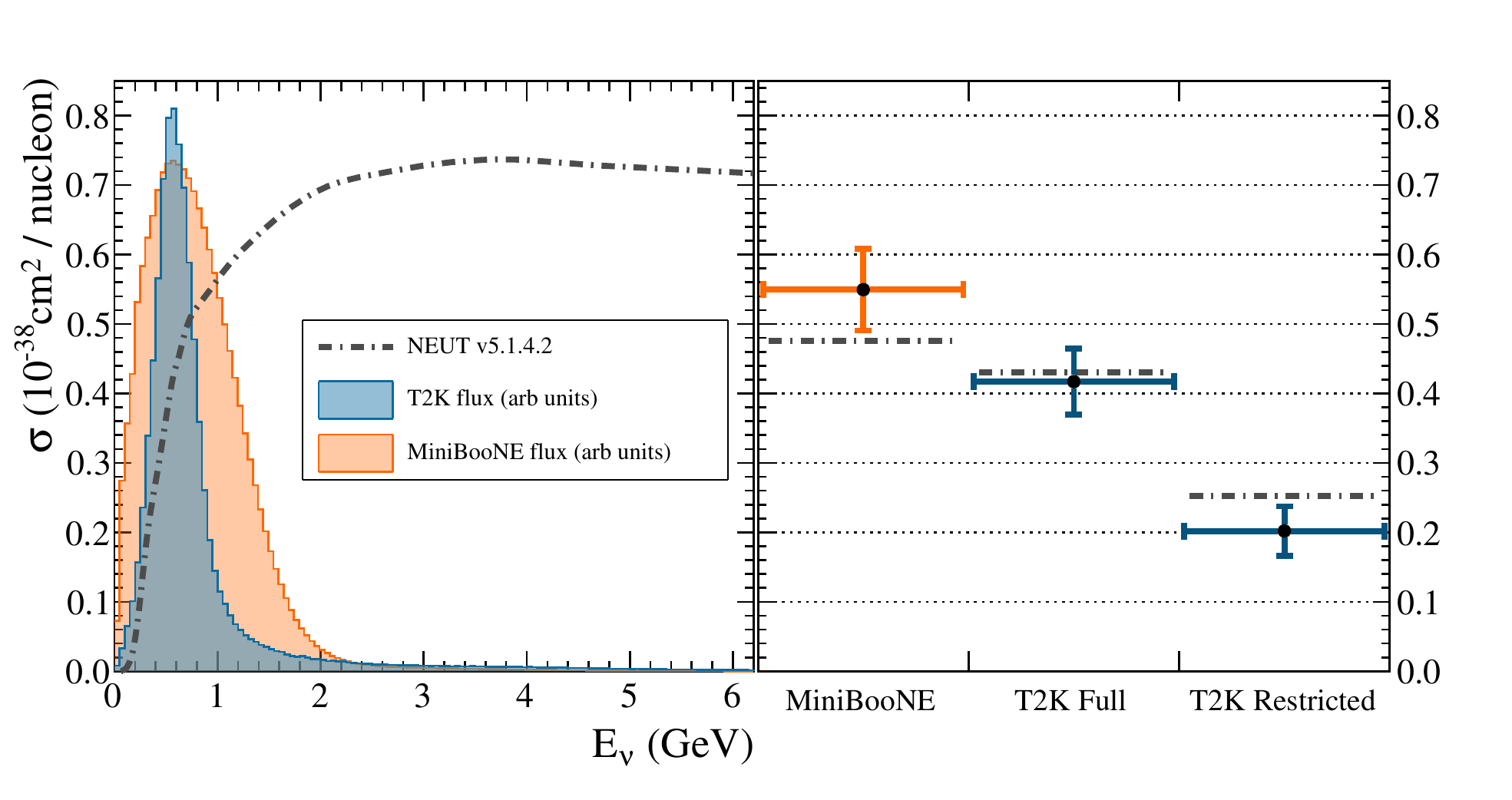}
  \caption{
    Left: The dash-dotted line is the predicted \cczeropi cross-section from NEUT as a function of the neutrino energy.
    The shapes of the neutrino fluxes for T2K and MiniBooNE are also shown.
    Right: A comparison of the measured \cczeropi flux integrated cross-section from MiniBooNE
    and the two T2K measurements presented in this paper,
    using the full (Analysis I) and restricted (Analysis II) phase space.
    As on the left, the dash-dotted lines represent the prediction from NEUT for each respective measurement.
  }
  \label{fig:T2KMBMinervaComparison}
\end{figure}

In Fig.~\ref{fig:T2KMBMinervaComparison}, the flux-integrated cross-sections measured in this paper and the one
measured by MiniBooNE~\cite{AguilarArevalo:2010zc}
are compared with the NEUT prediction. 
The MINER$\nu$A results~\cite{Fiorentini:2013ezn, Walton:2014esl}  are not included since comparisons 
to these measurements would depend on model-dependent assumptions; the analysis presented in~\cite{Fiorentini:2013ezn} 
includes only pure CCQE events after subtracting \cconepi events where the pion is absorbed by the nuclear medium,
and the analysis in Ref.~\cite{Walton:2014esl} only includes events with both the muon and proton being reconstructed, and thus is dependent on the
modeling of nucleon FSIs and nuclear effects. We look forward to new double-differential \cczeropi results from MINER$\nu$A.
In Fig.~\ref{fig:T2KMBMinervaComparison} the full-phase-space result from Analysis I and the restricted phase space result from Analysis II ($\cos\theta_\mu>0$ and $p_\mu > 200$ MeV) are reported and compared to the MiniBooNE measurement. 
The three measurements are compatible with NEUT predictions within their uncertainties, but there is a trend across the measurements as their acceptances vary; MiniBooNE, which has $4\pi$ acceptance, tends to measure a larger cross-section compared to the NEUT prediction than the T2K full-phase-space result, which in turn gives a larger cross-section, again compared to NEUT, than our restricted-phase-space result. 
The uncertainties are too large for any conclusive statement; however this may indicate that the models do not correctly describe 
the angular shape of the cross-section.
It is thus crucial for future measurements to significantly increase the high angle acceptance, as well as to report results in the restricted phase space where the experiments have good efficiency.
 The two analyses presented here have been designed to be robust against the dependence on the signal model assumed in the analysis. %
Future measurements using the T2K off-axis beam will include
more data and improved algorithms for backwards-going track and proton reconstruction, which will enable exclusive measurements of the muon and proton final state to further elucidate the nature of nuclear effects in neutrino interactions and possibly solve the 
present degeneracy between different models.

\begin{acknowledgments}
We thank the J-PARC staff for the superb accelerator performance and the CERN NA61 
collaboration for providing valuable particle production data.
We acknowledge the support of MEXT, Japan; 
NSERC (grant SAPPJ-2014-00031), NRC and CFI, Canada;
CEA and CNRS/IN2P3, France;
DFG, Germany; 
INFN, Italy;
National Science Centre (NCN), Poland;
RSF, RFBR and MES, Russia; 
MINECO and ERDF funds, Spain;
SNSF and SERI, Switzerland;
STFC, UK; and 
DOE, USA.
We also thank CERN for the UA1/NOMAD magnet, 
DESY for the HERA-B magnet mover system, 
NII for SINET4, 
the WestGrid and SciNet consortia in Compute Canada, 
and GridPP, UK.
In addition, participation of individual researchers
and institutions has been further supported by funds from: ERC (FP7), H2020 RISE-GA644294-JENNIFER, EU; 
JSPS, Japan; 
Royal Society, UK; 
DOE Early Career program, USA.%
\end{acknowledgments}

\appendix
\section{Comparisons with other models}
\begin{figure}
\begin{center}
 \includegraphics[width=6cm]{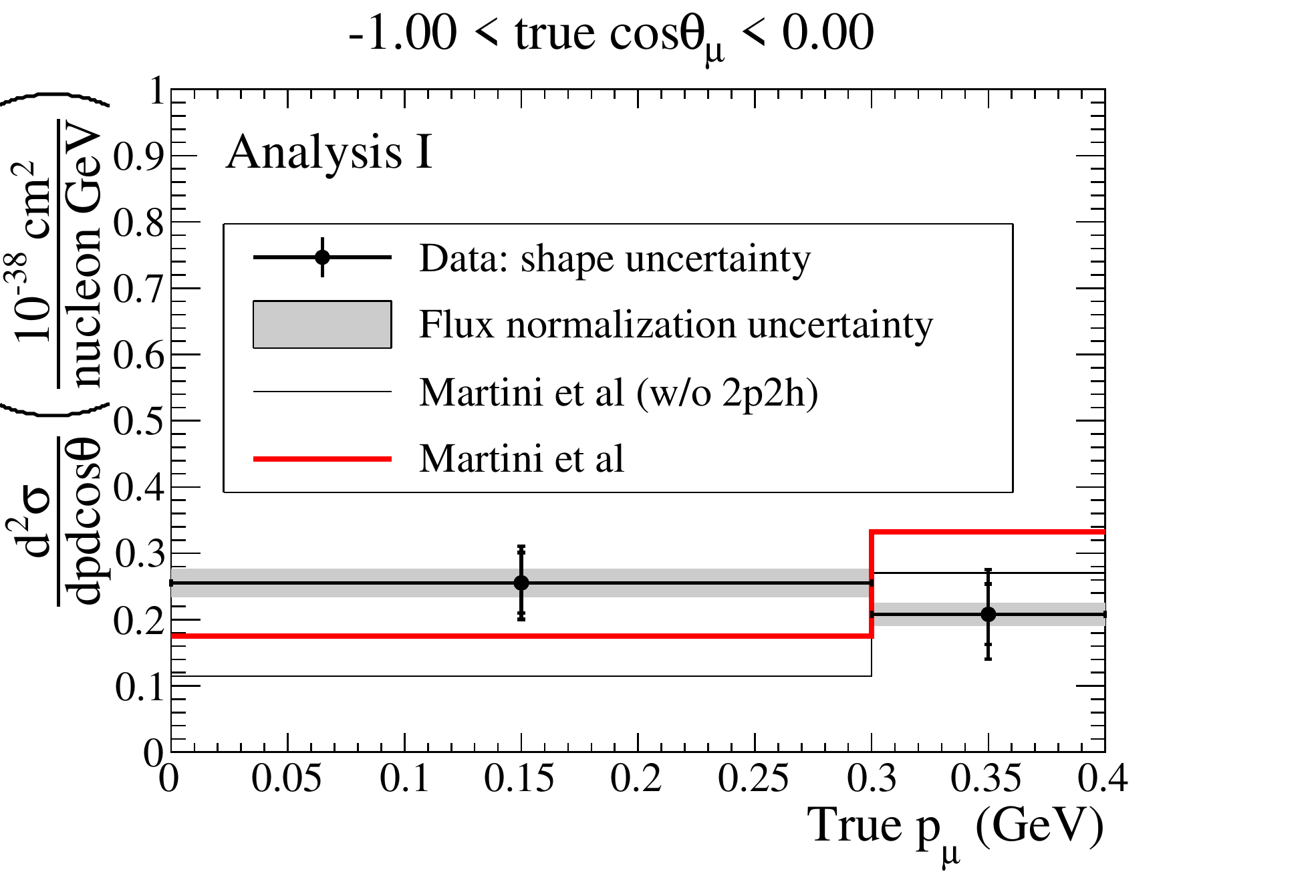}
 \includegraphics[width=6cm]{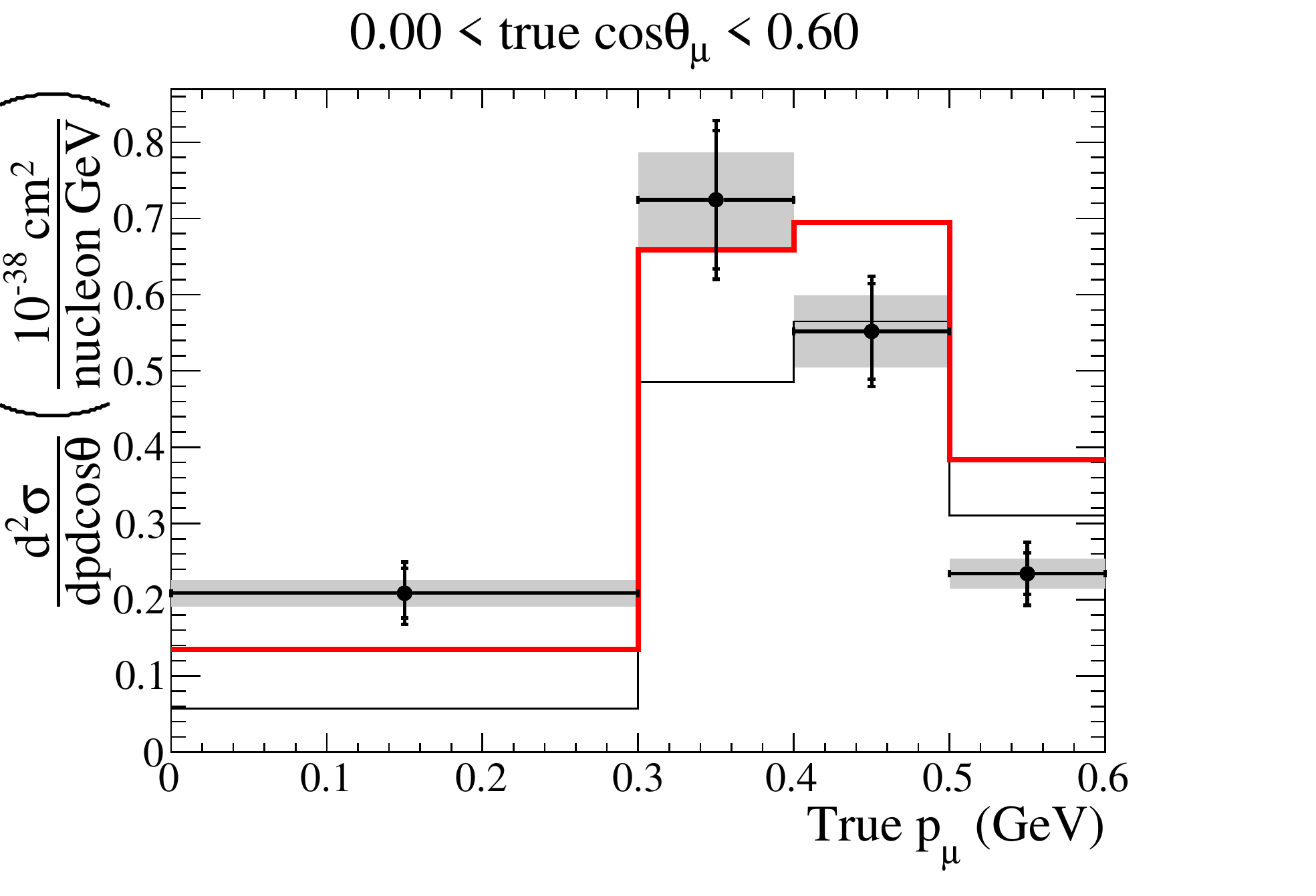}\\
 \includegraphics[width=6cm]{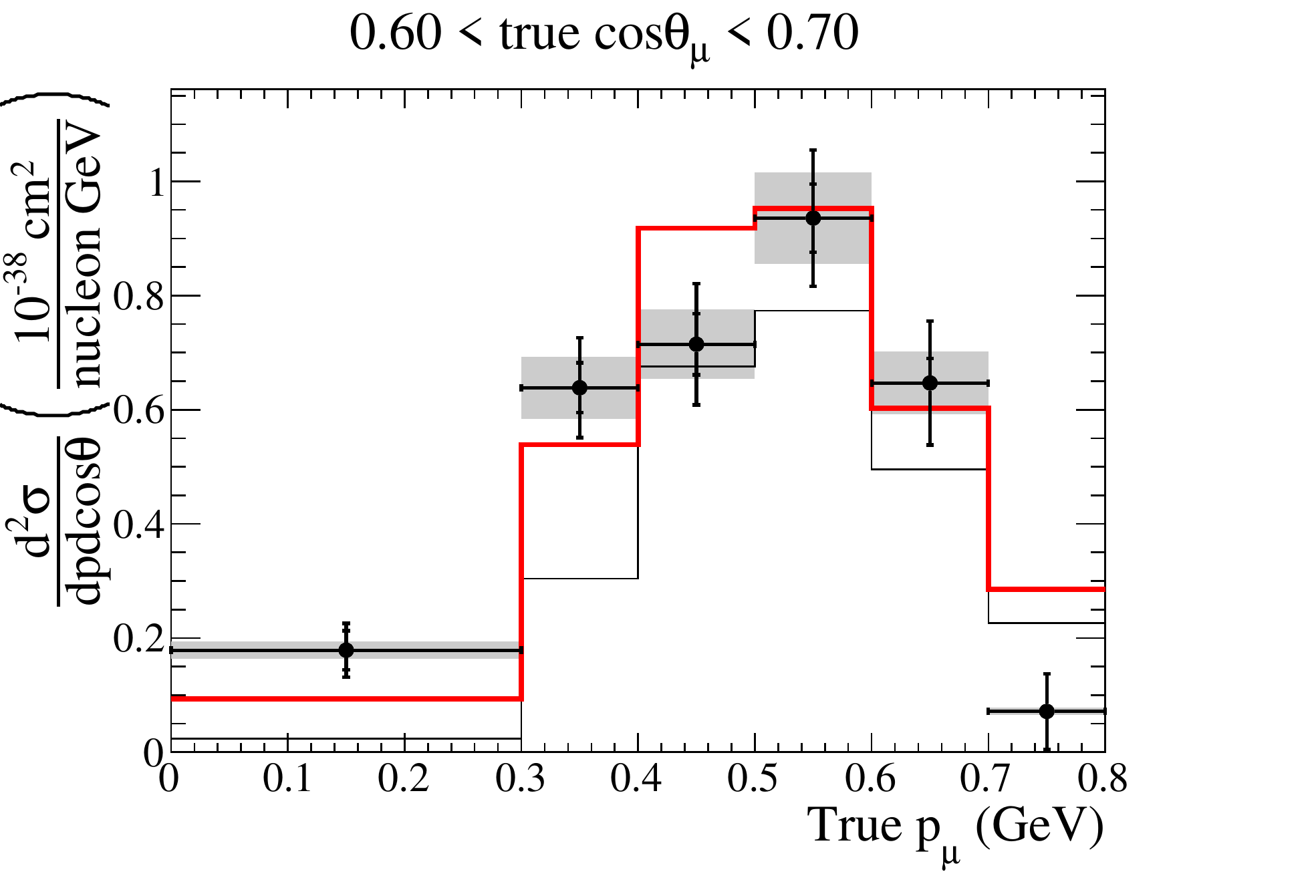}
 \includegraphics[width=6cm]{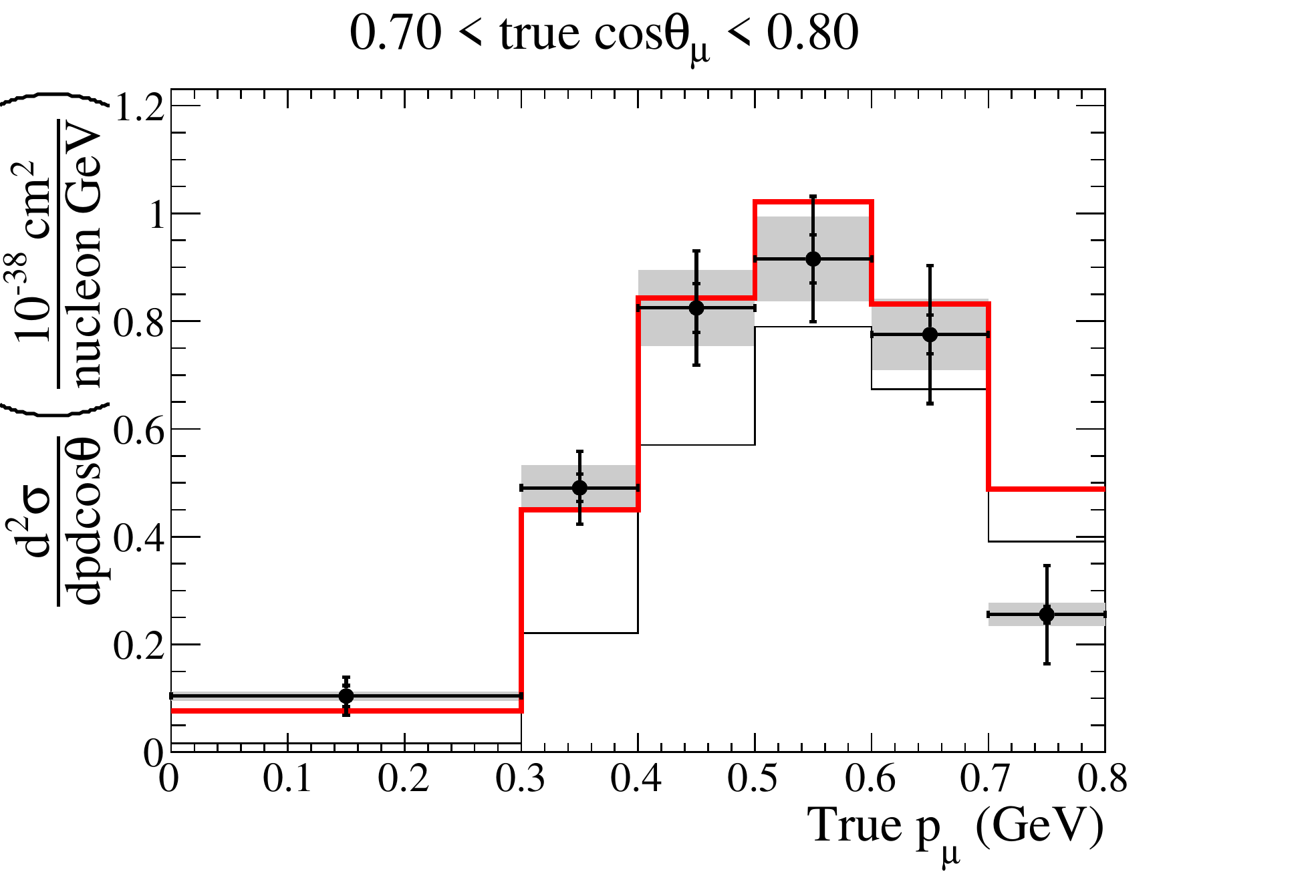}\\
 \includegraphics[width=6cm]{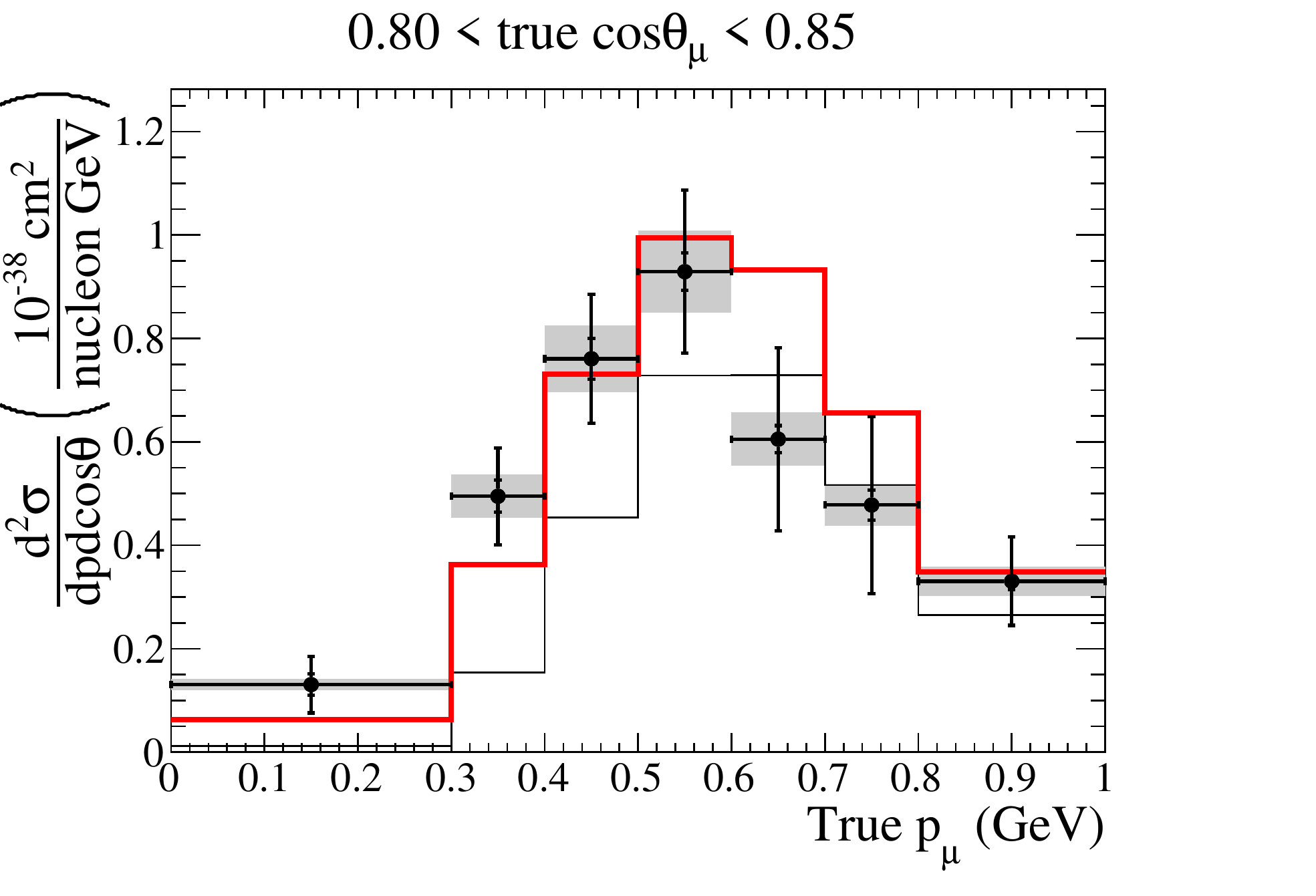}
 \includegraphics[width=6cm]{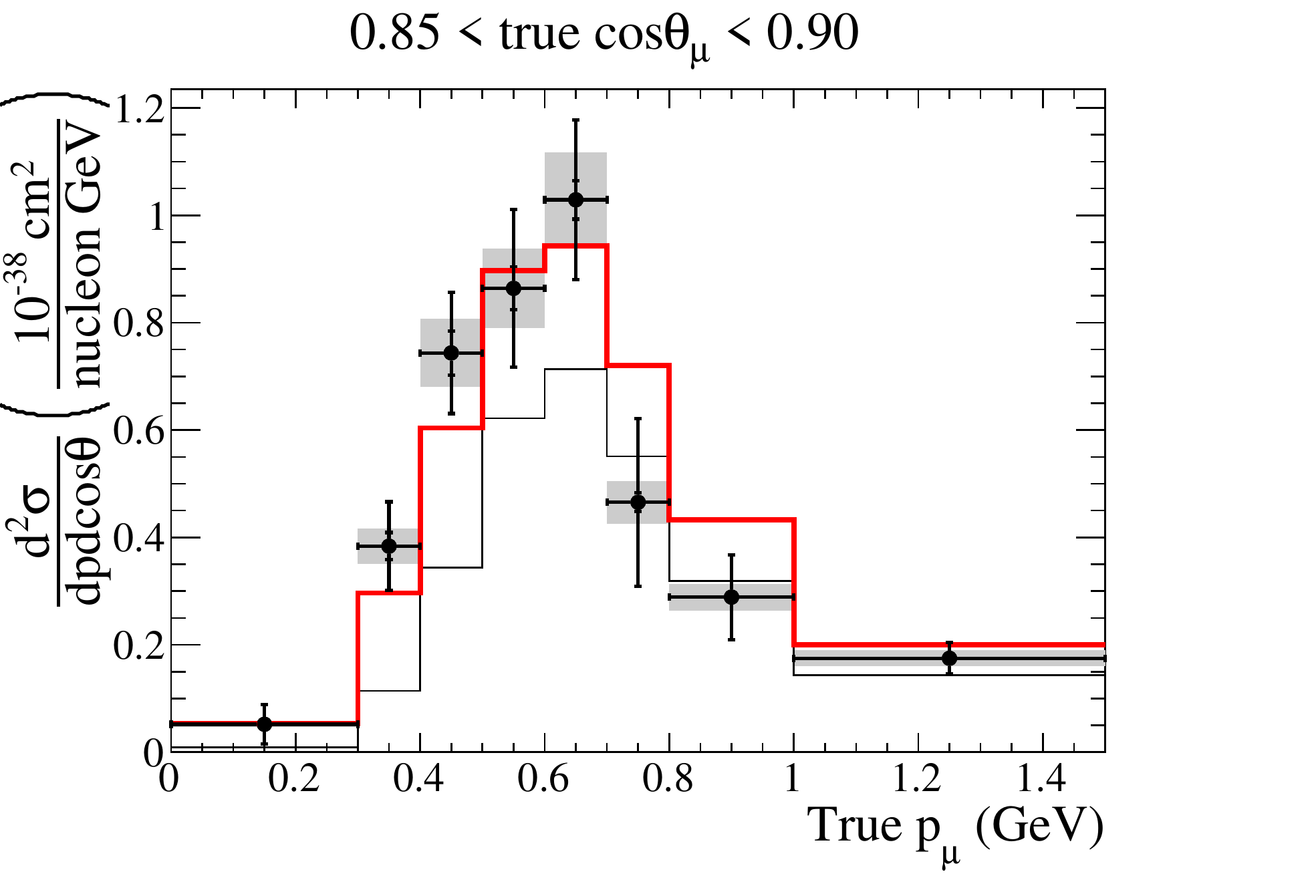}\\
 \includegraphics[width=6cm]{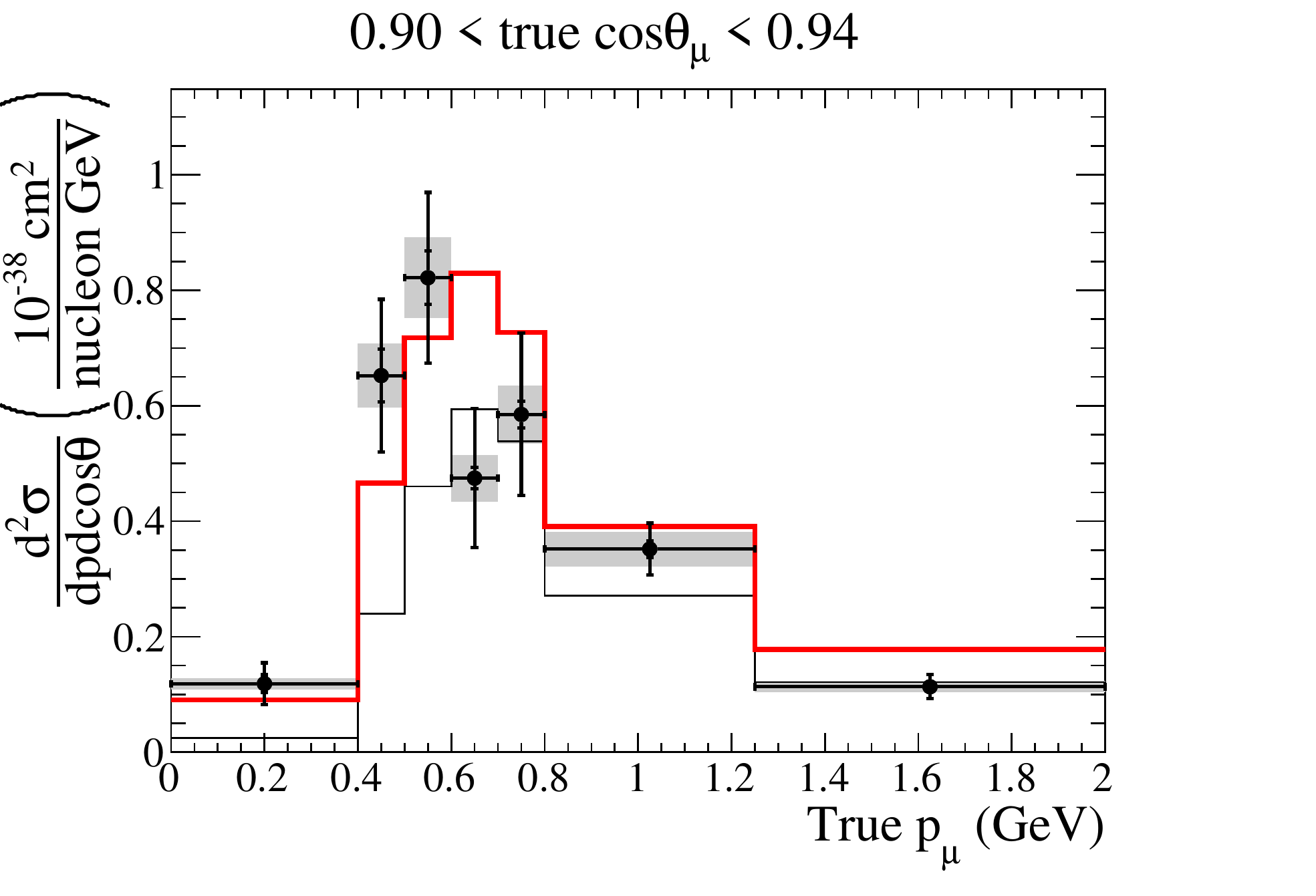}
 \includegraphics[width=6cm]{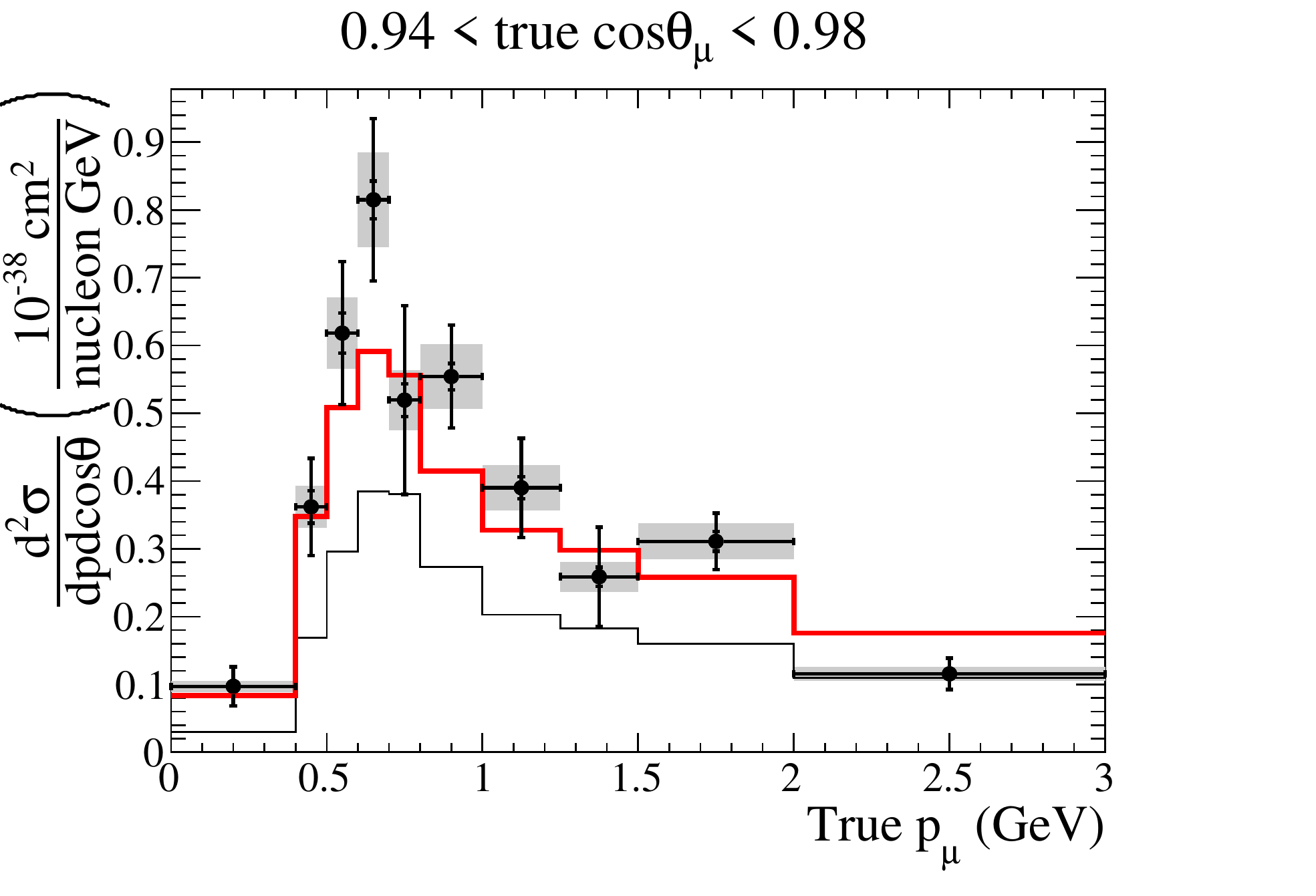}\\
 \includegraphics[width=6cm]{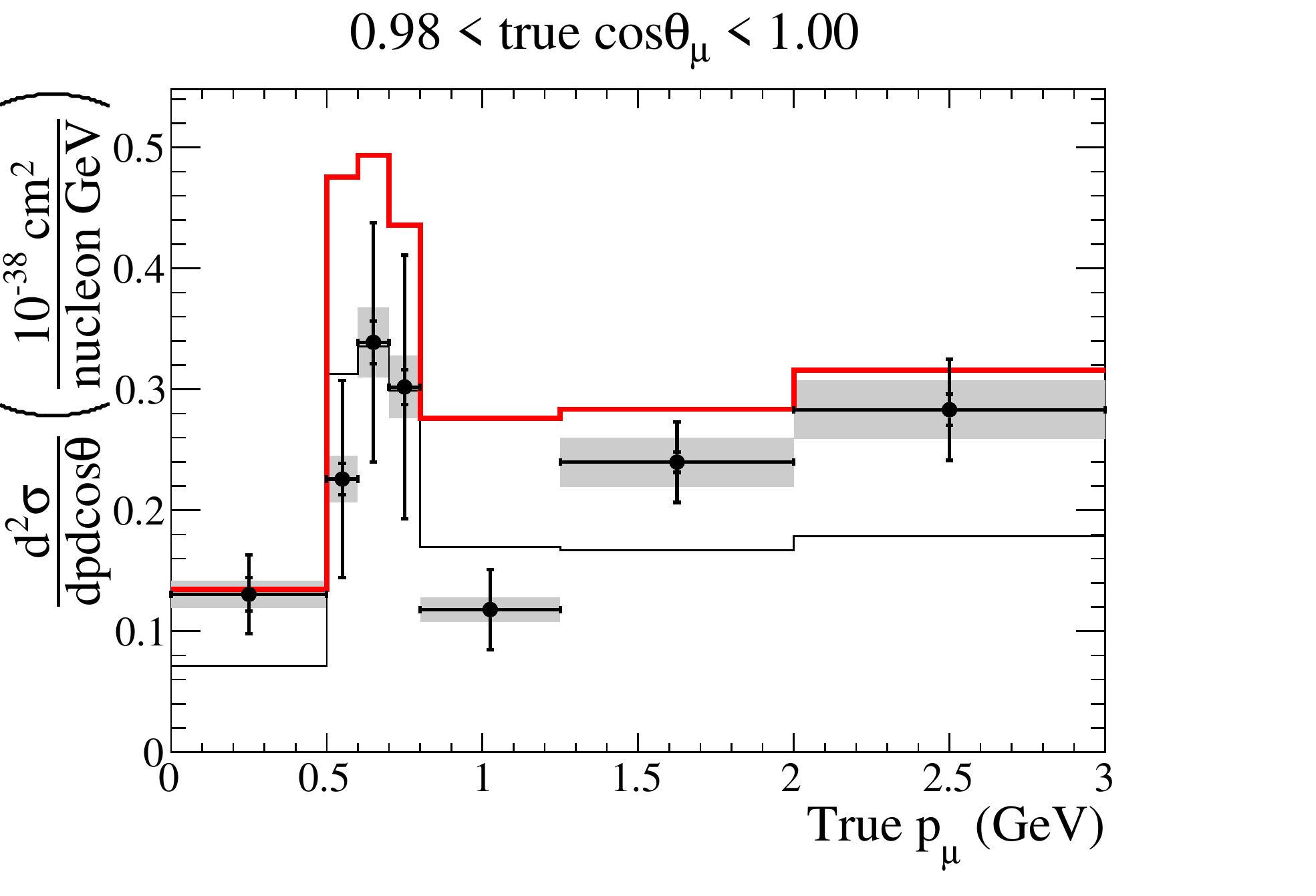}
\end{center}
\caption{Measured cross-section with shape uncertainties (error bars: internal systematics, external statistical) 
and fully correlated normalization uncertainty (gray band).
The results from the fit to the data are compared to predictions from Martini {\it et al} without 2p2h (black line), 
and with 2p2h (red line).}
\label{fig:xsecResultsLinMM}
\end{figure}

\begin{figure}
\begin{center}
 \includegraphics[width=6cm]{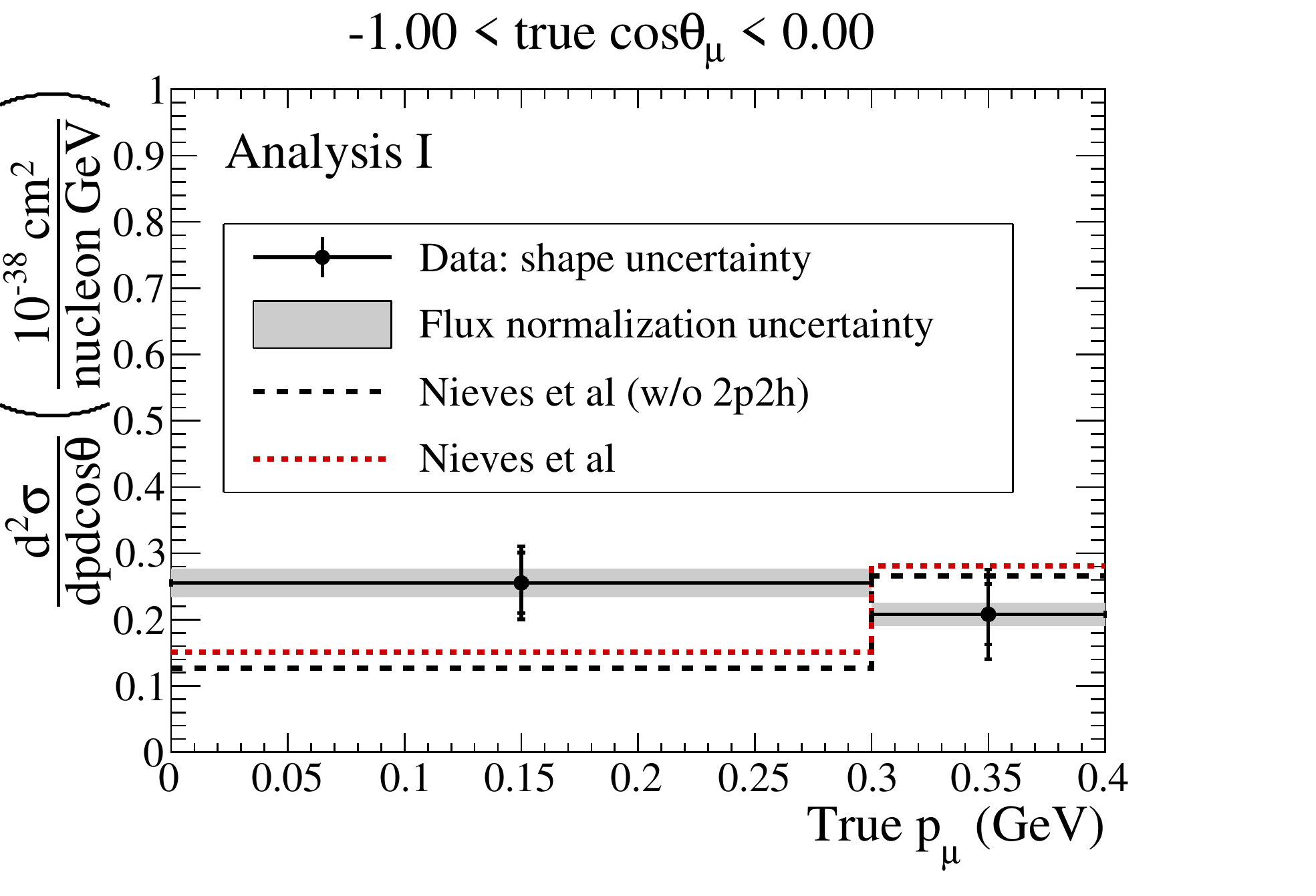}
 \includegraphics[width=6cm]{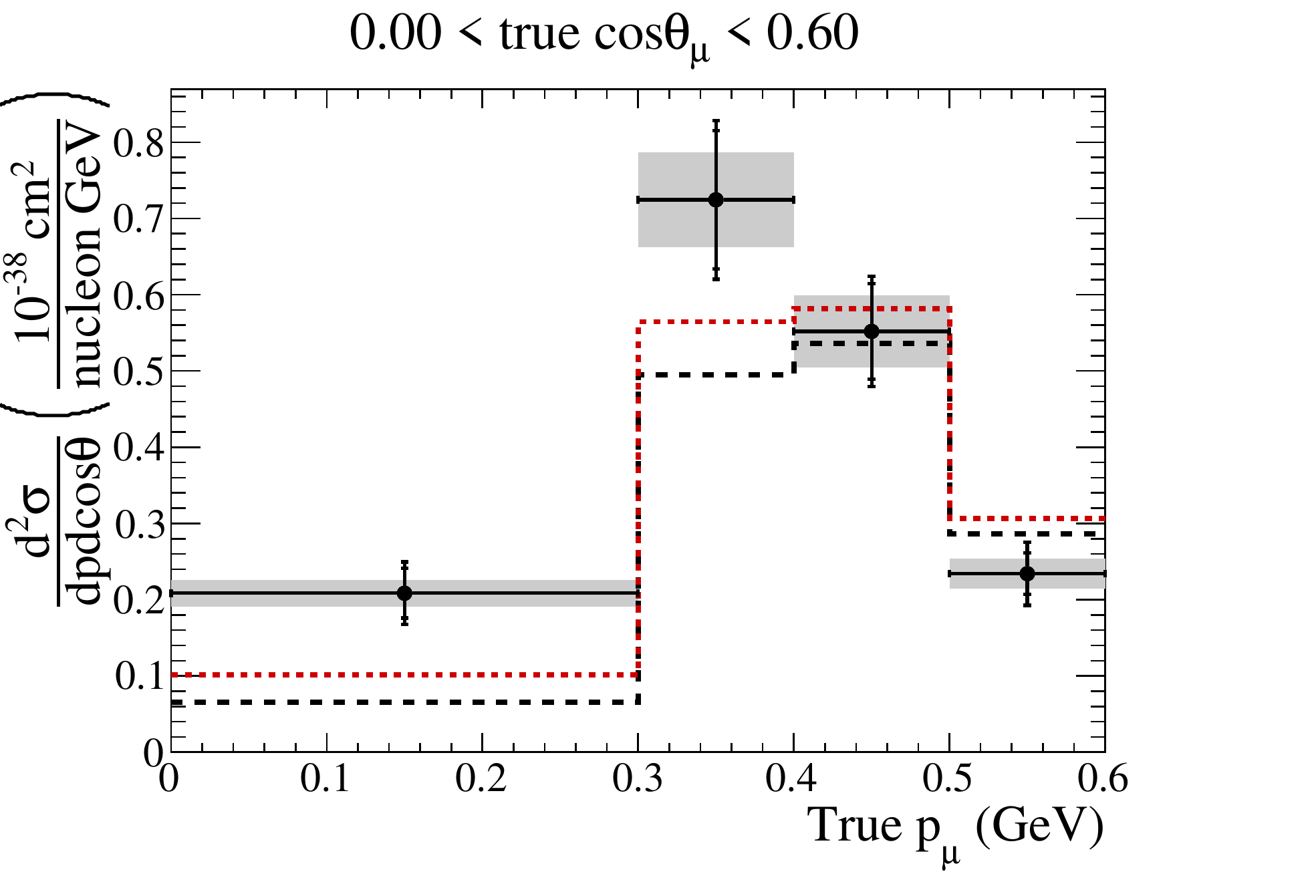}\\
 \includegraphics[width=6cm]{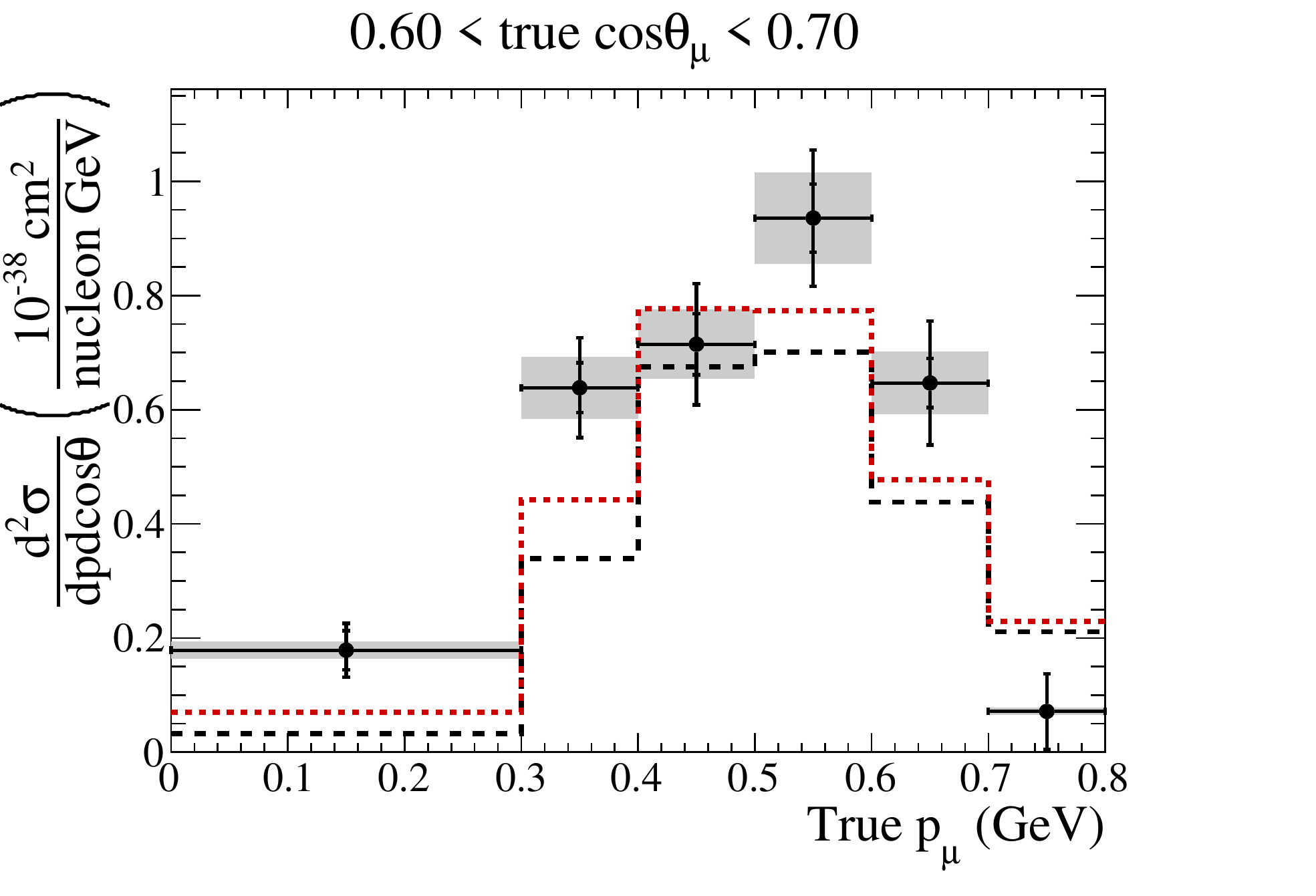}
 \includegraphics[width=6cm]{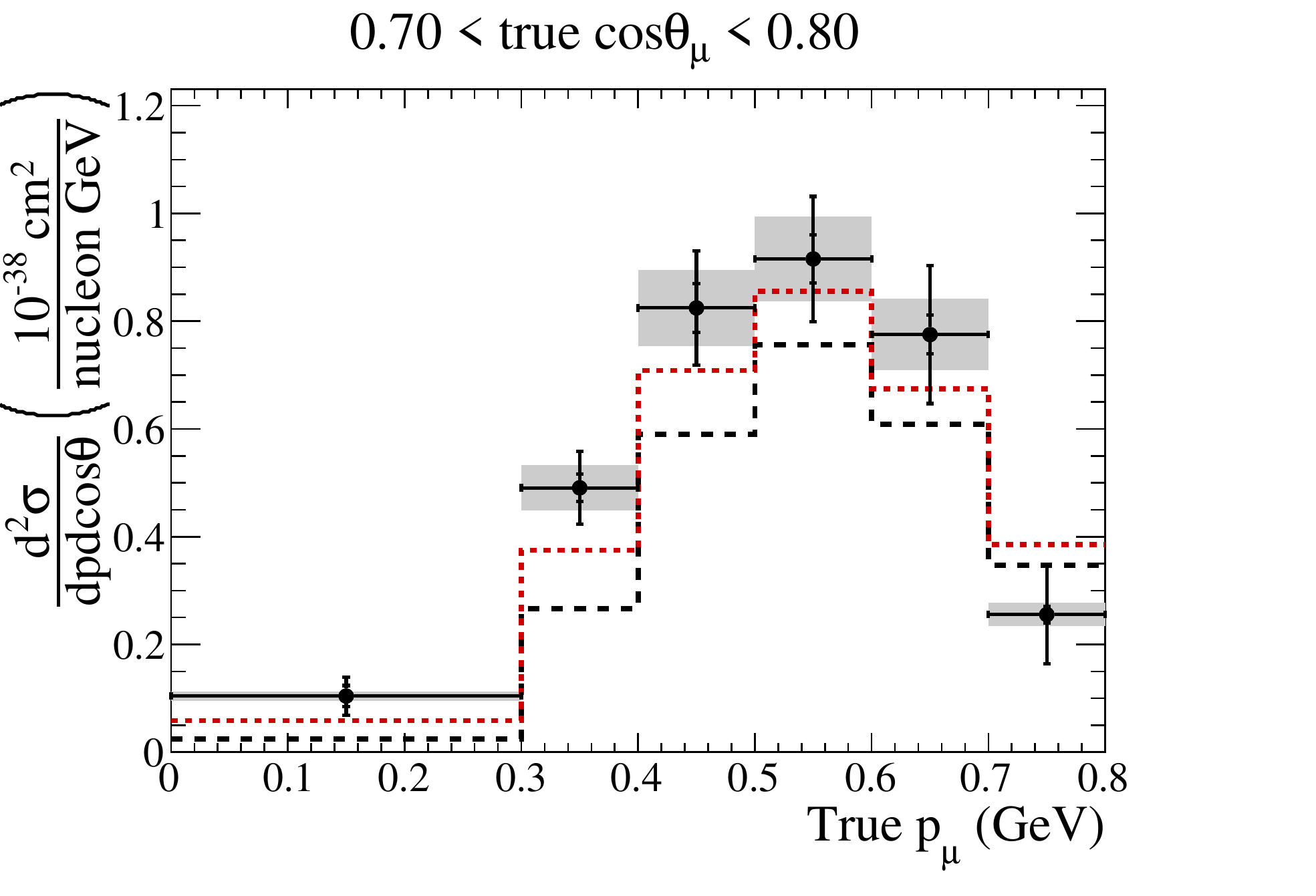}\\
 \includegraphics[width=6cm]{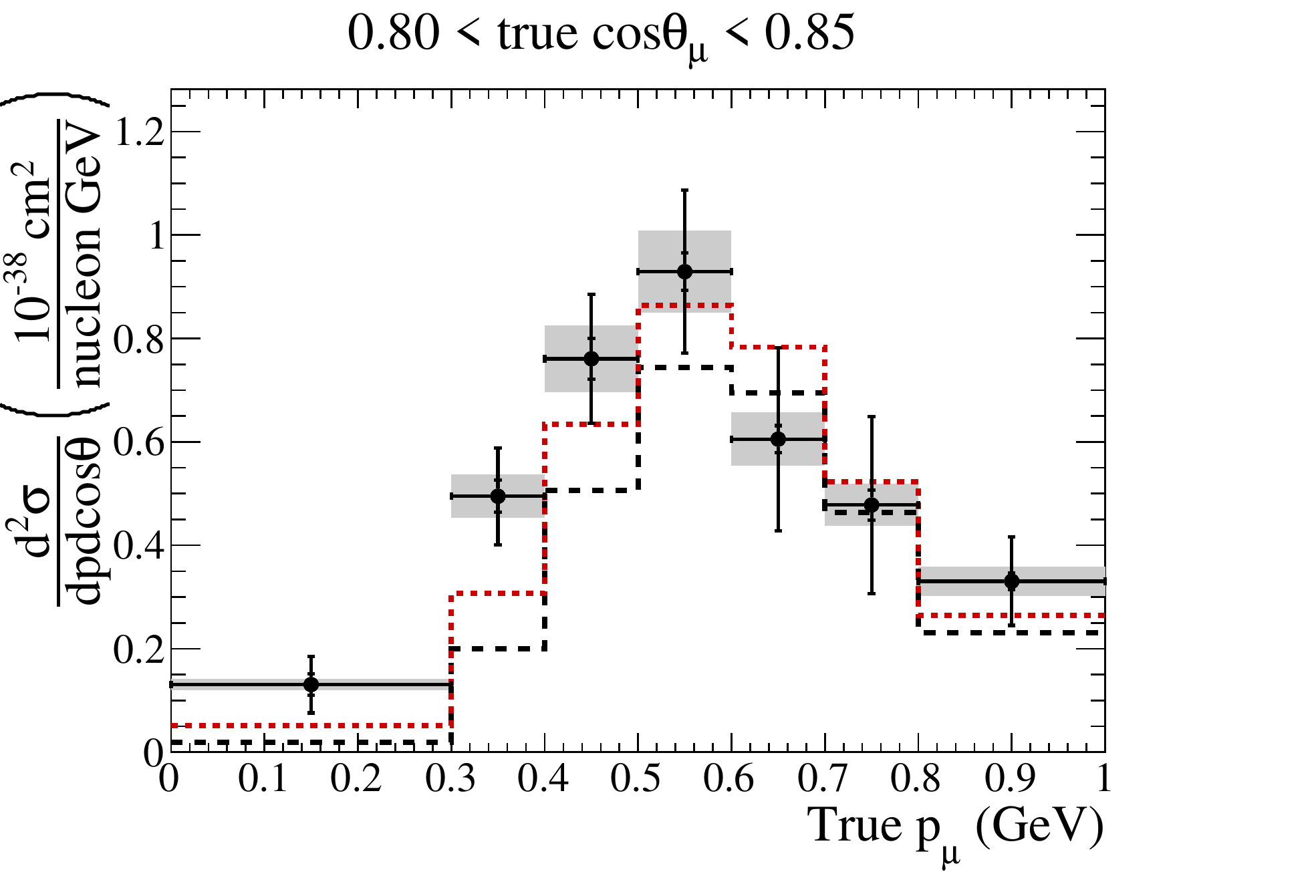}
 \includegraphics[width=6cm]{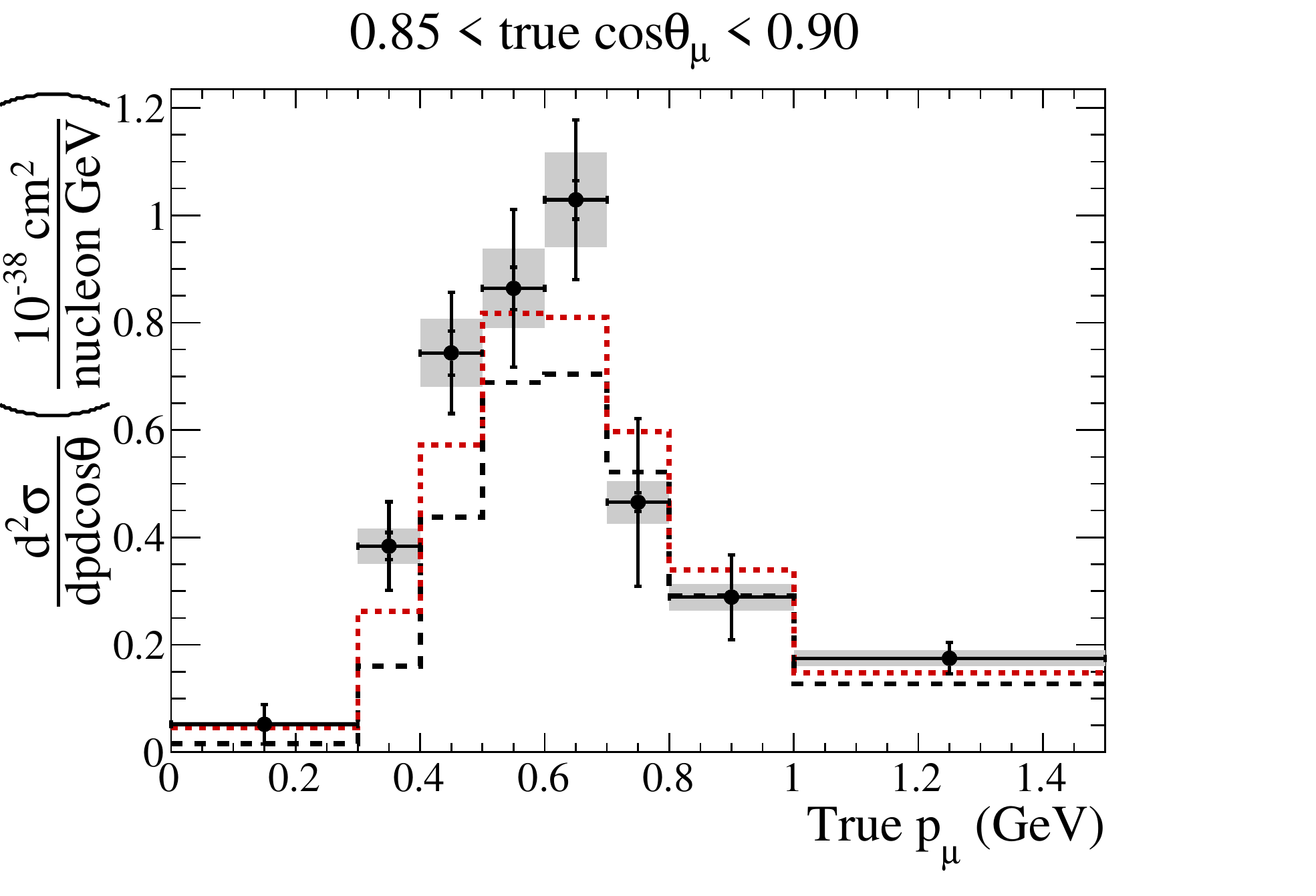}\\
 \includegraphics[width=6cm]{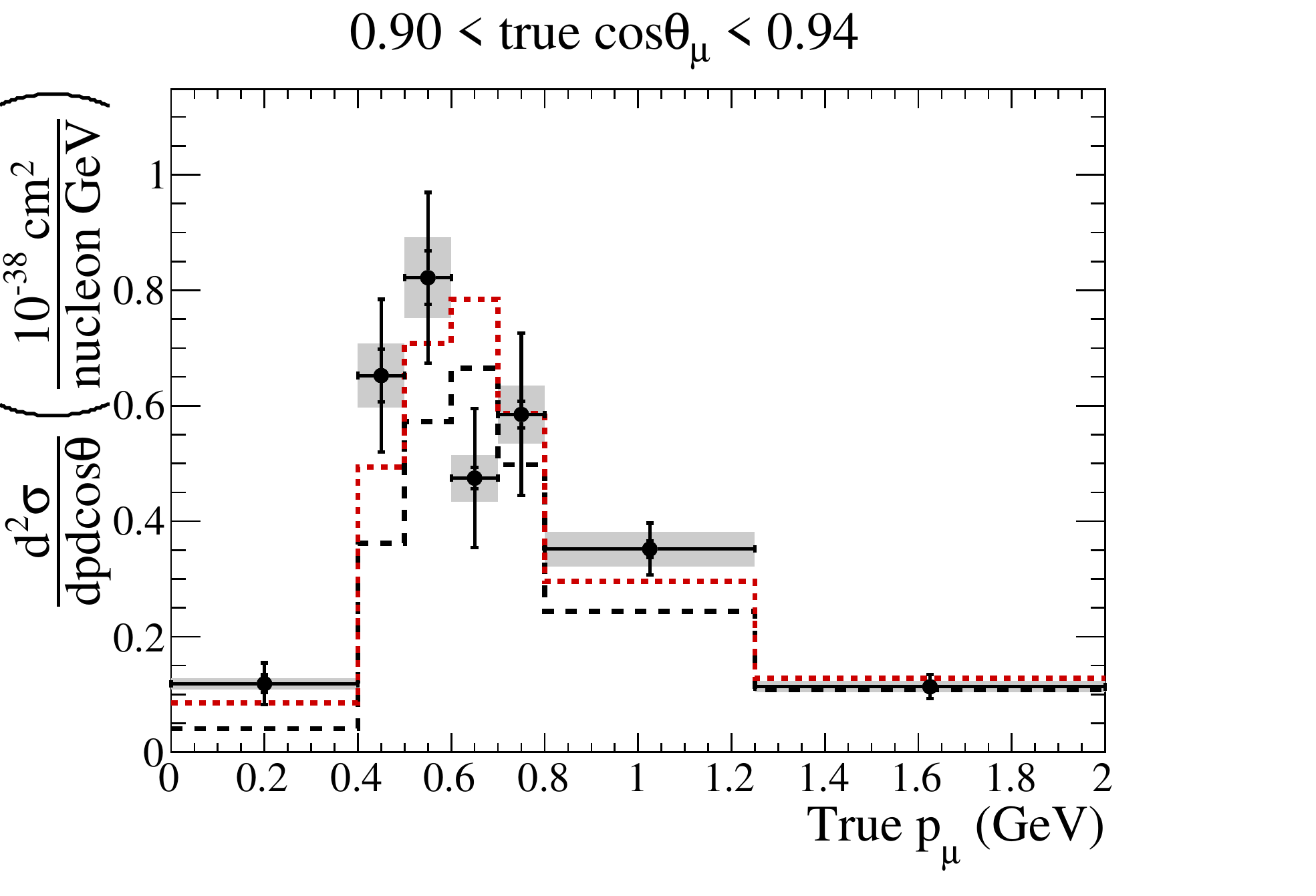}
 \includegraphics[width=6cm]{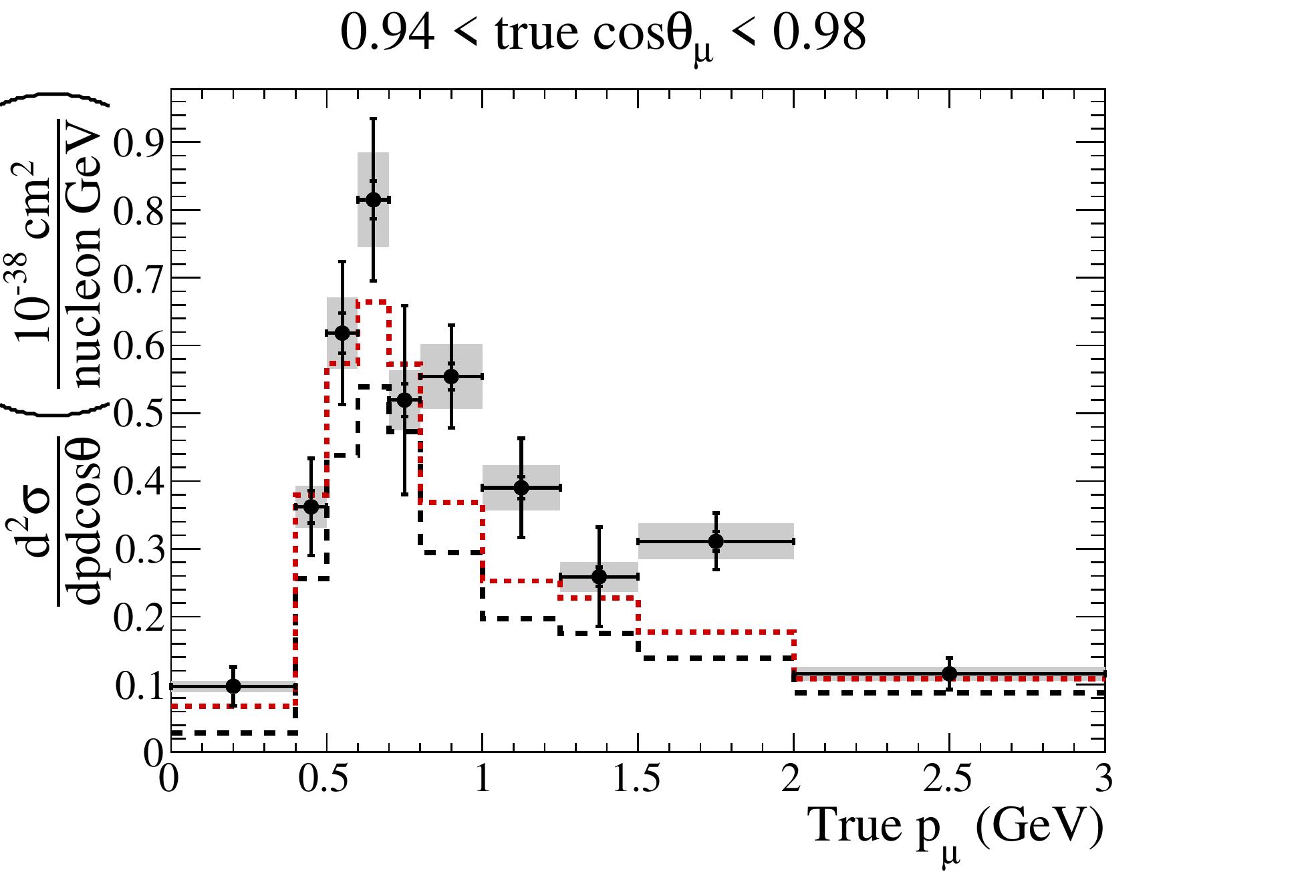}\\
 \includegraphics[width=6cm]{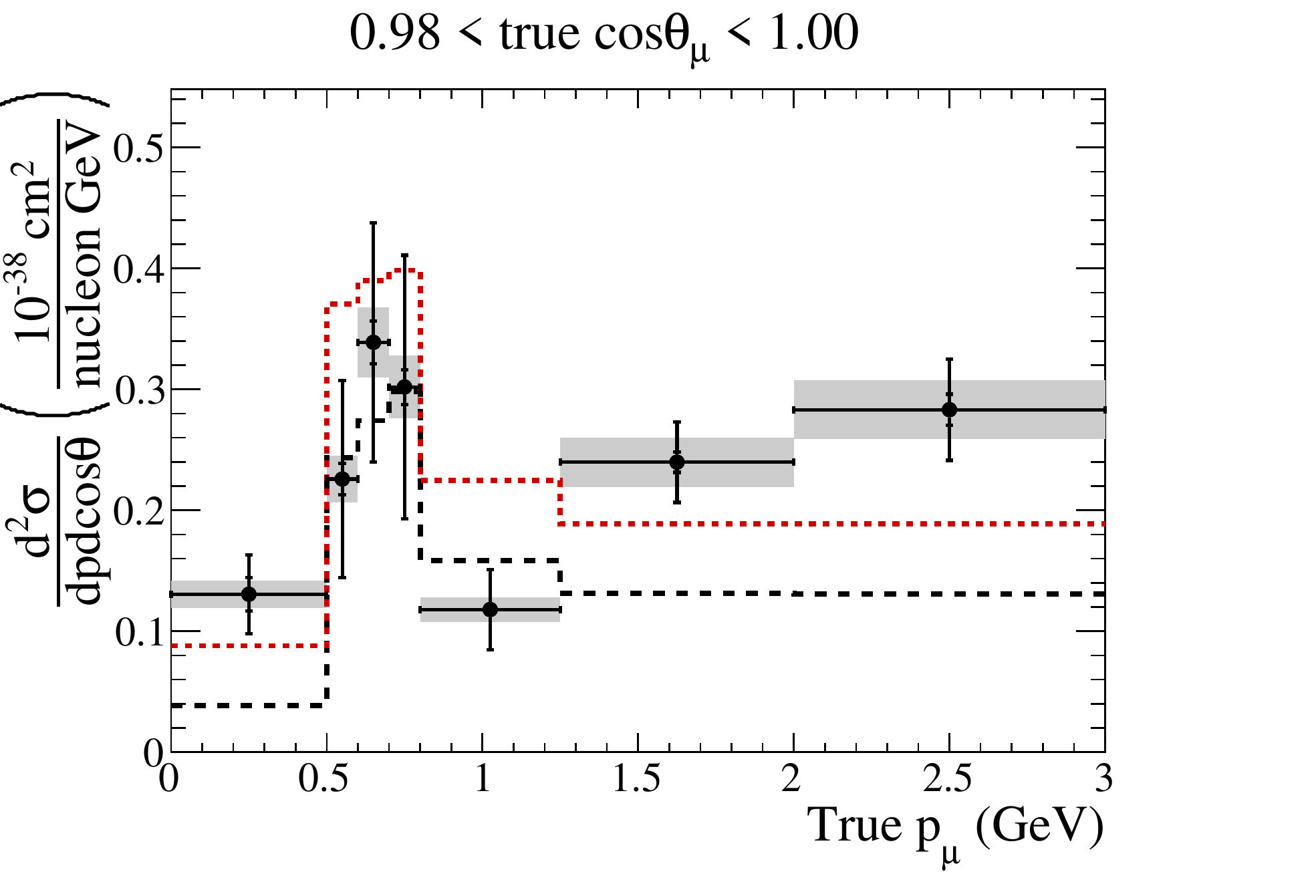}
\end{center}
\caption{Measured cross-section with shape uncertainties (error bars: internal systematics, external statistical) 
and fully correlated normalization uncertainty (gray band).
The results from the fit to the data are compared to predictions from Nieves {\it et al} without 2p2h (black line), 
and with 2p2h (red line).}
\label{fig:xsecResultsLinN}
\end{figure}

\begin{figure}
\begin{center}
 \includegraphics[width=6cm]{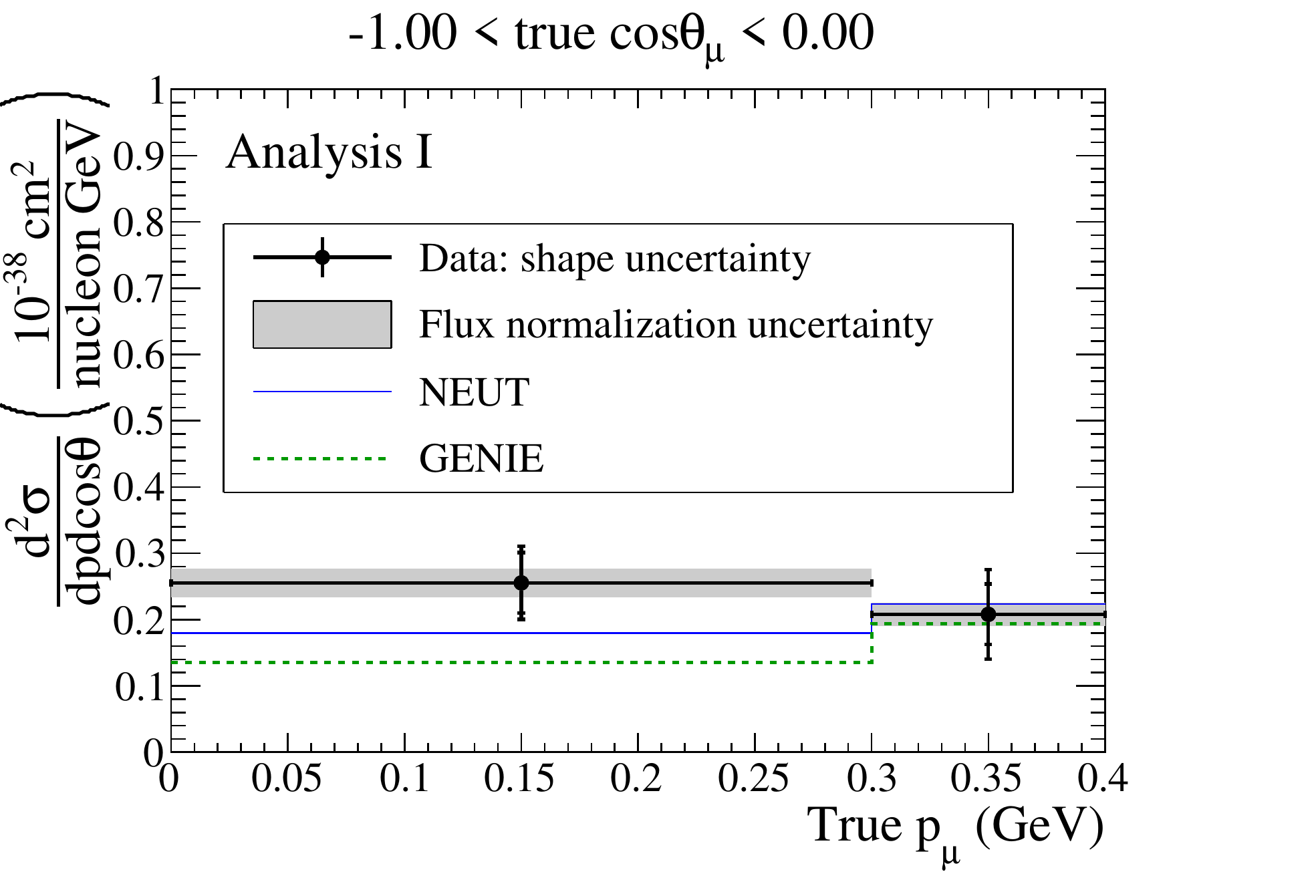}
 \includegraphics[width=6cm]{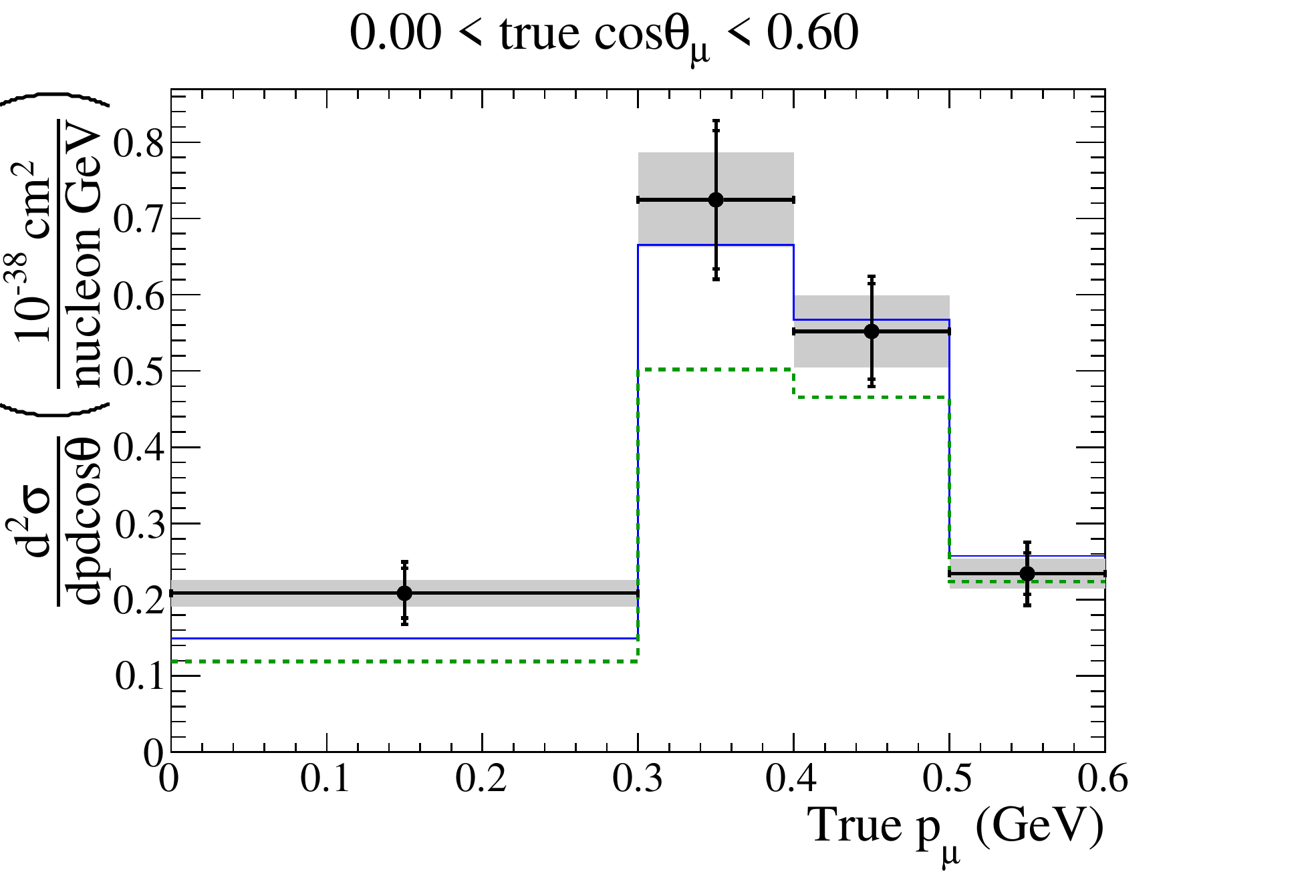}\\
 \includegraphics[width=6cm]{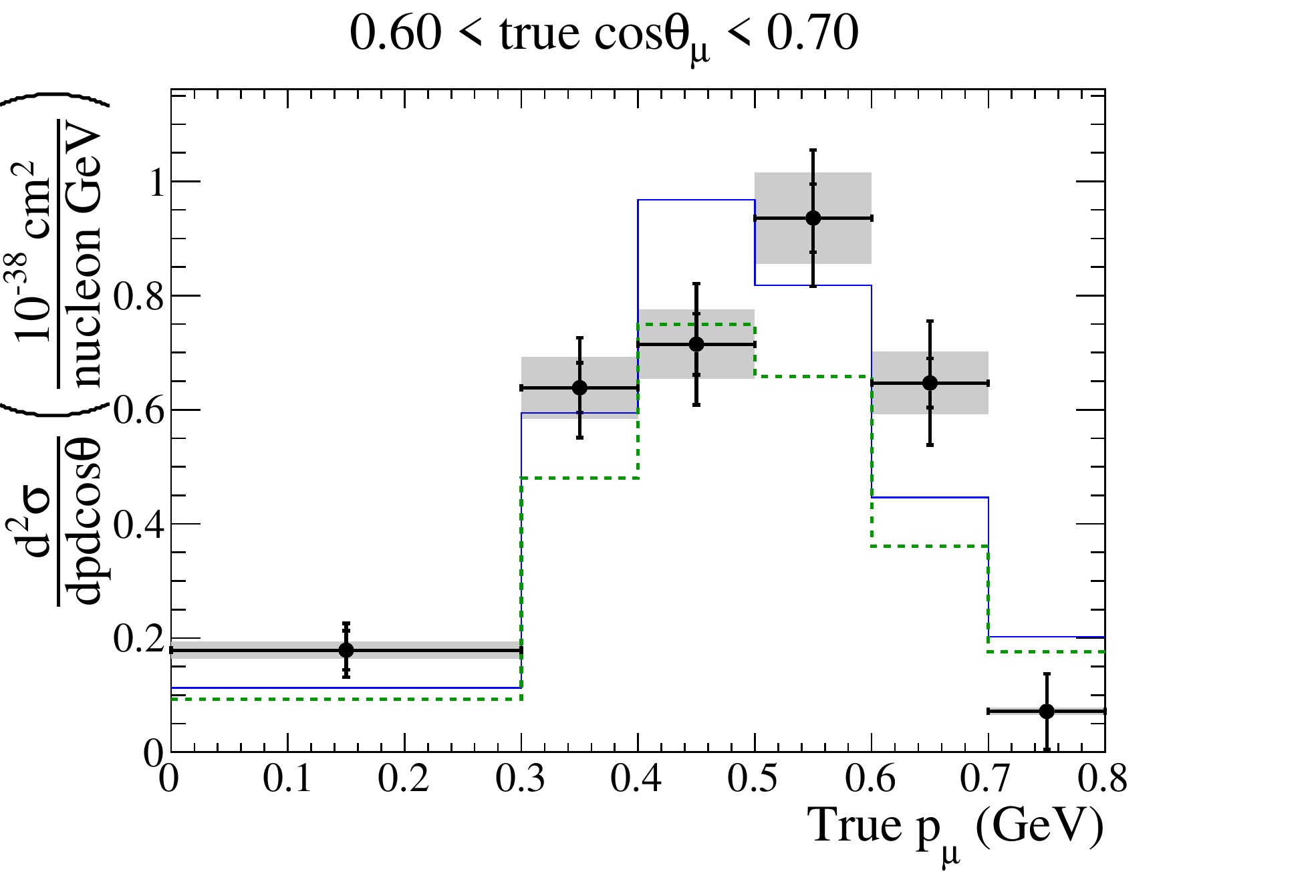}
 \includegraphics[width=6cm]{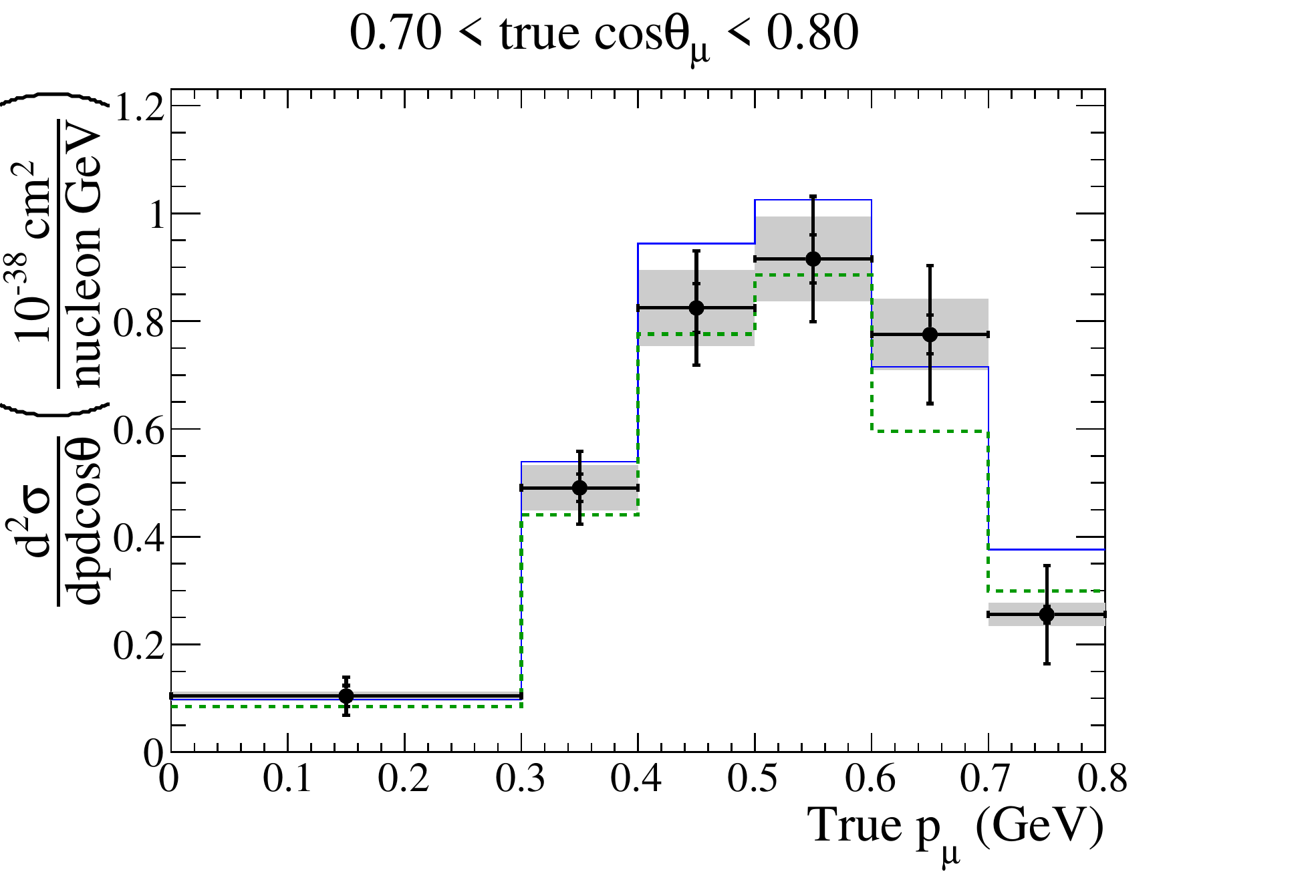}\\
 \includegraphics[width=6cm]{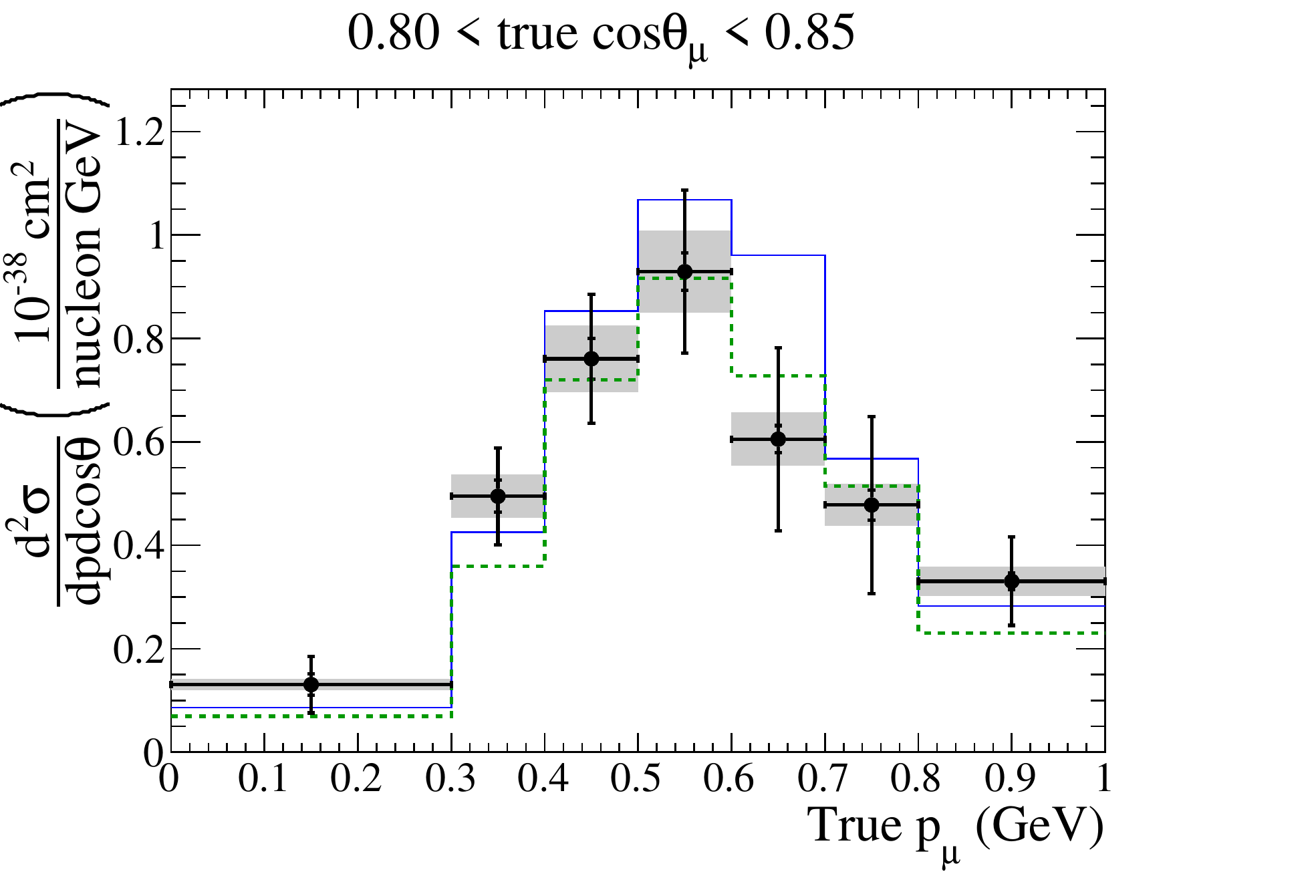}
 \includegraphics[width=6cm]{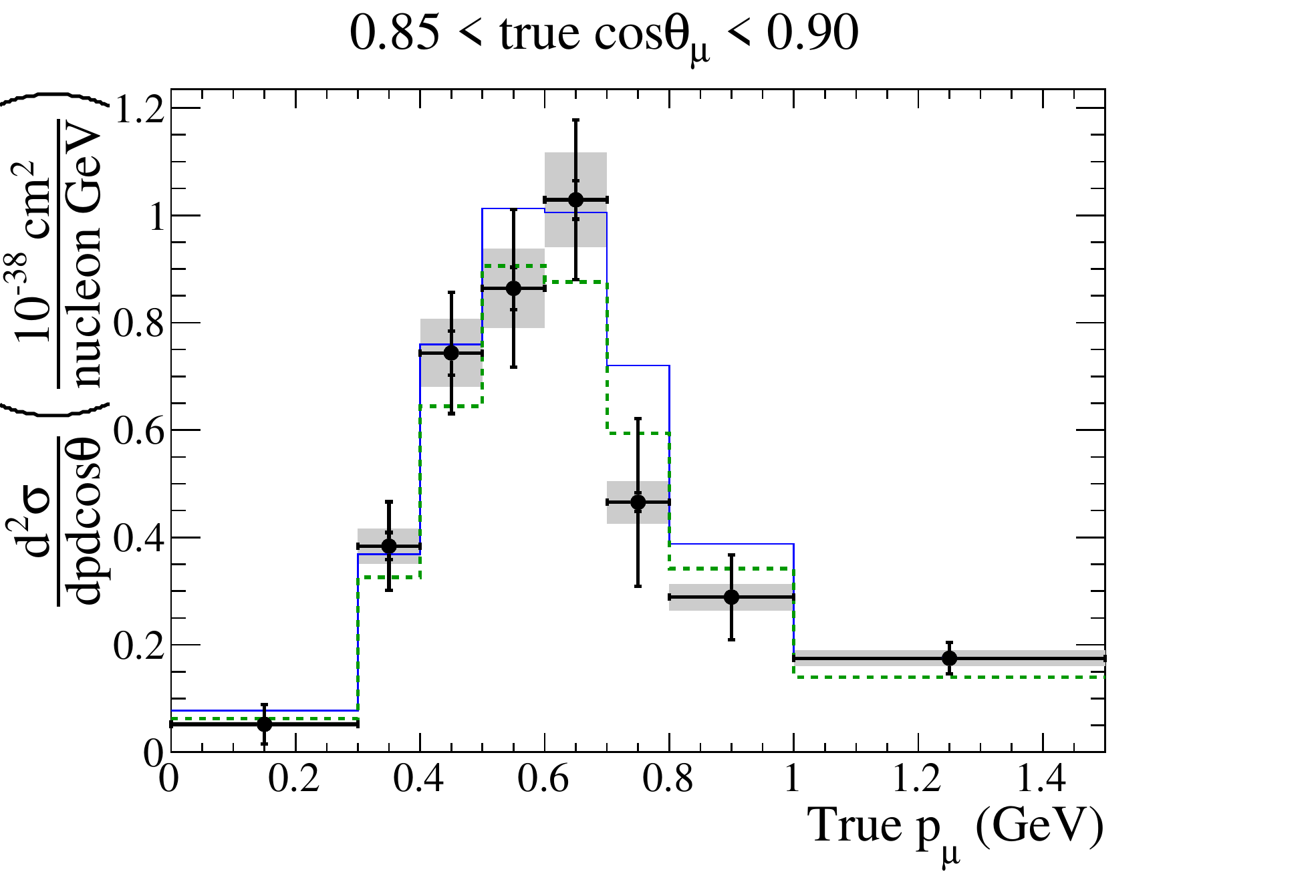}\\
 \includegraphics[width=6cm]{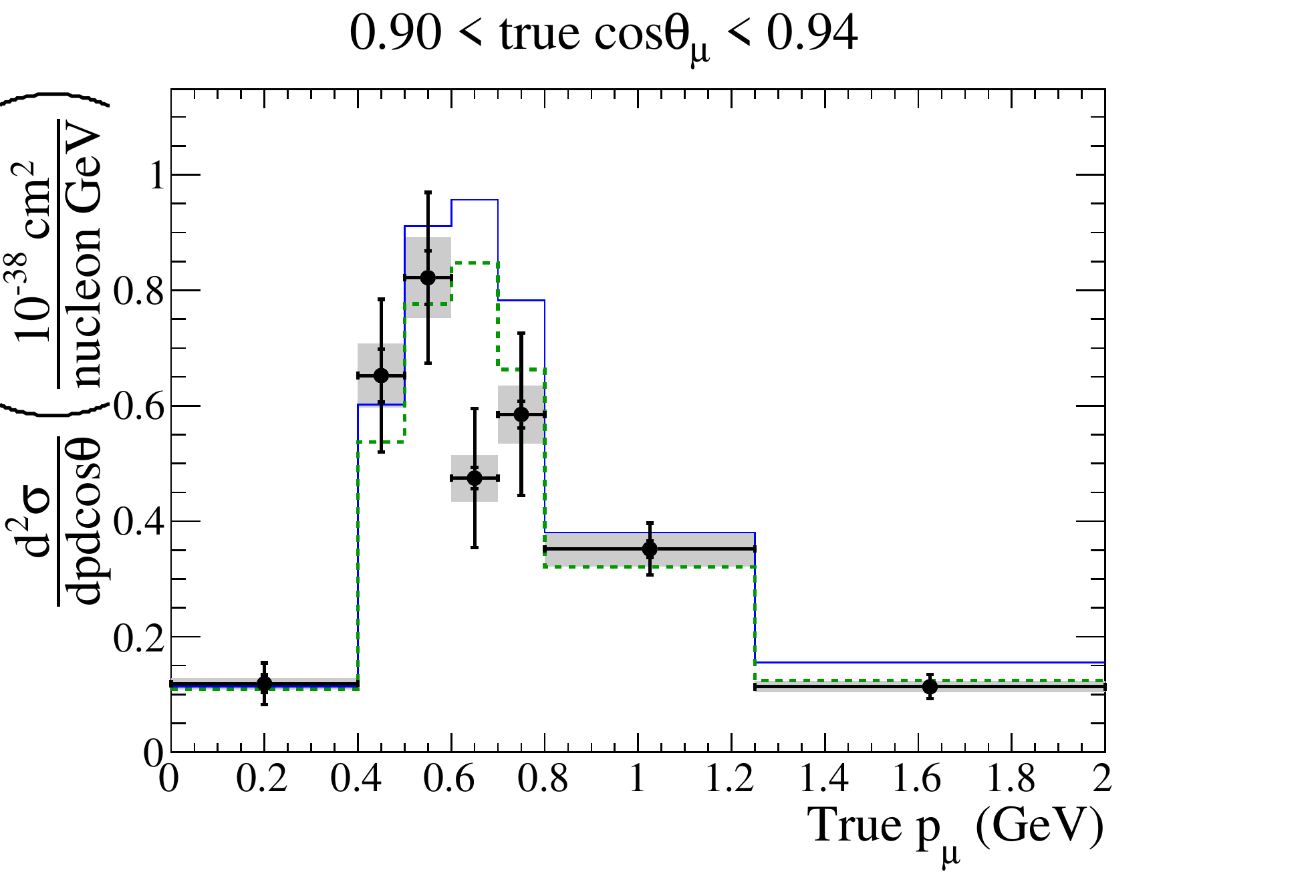}
 \includegraphics[width=6cm]{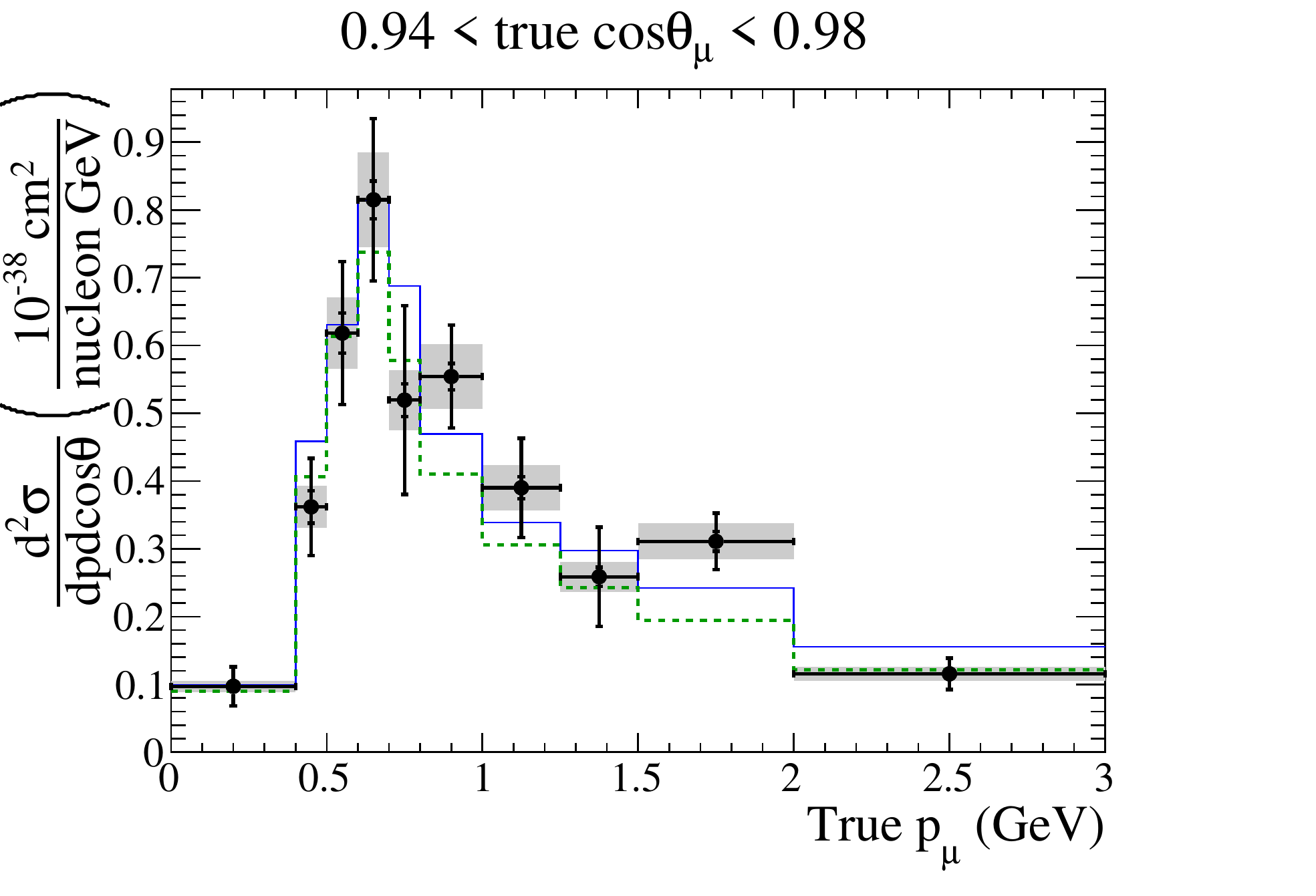}\\
 \includegraphics[width=6cm]{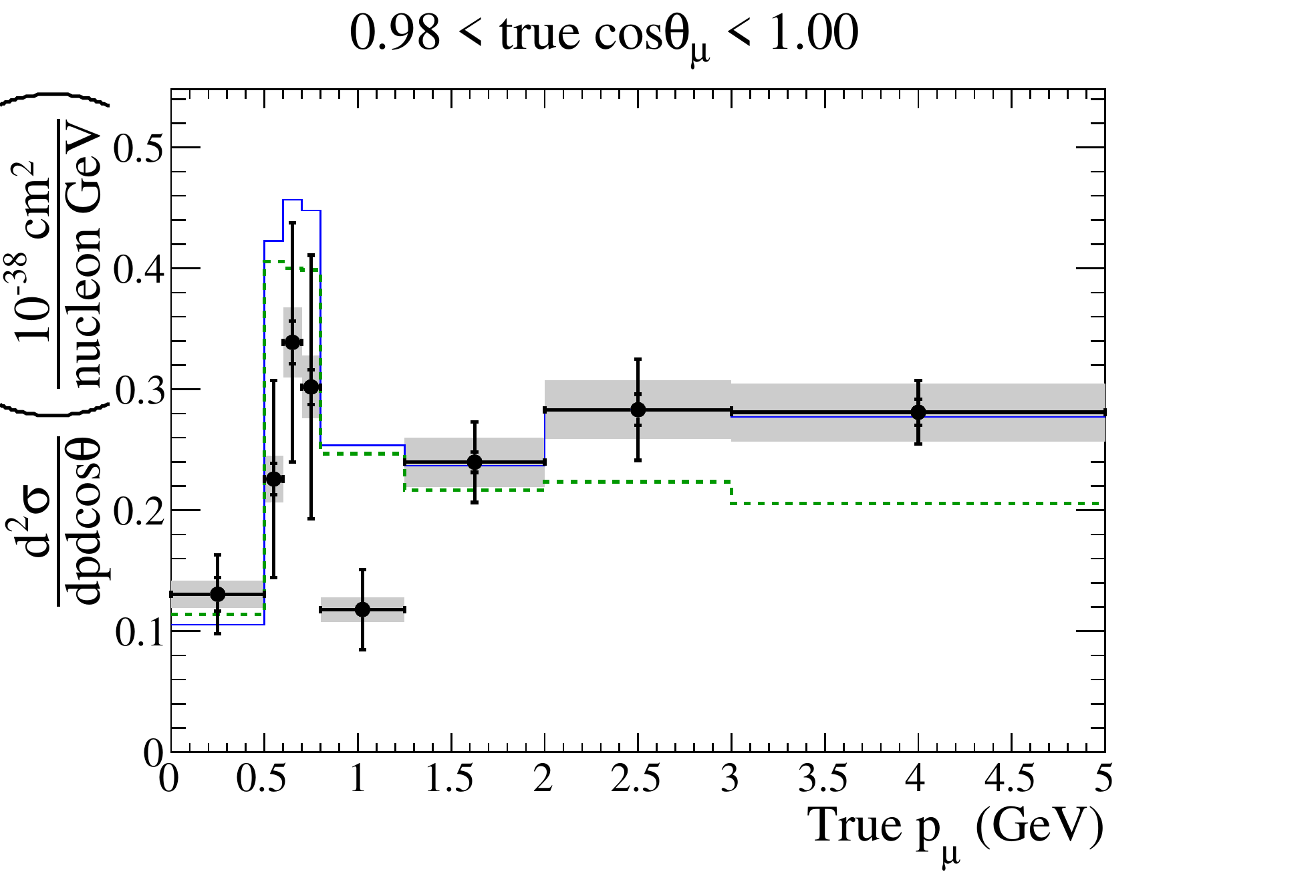}
\end{center}
\caption{Measured cross-section with shape uncertainties (error bars: internal systematics, external statistical) 
and fully correlated normalization uncertainty (gray band).
The results from the fit to the data are compared to predictions from NEUT (blue solid line), 
and from GENIE (green dashed line).}
\label{fig:xsecResultsLin}
\end{figure}

\begin{figure}
\begin{center}
 \includegraphics[width=6cm]{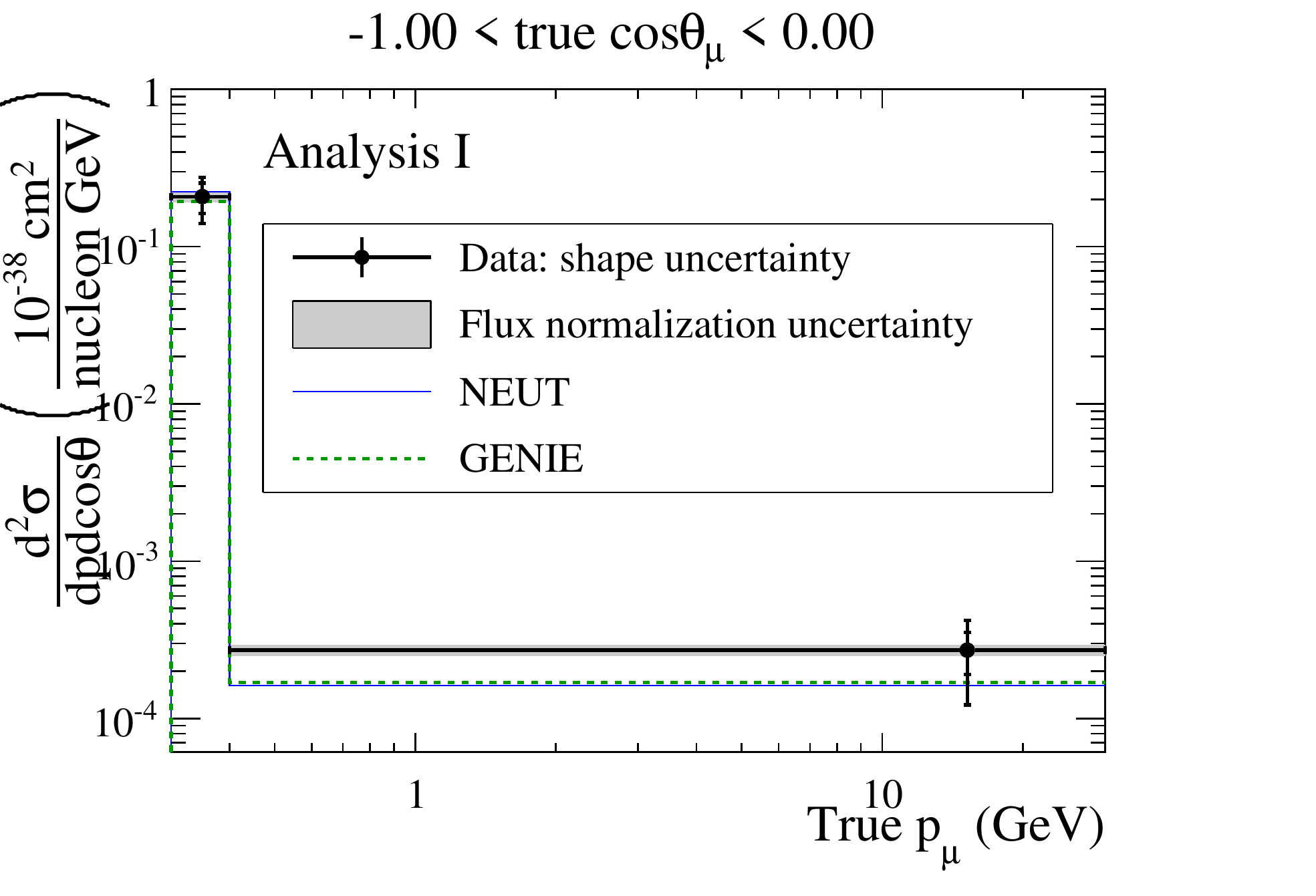}
 \includegraphics[width=6cm]{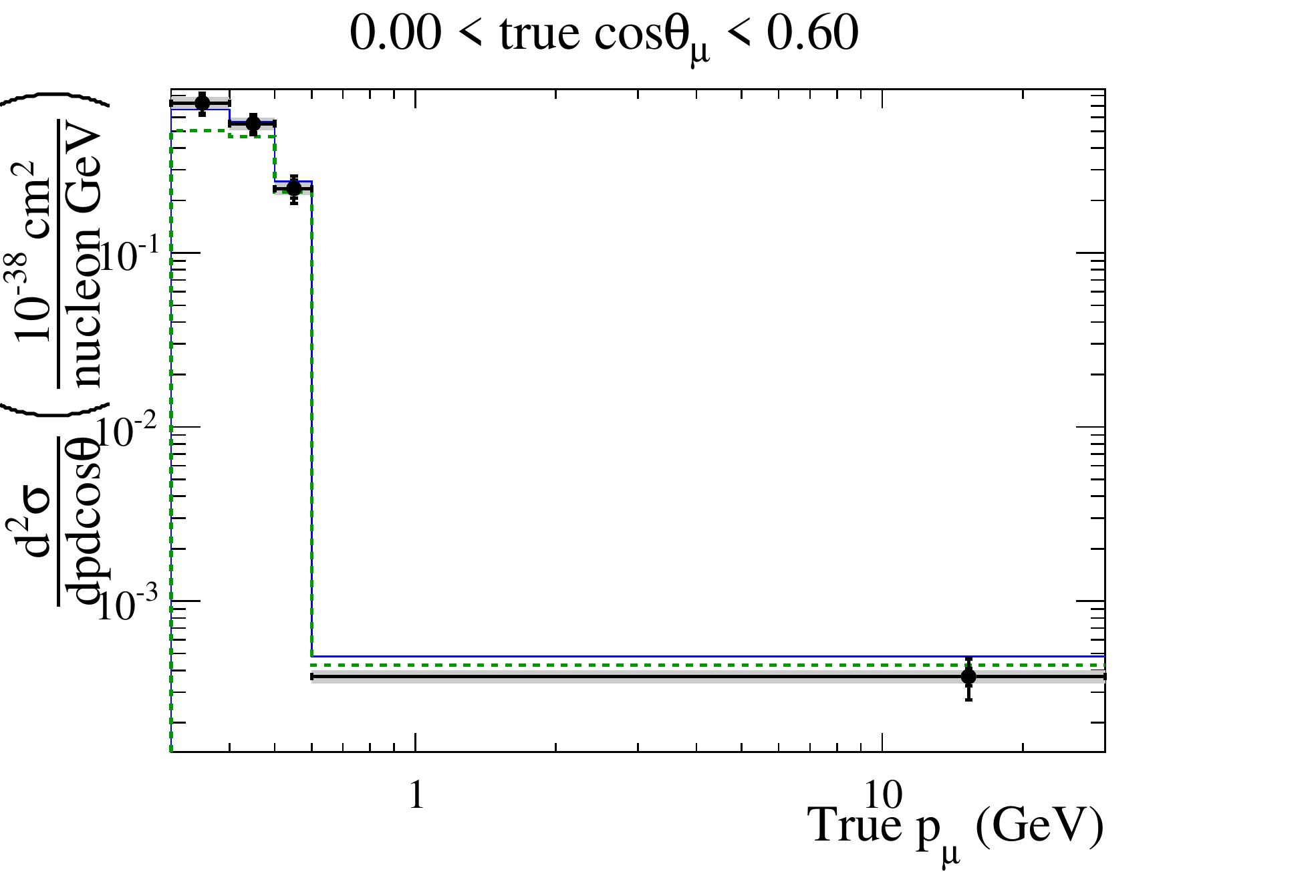}\\
 \includegraphics[width=6cm]{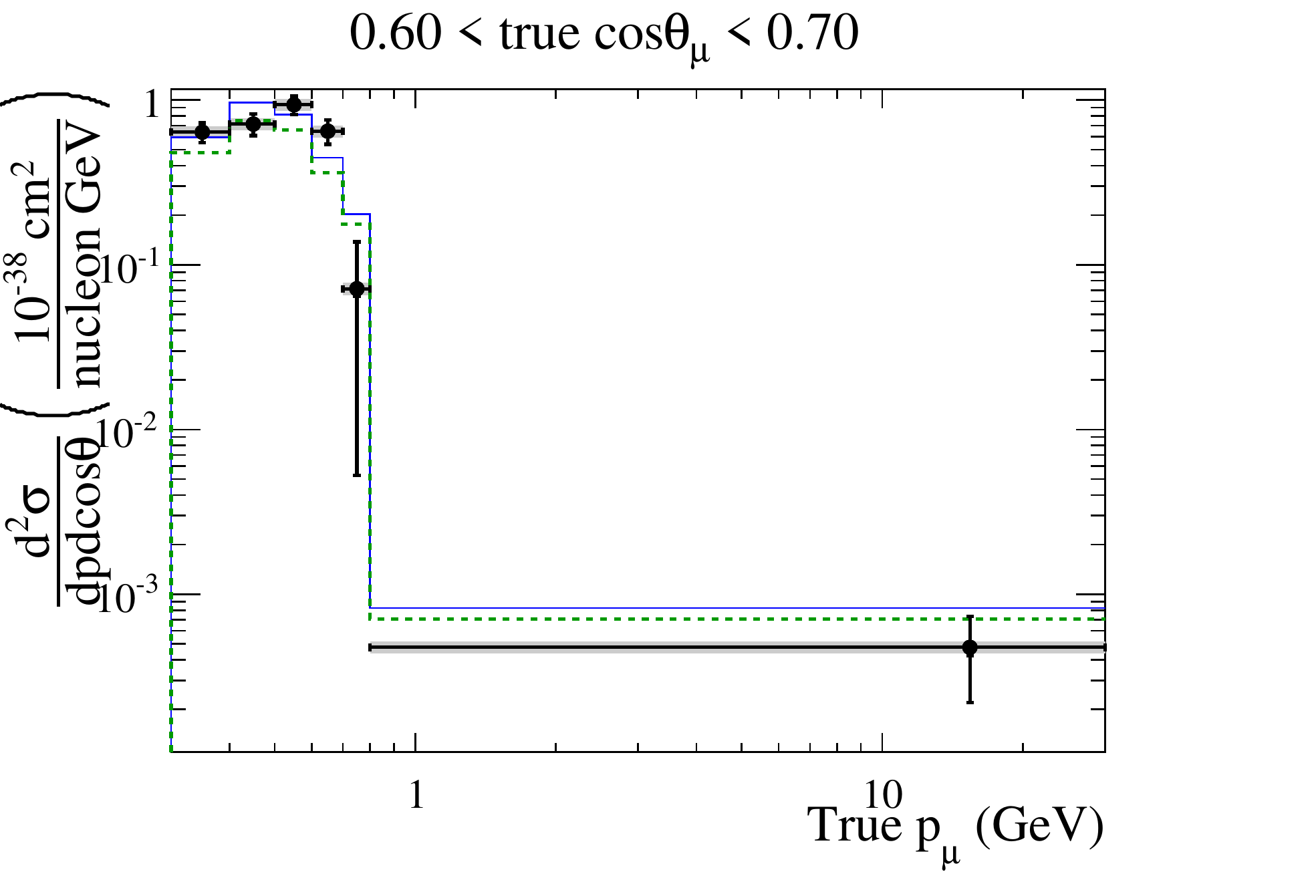}
 \includegraphics[width=6cm]{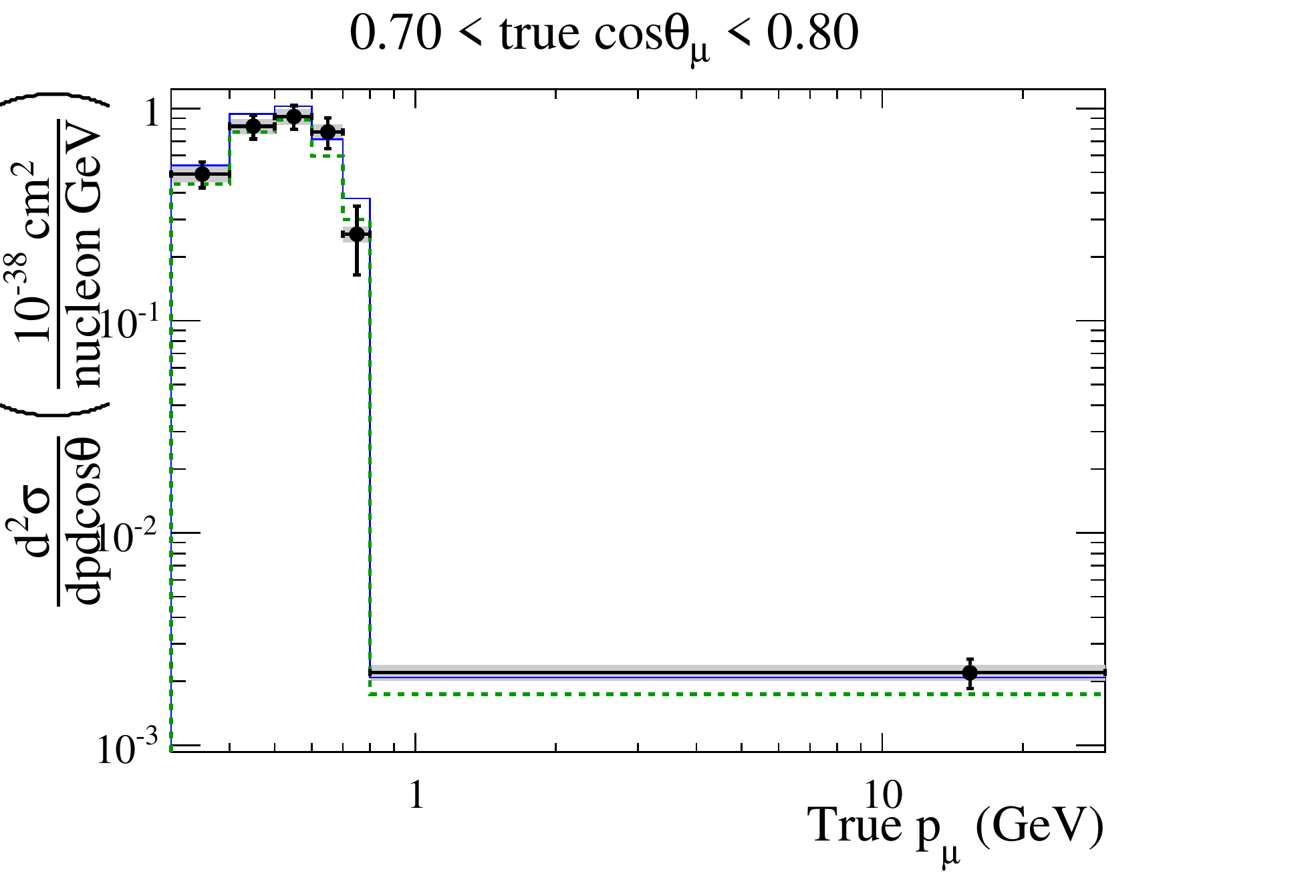}\\
 \includegraphics[width=6cm]{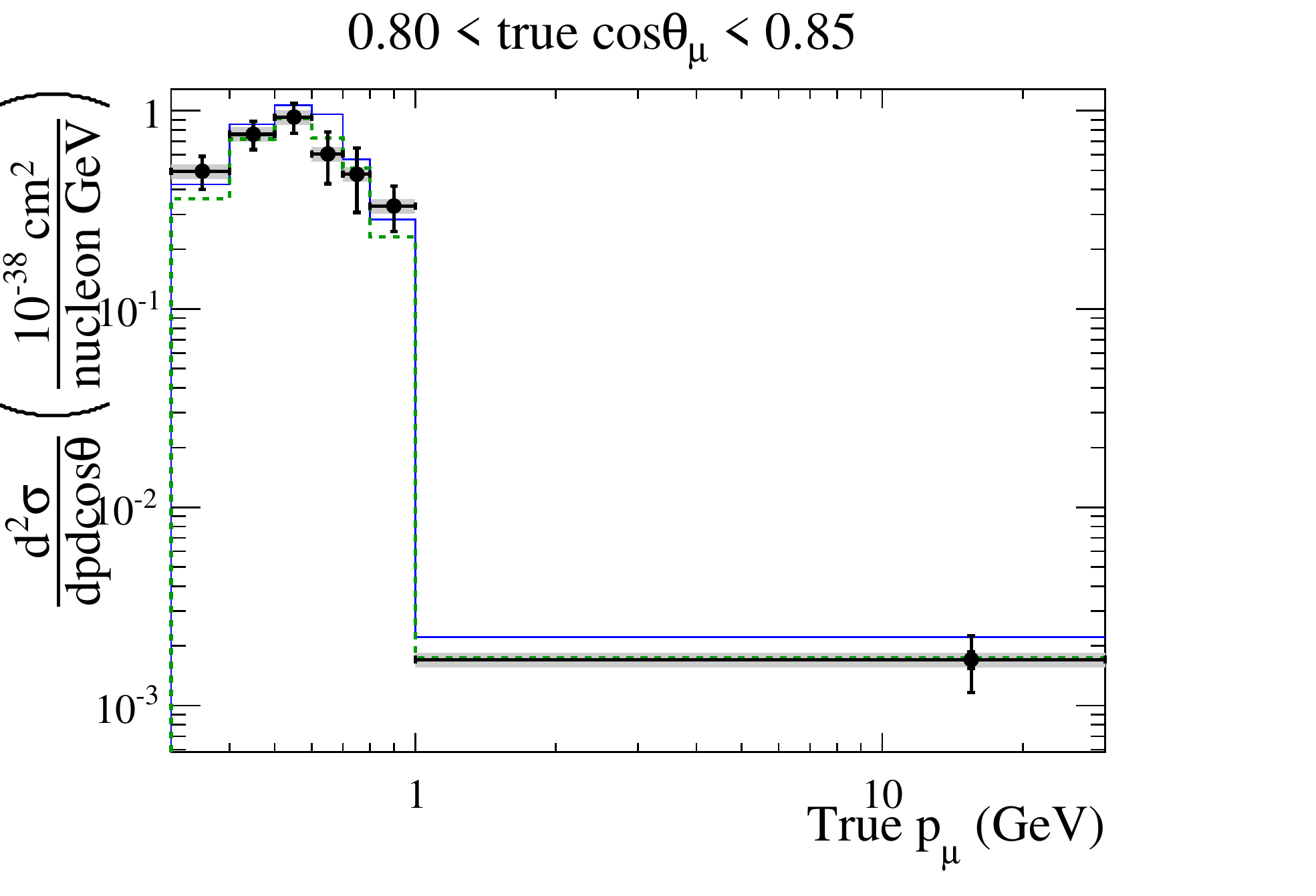}
 \includegraphics[width=6cm]{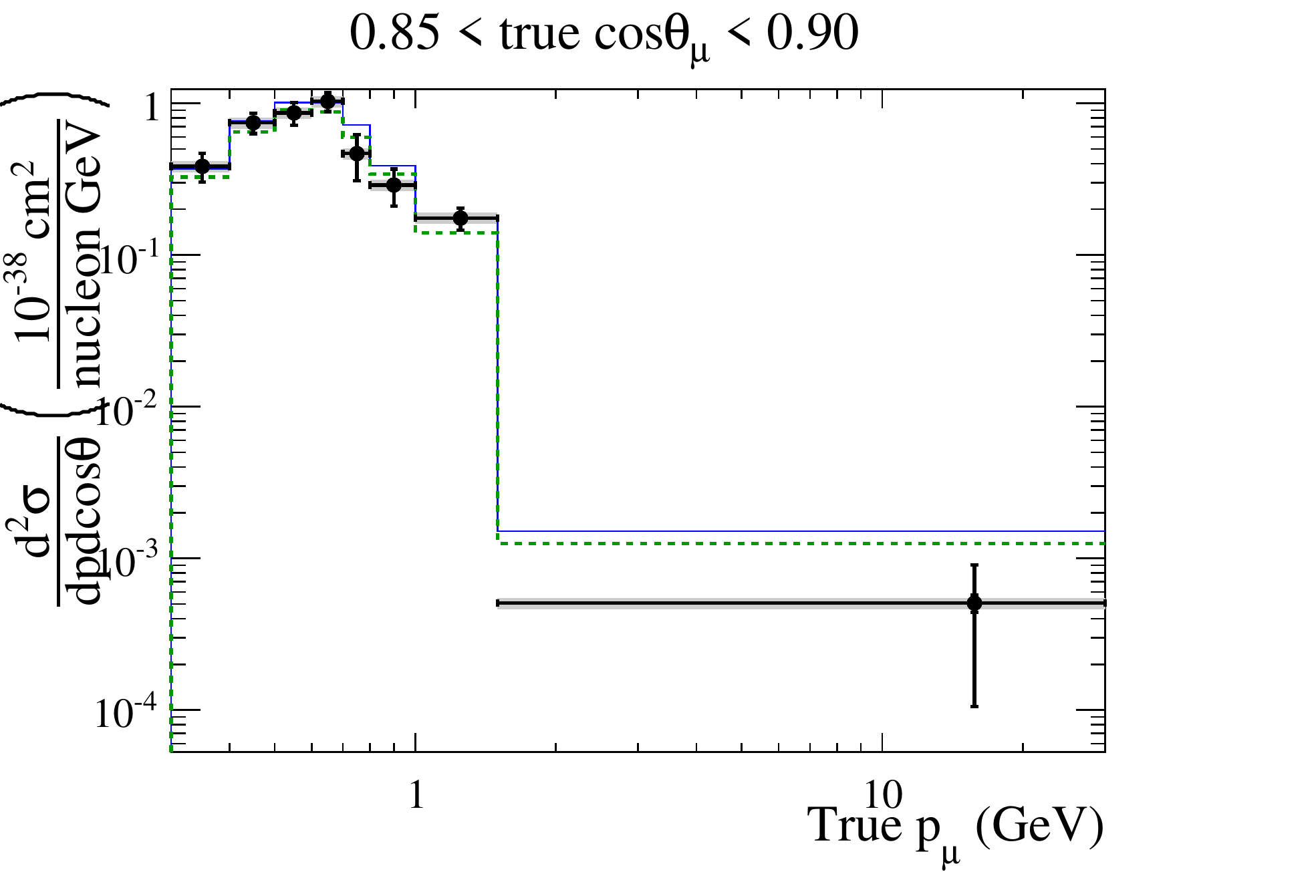}\\
 \includegraphics[width=6cm]{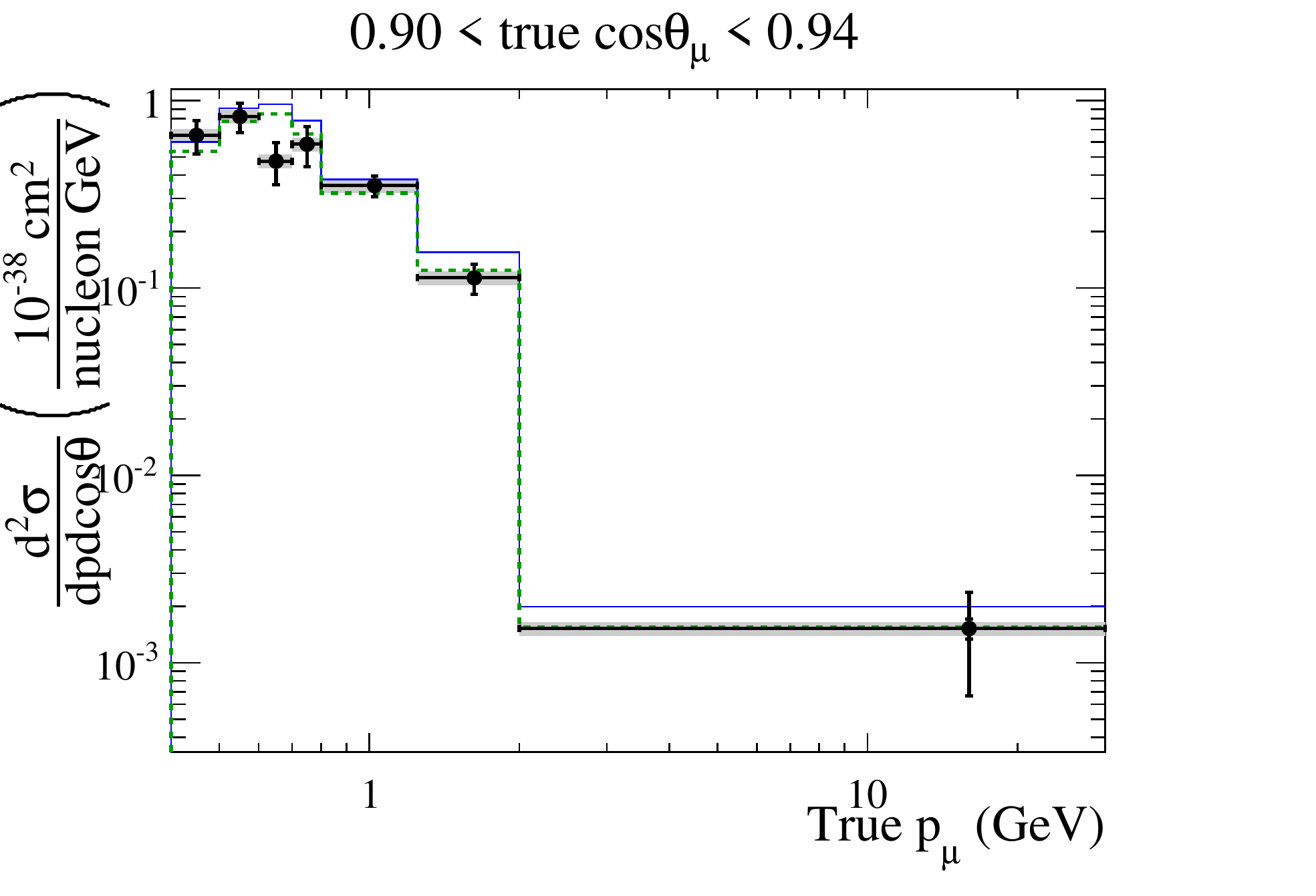}
 \includegraphics[width=6cm]{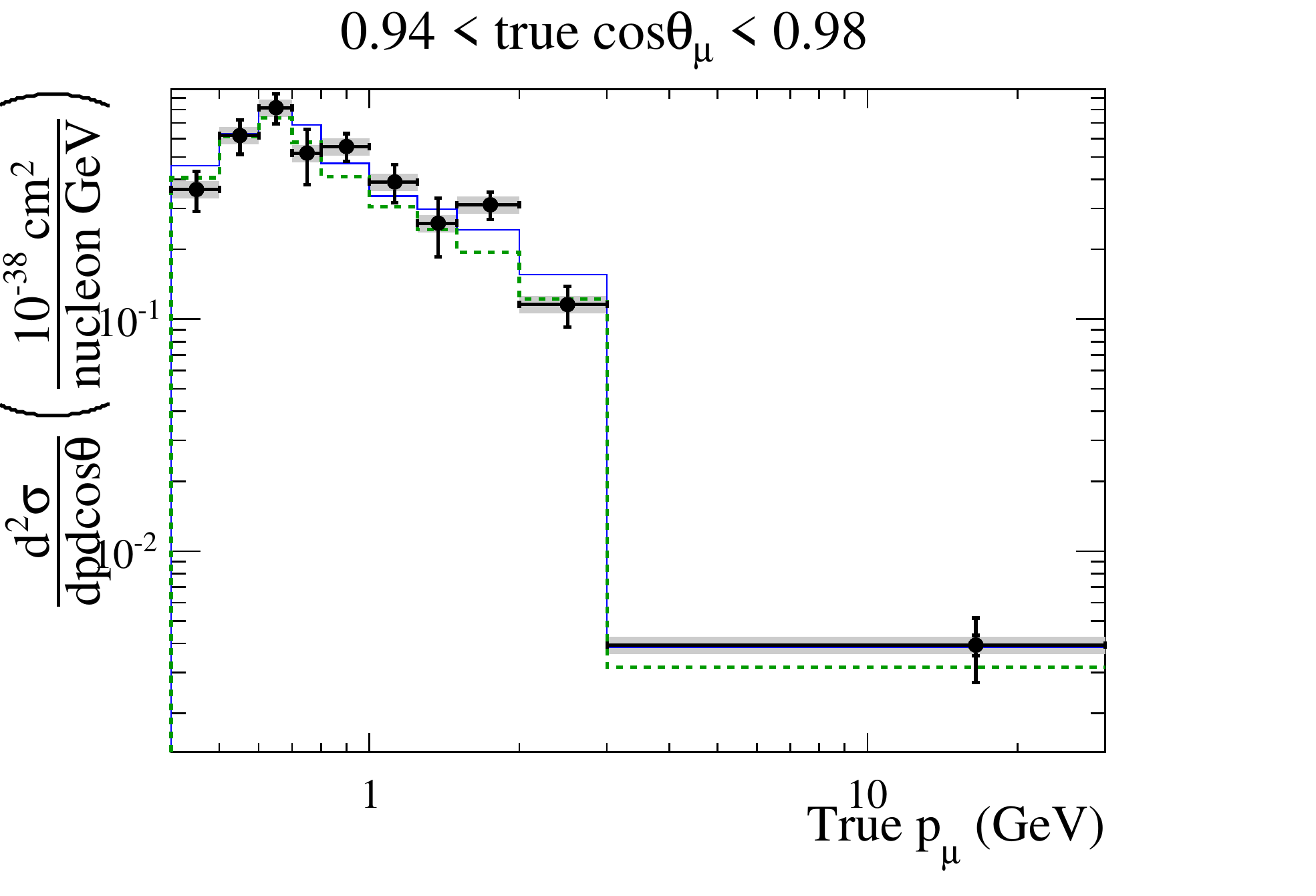}\\
 \includegraphics[width=6cm]{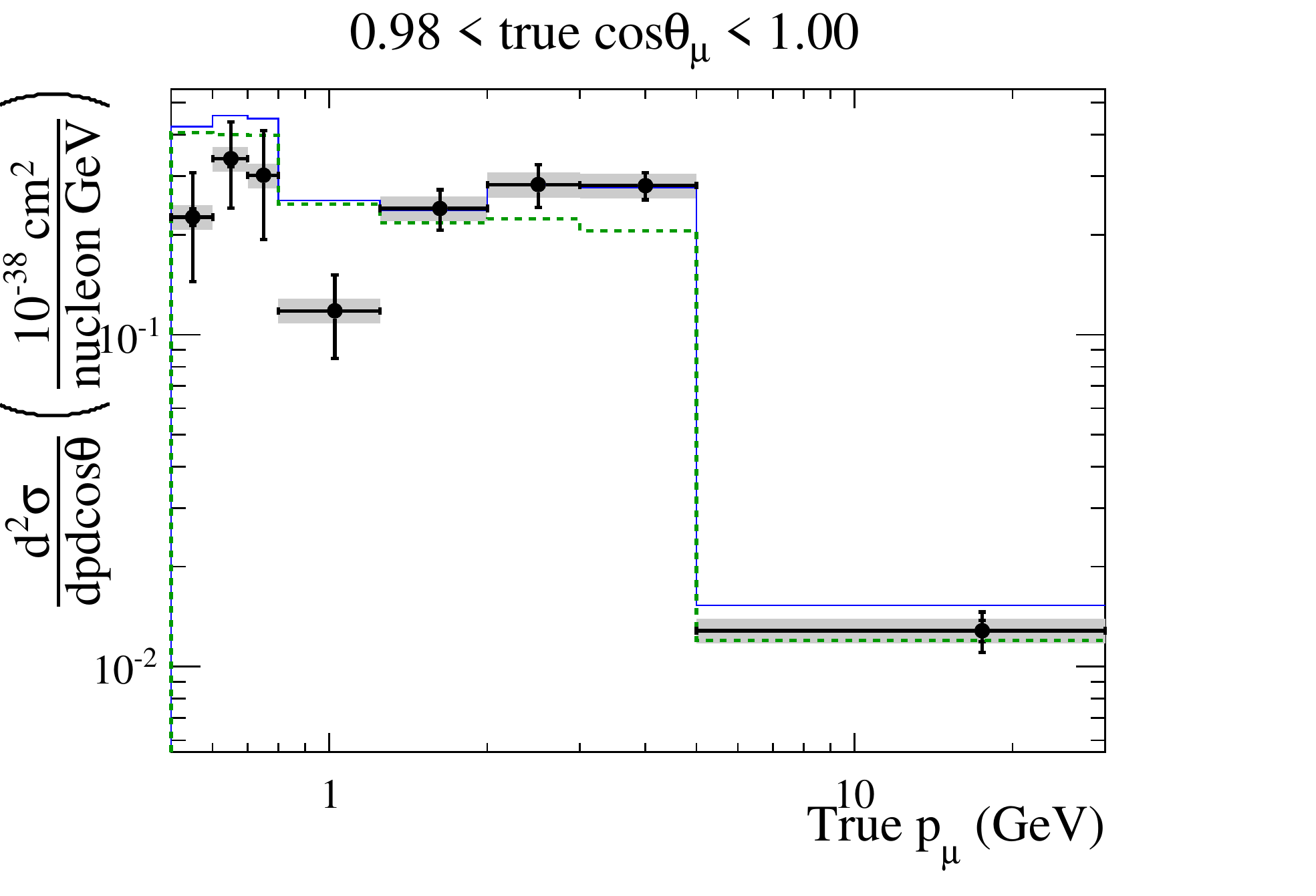}
\end{center}
\caption{Same as in Fig.~\ref{fig:xsecResultsLin}, but with a logarithmic scale.}
\label{fig:xsecResultsLog}
\end{figure}

The double-differential flux-integrated cross-section extracted from Analysis I
is compared to the Martini {\it et al}~\cite{Martini:2009,Martini:2010}
(Fig.~\ref{fig:xsecResultsLinMM}) and Nieves {\it et al}~\cite{Nieves:2012,Nieves:2012yz} 
(with a cut of $q_3<1.2$ GeV; Fig.~\ref{fig:xsecResultsLinN}) models with and without 2p2h contributions.
The internal error bars indicate the shape systematic uncertainties, and the external error bars indicate the quadratic sum of
the shape systematic uncertainties and the statistical uncertainties in the data.
The flux normalization uncertainty is indicated by the gray band.
The results are compared 
to NEUT and GENIE Monte Carlo in Fig.~\ref{fig:xsecResultsLin} (linear scale)
and Fig.~\ref{fig:xsecResultsLog} (logarithmic scale).

\section{Analysis II background comparisons}
\begin{figure}
\begin{center}
 \includegraphics[width=6cm]{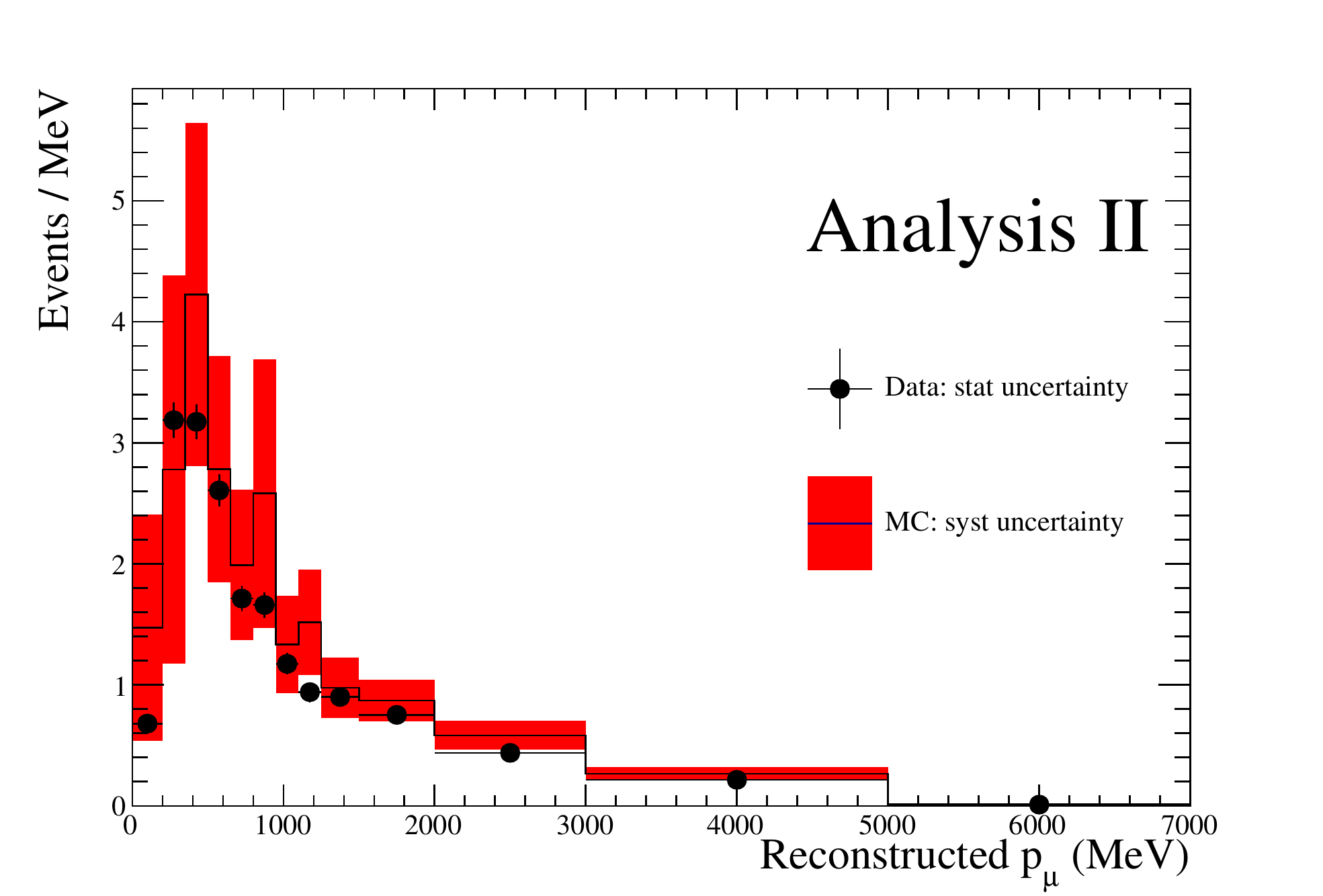}
 \includegraphics[width=6cm]{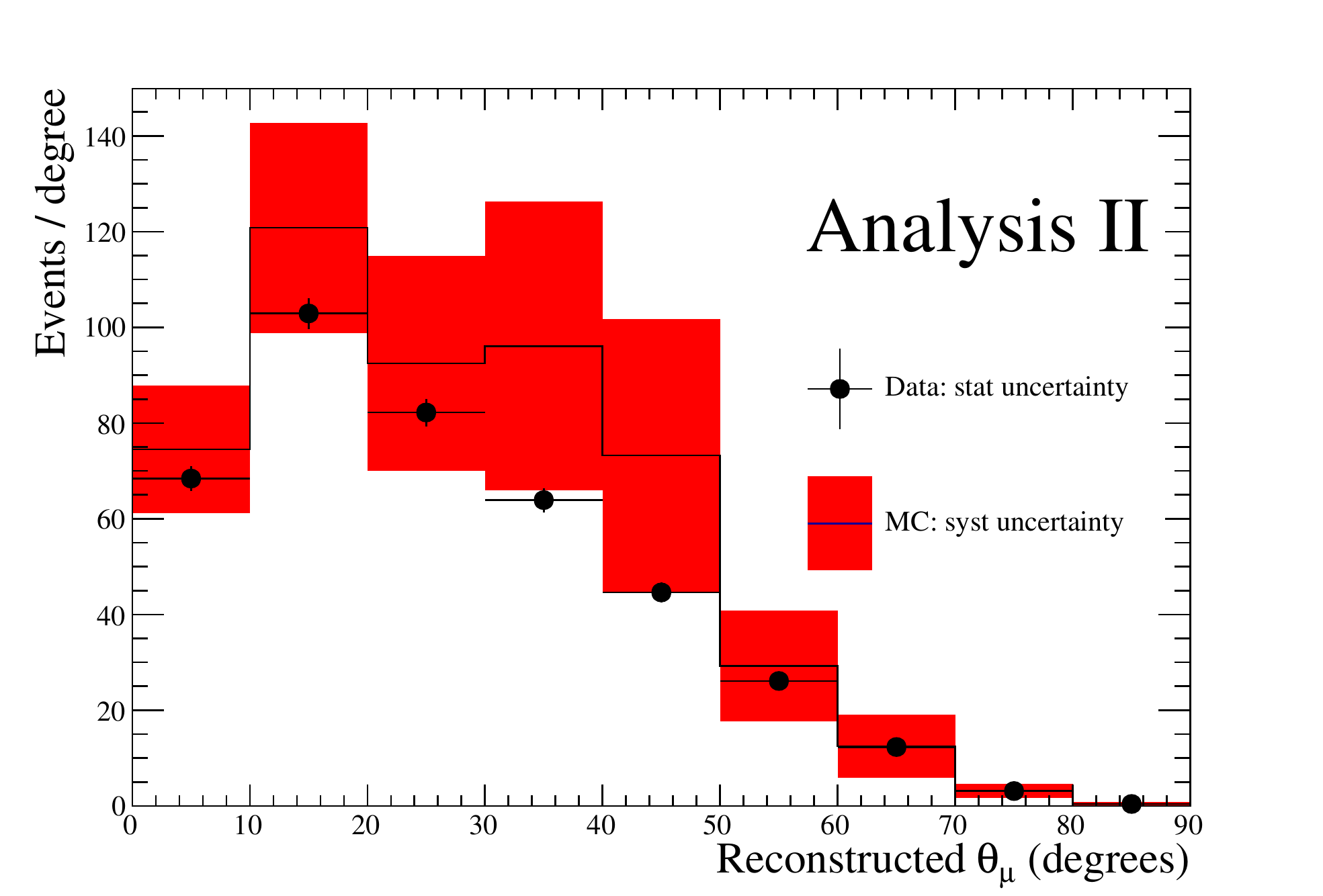}\\
 \includegraphics[width=6cm]{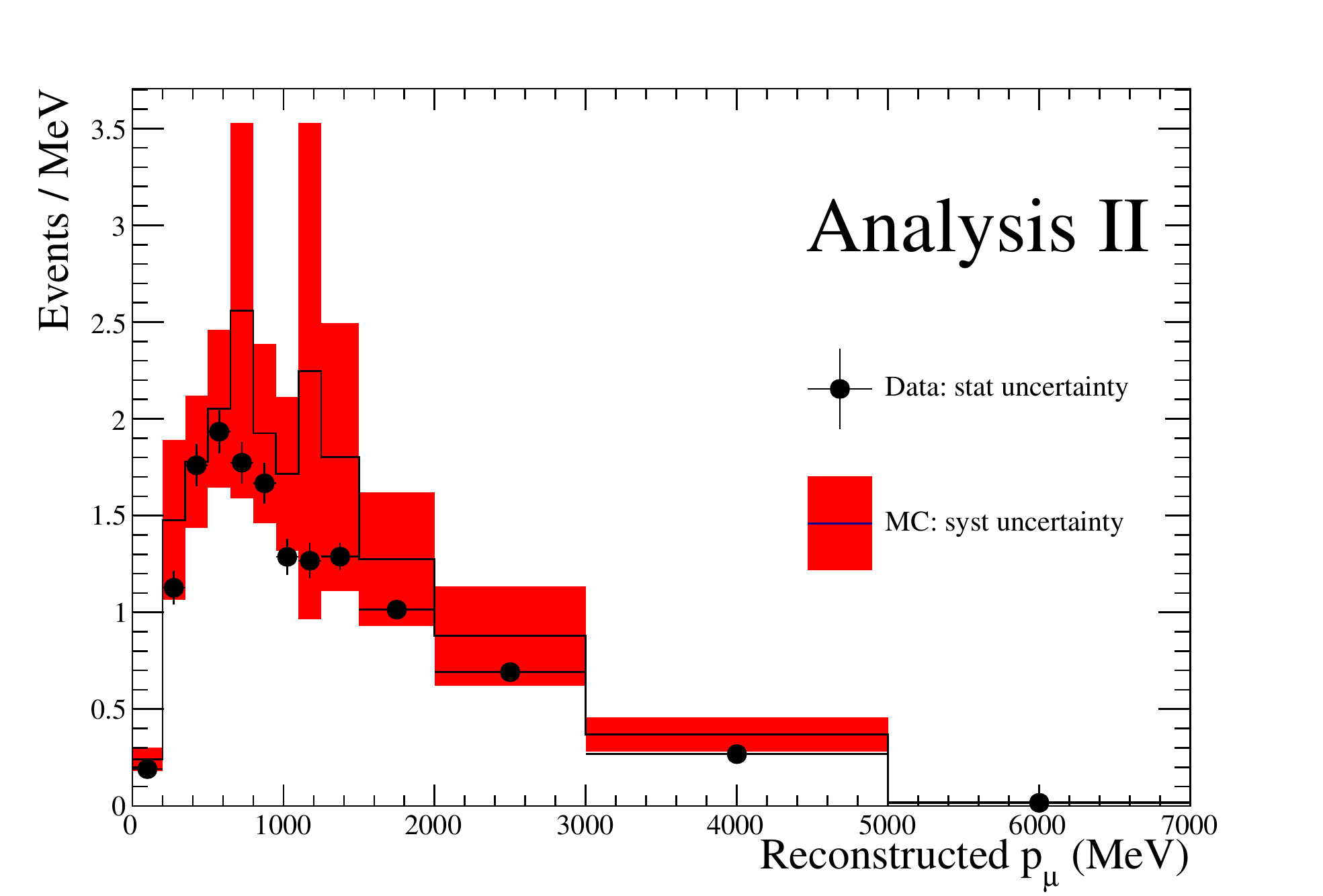}
 \includegraphics[width=6cm]{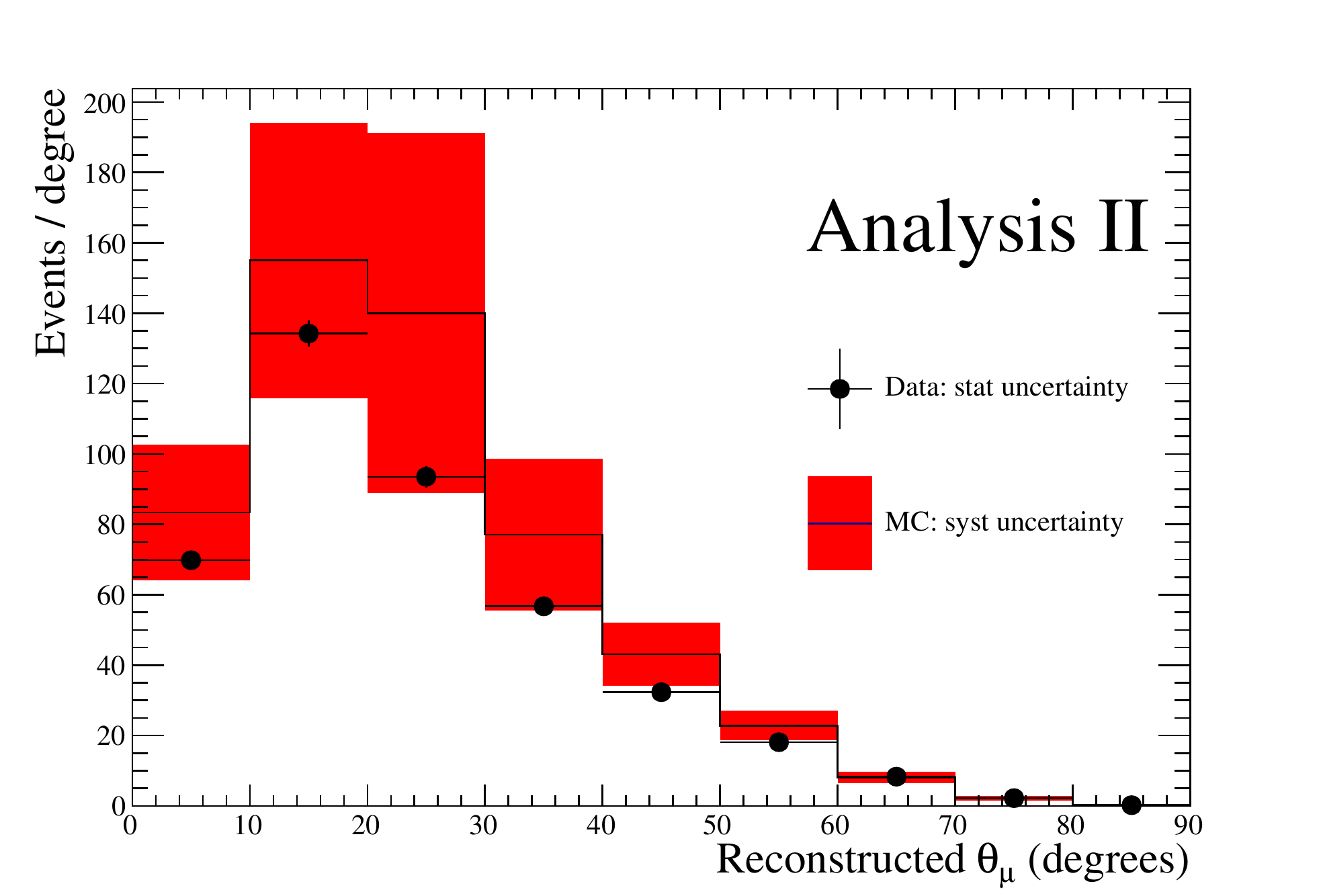}\\
\end{center}
\caption{Background comparison for a selection enhanced in CC1$\pi^+$ (top), and a selection enhanced in multi-pion and deep inelastic scattering (bottom). The data and MC agree within the uncertainties assigned to the MC.}
\label{fig:an2Background}
\end{figure}

Figure \ref{fig:an2Background} shows a comparison between data and MC for background-enhanced selections in Analysis II.
The selections are orthogonal to the signal selection, with one enhanced in CC1$\pi^+$ events, and the other enhanced in multi-pion and deep inelastic scattering events.

\bibliography{CC0pi2015}

\end{document}